\documentclass[fleqn,usenatbib]{mnras}

\usepackage{newtxtext,newtxmath}

\usepackage[T1]{fontenc}

\DeclareRobustCommand{\VAN}[3]{#2}
\let\VANthebibliography\thebibliography
\def\thebibliography{\DeclareRobustCommand{\VAN}[3]{##3}\VANthebibliography}

\usepackage{graphicx}	
\usepackage{amsmath}	
\usepackage{xcolor}

\title[EDisCS H$\alpha$ Star Formation Rates]{H$\alpha$-based Star Formation Rates in and around z $\sim$ 0.5 EDisCS clusters}

\author[J.R. Cooper et al.]{
Jennifer R. Cooper,$^{1}$\thanks{E-mail: jrc323@cornell.edu}
Gregory H. Rudnick,$^{1}$\thanks{E-mail: grudnick@ku.edu}
Gabriel G. Brammer,$^{2,3}$
Tyler Desjardins,$^{4}$
\newauthor
Justin L. Mann,$^{1}$
Benjamin J. Weiner,$^{5}$
Alfonso Arag\'{o}n-Salamanca,$^{6}$
Gabriella De Lucia,$^{7}$
\newauthor
Vandana Desai,$^{8}$
Rose A. Finn,$^{9}$
Pascale Jablonka,$^{10,11}$
Yara L. Jaff\'{e},$^{12}$
John Moustakas,$^{9}$
\newauthor
Damien Sp\'{e}rone-Longin,$^{10}$
Harry I. Teplitz,$^{8}$
Benedetta Vulcani,$^{13}$
and Dennis Zaritsky$^{5}$
\\
$^{1}$Department of Physics and Astronomy, The University of Kansas, 1251 Wescoe Hall Drive, Lawrence, KS 66045, USA\\
$^{2}$Niels Bohr Institute, University of Copenhagen, Jagtvej 128, København N, DK-2200, Denmark\\
$^{3}$Cosmic Dawn Center (DAWN)\\
$^{4}$Space Telescope Science Institute, 3700 San Martin Dr, Baltimore, MD 21218\\
$^{5}$Steward Observatory, University of Arizona, 933 North Cherry Avenue, Tucson, AZ 85721, USA\\
$^{6}$School of Physics and Astronomy, University of Nottingham, University Park, Nottingham NG7 2RD, UK\\
$^{7}$INAF - Osservatorio Astronomico di Trieste, via G. B. Tiepolo 11, I-34143 Trieste, Italy\\
$^{8}$Infrared Processing and Analysis Center, California Institute of Technology, MS 100-22, Pasadena, CA 91125, USA\\
$^{9}$Siena College, 515 Loudon Rd., Loudonville, NY 12211\\
$^{10}$Institute of Physics, Laboratory of Astrophysics, \'{E}cole Polytechnique F\'{e}d\'{e}rale de Lausanne (EPFL), 1290 Sauverny, Switzerland\\
$^{11}$ GEPI, Observatoire de Paris, Universit\'{e} PSL, aNRS, Place Jules Janssen, F-92190 Meudon, France\\
$^{12}$Instituto de F\'{i}sica y Astronom\'{i}a, Universidad de Valpara\'{i}so, Avda. Gran Breta\~{n}a 1111 Valapra\'{i}so, Chile\\
$^{13}$INAF - Osservatorio Astronomico di Padova, Vicolo Osservatorio 5, IT-35122 Padova, Italy\\
}

\date{Accepted 2021 October 26. Received 2021 October 21; in original form 2021 March 1.}

\pubyear{2021}

\begin{document}
\label{firstpage}
\pagerange{\pageref{firstpage}--\pageref{lastpage}}
\maketitle

\begin{abstract}
We investigate the role of environment on star-formation rates of galaxies at various cosmic densities in well-studied clusters. We present the star-forming main sequence for 163 galaxies in four EDisCS clusters in the range 0.4 $<$ z $<$ 0.7. We use {\em Hubble Space Telescope}/Wide Field Camera 3 observations of the H$\alpha$ emission line to span three distinct local environments: the cluster core, infall region, and external field galaxies. The main sequence defined from our observations is consistent with other published H$\alpha$ distributions at similar redshifts, but differs from those derived from star-formation tracers such as 24$\mu$m. We find that the H$\alpha$-derived star-formation rates for the 67 galaxies with stellar masses greater than the mass-completeness limit of M$_*>$ 10$^{9.75}$M\textsubscript{\(\odot\)} show little dependence on environment. At face value, the similarities in the star-formation rate distributions in the three environments may indicate that the process of finally shutting down star formation is rapid, however, the depth of our data and size of our sample make it difficult to conclusively test this scenario. Despite having significant H$\alpha$ emission, 21 galaxies are classified as {\em UVJ}-quiescent and may represent a demonstration of the quenching of star formation caught in the act.

\end{abstract}




\section{Introduction}
Star formation governs the conversion of a galaxy's gas into stars and is characterized by the balance of cold gas accretion and feedback \citep{dutton,bouche}. Global star formation peaked at z $\sim$ 2 and has been in a rapid decline to the present day \citep{md,bouwens15,cucc12}. Across all redshifts and masses, a tight correlation illustrates that more massive star-forming galaxies are forming stars at a quicker rate than those at lower stellar masses; this relation is referred to as the star-forming main sequence \citep{noeske,peng10,whitaker12}. The normalization of the star-forming main sequence evolves with time, where galaxies at z $\sim$ 2 and $z\sim$ 1 have a main sequence that is 20x \citep{daddi07} and 7x higher \citep{elbaz07}, respectively, than at z $\sim$ 0 \citep{brinch07}. This decline in overall star formation to the present day poses many questions surrounding the nature and fate of the universe.
\par 
While the main sequence is generally presented with a slope ranging from 0.2 -- 1.2 \citep{speagle14}, there have been numerous studies that show that bulge-dominated massive galaxies contribute towards a flattening in the star-formation rate (SFR) at higher masses \citep{karim11,whitaker12,whitaker14,schr15,erf16}. This `internal' quenching mechanism is directly related to the morphology of the galaxy and results in a less efficient conversion of gas to stars \citep{martig09}. Processes that suppress star formation, such an expulsion of the gas through feedback, a cutoff in gas accretion \citep{larson80}, or the removal of the gas via ram-pressure stripping \citep{quilis2000} can also cause a deviation to SFRs lower than the main sequence. Additionally, some ram-pressure stripping and merger events have been observed to first create enhanced SFR activity, followed by a suppression phase \citep{pogg16,jaffe16,vul18}. The overall observed scatter in the main sequence is likely due to varying star formation histories of each galaxy \citep{hopkins14,dom14}, where  this scatter is consistent across stellar mass and redshift at $\sim$ 0.3 dex \citep{whitaker12,tacch16}.
\par 
Further exploration of the SFR -- stellar mass relation is expanded by investigating distinct cosmic environments. The densest regions of the universe consist of galaxy clusters with thousands of members that are gravitationally bound and contain a hot intracluster medium (ICM). Many of the most massive galaxies reside in the cluster cores, with the brightest cluster galaxy (BCG) generally being at the minimum or center of the cluster potential well and elliptical in shape. However, cluster membership extends far beyond the virial radius and encompasses the infall region where galaxies are initially accreted into the cluster environment. This infall region of galaxy clusters has the potential to host the sites of galaxy transformation and quenching processes {\em in situ} that may differ from those in the core. \cite{Just_2019} found that 30 - 70\% of the galaxies in local clusters were located in the infall region at z $\sim$ 0.6, meaning that these galaxies may become the majority of cluster galaxies at z $\sim$ 0. This finding reinforces the importance of the cluster infall region with respect to the environmentally-driven transformation of galaxies.
\par 
There have been numerous studies of star formation in clusters, conducted using various emission lines. Studies which consider the galaxy population as a whole see a clear suppression of star formation in dense environments \citep[e.g.][]{bal97,kauff04, patel09}. It is now recognized that this difference is primarily driven by the higher fraction of passive galaxies in clusters compared to the field. When considering only star-forming galaxies, studies yield conflicting pictures as to the effect of galaxy environment on the SFR of star-forming galaxies. For example, various authors find no difference in the SFRs of star-forming galaxies in low and high density environments \citep[e.g.][]{pogg08,peng10,koyama13,tiley20}. However, other studies find evidence for a suppression of star formation in cluster galaxies relative to the field, mostly manifested in a tail to low SFRs \citep{Wolf09,finn10,vul10,pacc16,old19}. These studies indicate that the cluster environment is indeed suppressing SFRs of star-forming galaxies, though perhaps only for a subset of the population. The apparent contradiction between these studies is somewhat difficult to reconcile for multiple reasons. The studies use various tracers, have different sensitivities to low SFR, and do not all probe the same dynamic range in density. However, most modern studies that examine clusters and are sensitive to low SFRs do find an excess population in clusters with suppressed star formation \citep[e.g.][]{pacc16}. These results indicate that star formation is being quenched in clusters.  The timescale needed to quench star formation is highly dependent on the exact distribution of SFRs below the main sequence.  For example, a lack of galaxies below the main sequence would argue for a fast quenching timescale ($<1{\rm~Gyr}$). This is necessary to avoid a substantial population of galaxies with significantly reduced, but non-zero SFRs.  

In contrast to this fast timescale, studies which model the buildup of quiescent galaxies in clusters over time require a significantly longer timescale between when galaxies cross the virial radius and when they quench, on the order of five Gyr at $z\sim0$ and shorter at higher redshift \citep{mcgee11,DeLucia12,muzzin14,taranu14,haines15,foss17}. A proposal for reconciling these different timescales is one in which galaxies follow a delayed-then-rapid quenching process as they fall into a more massive halo \citep{wetzel13}. In this picture, galaxy SFRs are unaffected for the first two -- four Gyr \citep{wetzel13}, followed by a rapid quenching period. This proposal has been remarkably successful at explaining both the evolution in the quenched fraction and the distribution of galaxy SFRs. Despite this success, the physical processes acting during the `delay' phase, and the process responsible for the ultimate `rapid' quenching remain ambiguous. Indeed, some recent studies indicate that this phase is one in which the spatial extent of the star formation within star-forming cluster galaxies is being slowly reduced, thus indicating that the `delay' phase is really a slow quenching phase \citep{Finn18}.

Making progress in our understanding of galaxy quenching in dense environments requires studies that probe the distribution of SFRs for star-forming galaxies to low levels of SFR and over a large dynamic range in densities and with a single tracer. It is also important that studies extend beyond the local universe, as quenching timescales evolved to longer times at lower redshift \citep{Balogh16,foltz18}. To probe the full evolution of galaxy SFRs as galaxies fall into clusters, a final ingredient is that studies probe beyond the virial radius into the infall regions, as that is where environmental transformation may first occur \citep{lewis02,gomez03}. H$\alpha$ is an excellent tracer of star formation as it is less susceptible to extinction or metallicity than other optical emission lines, such as   [\ion{O}{ii}] \citep[e.g.][]{Moustakas06}, and because it probes the instantaneous SFR \citep{Kennicutt98}. There have been a small number of wide-field H$\alpha$ studies of clusters beyond the local universe \citep{kodama04,koyama11,sobral11}. These have focused on very massive clusters and have included only one cluster per study. However, they do not present a consistent picture of the effect of the infall region. For example, \cite{koyama11} find that the fraction of H$\alpha$ emitters with red colors peaks in groups, but is elevated in groups in the infall region with respect to the core. On the other hand, \cite{sobral11} finds that the SFR of H$\alpha$ emitters climbs significantly from low to intermediate densities, but declines again at the highest densities that correspond to cluster cores. However, \cite{sobral11} also find that this boosting of the SFR is dominated by galaxies with stellar masses lower than $10^{10.6}M_\odot$. These varied results highlight the need for studies of the SFR in the infall regions of multiple clusters with the same tracer and survey selection. Having larger samples of clusters is especially important given the significant cluster-to-cluster variation in galaxy properties \citep[e.g.][]{pogg06,moran07,patel11,oemler13}

At z $\gtrsim$ 0.5, the H$\alpha$ line is located at $\lambda_{obs}>1.0 \mu$m which is difficult to observe from the ground. The {\em Hubble Space Telescope} ({\em HST}) provides access to H$\alpha$ through slitless spectroscopy using The Wide Field Camera 3 (WFC3). The 3DHST survey \citep{vandokkum11,mom16} demonstrated the power of this mode by observing more than 100,000 galaxies in the CANDELS fields. The grism spectra, coupled with broad-band imaging, allowed the 3DHST team to produce and release robust redshifts, emission line fluxes, and 2D emission line spatial maps. This showcased the power of the grism and led to numerous publications regarding sizes \citep{nelson12,van14}, the main sequence \citep{whitaker12,whitaker14} and assembly of galaxies \citep{van13a,barro14,lang14}. The success of this study led to other surveys utilizing the same combination of observation modes such as the Grism Lens-Amplified Survey from Space (GLASS; \citealt{treu15}), which was able to observe galaxies in varying cosmic environments \citep{vul15,vul16,vul17,abra18}. GLASS and other surveys \citep{Lotz13,LeeBrown17} demonstrated the abilities of the {\em HST} grism even in crowded cosmic regions.

\par
This study aims to characterize the distribution of SFRs in galaxies in the infall and core regions of four $z\sim 0.5$ clusters, and to compare them to a consistently measured field sample. We seek to quantify how SFRs are affected during a galaxy's journey into the cluster environment. This paper is organized as follows. In \S~\ref{sec:meth}, we describe the sample properties, observations and reduction methodologies. In \S~\ref{sec:results} and \S~\ref{sec:dis}, we present the SFR -- stellar mass results and comparison to the literature. In \S~\ref{sec:fut} we discuss future work and analysis possible with this dataset. The virial radius (R$_{200}$) is defined as the radius of the enclosed circle that has a density $\rho$ 200x that of the critical density $\rho_c$ of the Universe at a given redshift. All magnitudes are given in the AB system, and we assume a Chabrier IMF \citep{chabimf}. We adopt a $\Lambda$CDM cosmology with $\Omega_m$ = 0.307, $\Omega_{\Lambda}$ = 0.693, and  H$_{0}$ = 67.7 km s$^{-1}$ Mpc$^{-1}$ \citep{planck}.

\section{Methodology \& Data}
\label{sec:meth}
\subsection{ESO Distant Cluster Survey}
\label{sec:eso} 

\begin{table*}

 \begin{center}
 \begin{tabular}[t]{ cccccccc }
 \hline
 Cluster ID & RA & Dec  & z$_{spec}$ & R$_{200}$  &  R$_{infall}$ & M$_{200}$ & $\sigma_v$ \\
  & (hours) & (degrees) & & (Mpc) & (Mpc) &  (10$^{14}$ M\textsubscript{\(\odot\)}) & (km s$^{-1}$) \\
  (1)       &   (2)     & (3)       & (4)       & (5)             & (6)    & (7)    & (8)\\
 \hline
 Cl1059.2-1253   & 10:59:07.1    & -12:53:15 &   0.4564 &  $0.99\substack{+0.10 \\ -0.11}$\ & 3.19 & $1.78\substack{+0.60 \\ -0.53}$\ & $510\substack{+52 \\ -56}$\ \\
 Cl1138.2-1133   &   11:38:10.3  & -11:33:38   & 0.4796 & $1.40\substack{+0.14 \\ -0.15}$\ & 4.62 &  $5.20\substack{+1.69 \\ -1.46}$\ & $732\substack{+72 \\ -76}$\ \\
 Cl1227.9-1138 & 12:27:58.9 & -11:35:13 &  0.6357 & $1.00\substack{+0.13 \\ -0.13}$\ & 3.76 & $2.29\substack{+0.97 \\ -0.78}$\ & $574\substack{+72 \\ -75}$\ \\
Cl1301.7-1139 & 13:01:40.1 & -11:39:23 & 0.4828 & $1.31\substack{+0.16 \\ -0.16}$\ & 4.34 & $4.29\substack{+1.73 \\ -1.42}$\ & $687\substack{+82 \\ -89}$\ \\

 \hline
 \end{tabular} \par

 \end{center}

\caption{Parameters for each of the clusters in this study from \protect\cite{Just_2019}. 1. EDisCS Cluster ID 2. Right ascension in hours 3. Declination in degrees 4. Cluster redshift 5. Virial radius in Mpc 6. Infall radius in Mpc 7. Virial mass 8. Velocity dispersion. The range in velocity dispersions between the clusters is small in order to reduce cluster to cluster variation. Each of the clusters has an infall radius between three -- four Mpc from the BCG-defined center. The multiband wide-field observations in each cluster extend past the infall region for sufficient cluster coverage \protect\citep{Just_2019}. }
 \label{tab:clusters}
 \end{table*}
The ESO Distant Cluster Survey (EDisCS; \citealt{white05}). is an ESO Large Program derived from the optically brightest objects of the Las Campa\~{n}as Distant Cluster Survey \citep{gonz01} and comprises 20 clusters within 0.4 $<$ z $<$ 0.8. The velocity dispersion ($\sigma_v$) of these clusters ranges from 200 - 1200 km s$^{-1}$ \citep{halliday04,mj08} and is characteristic of local cluster progenitors due to mid-mass halo sizes \citep{mj08}. The main goal of EDisCS is to examine the evolution of cluster populations over a large span of cosmic time and compare results to with respect to halo mass and local cluster populations. 

For the purpose of this study we decide to separate our galaxies into three distinct, but broadly defined, environments: (i) the cluster core, within $R_{200}$, (ii) the infall region, which corresponds to all galaxies at the cluster redshift but beyond the virial radius, and (iii) the field, which corresponds to foreground and background galaxies. We define the cluster center in all cases as the location of the BCG and measure clustercentric radii from that location. The BCG lies at the approximate center of the member distribution \citep{white05} for our clusters and \cite{Just_2019} showed that any offsets of the BCG location from the center are $<10\%$ of the infall radius for the clusters in our sample.  However, some of our clusters exhibit significant substructure \cite{DeLucia09}, indicating that defining environment purely by clustercentric radius may wash out some trends with local density.
\par
\subsubsection{Cluster Core}
The core regions, which are typically on the order of 0.5 -- 2 Mpc across, are defined as the area within the virial radius and typically include the BCG. For EDisCS, the cores have been extensively studied with deep optical imaging and spectroscopy on VLT \citep{white05,halliday04,mj08,vul12} and NIR observations on the New Technology Telescope \citep{white05,rudnick09} which has allowed further EDisCS studies such as brightest cluster galaxy identification \citep{white05,whiley08}, morphologies \citep{desai07,simard09,vul11a,vul11b}, fundamental-plane parameters \citep{saglia10}, red-sequence identification \citep{delucia04}, weak lensing \citep{clowe06}, 24$\mu m$ MIPS SFRs \citep{finn10} and [\ion{O}{ii}] SFRs \citep{pogg06,pogg09,vul10}. Cluster cores are dense regions that are attractive for studying cluster properties and are well-suited for observations due to high contrast with the background and density of objects within a given FOV. However, physical processes affecting the evolution of a galaxy appear to occur as these sources enter a cluster environment, and thus the cores likely only provide information on their fate. This is reinforced through observations that the cores typically include a higher fraction of massive red disk or quiescent galaxies  \citep{dressler80,bell04,kauff04,erf16}. A spatially-expanded view of clusters is required to gather information on environmentally-driven quenching mechanisms across a representative sample of galaxies within a cluster. 

\subsubsection{Wide Field Follow-up Surveys}
\label{sec:wff} 
Galaxy clusters extend far beyond their cores and virial radii, and in order to achieve a more informed understanding of the role of environment on galaxy evolution, it is important to extend analyses to projected radii greater than R$_{200}$. This is a challenging task, as the reduced density of the cluster density profile results in a decreased contrast with the foreground and background \citep{Newman13}, so large and wide-field spectroscopic studies are required to conclusively establish membership in these regions.

For this reason, we undertook a wide-field imaging and spectroscopic follow-up of the EDisCS clusters. The imaging consisted of $BVRIzK$ data covering approximately $30\arcmin\times30$\arcmin \ around the cluster. The {\em VRI} photometry was observed with the Wide Field Imager (WFI) on the 2.2m Max Planck Gesellschaft/European Southern Observatory (MPG/ESO) telescope \citep{baade99}, while {\em Bz} observations were completed on the MOSAIC instrument on the Cerro Tololo Inter-American Observatory (CTIO) Blanco or Mayall 4-meter telescope. The $K$-band data were taken with the NEWFIRM instrument on the Mayall telescope. These imaging observations are described in detail in \cite{Just_2019} and Mann et al. (in prep). The spectroscopic component of the survey was conducted with the Low-Dispersion Prism (LDP) on IMACS/Magellan, which covers out to 6$R_{200}$ for our clusters. These observations produced a deep catalog of 25,000 redshifts with an accuracy of $\sigma$ = 0.007 and a high spectroscopic completeness up to R$_{AUTO}$ $<$ 23.3 \citep{Just_2019}. This information is crucial towards establishing cluster membership beyond the central core as in previous EDisCS studies and allows for targeted followup observations of groups or infalling populations.  

As described in \cite{Just_2019}, we derived rest-frame $U-V$ and $V-J$ (hereafter $UVJ$) colors for all of our galaxies. Due to residual zeropoint calibration issues, these colors required secondary adjustments to bring them in line with the $UVJ$ colors as measured for core galaxies from the EDisCS survey. This process is described in Appendix~\ref{sec:app}. Following those adjustments, we have reliable $UVJ$ colors that can be used to separate galaxies into quiescent and star-forming \citep[e.g.][]{Wuyts07, will09}. However, the way in which we performed the adjustments impacted the reliability of our $U$-through-$K$ SEDs and added an unacceptable level of systematic uncertainty to SED-based stellar mass estimates. In \S\ref{sec:corrs} we describe our alternate method for our computation of the stellar mass using just the calibrated $UVJ$ colors.

 \cite{Just_2019} utilized the theory of secondary infall to identify the infall region of the galaxy clusters with the equations given in \cite{white92}. This theory describes how shells of mass evolve with redshift when centered on a cosmic perturbation; shells that are contained within a critical mass will eventually follow a gravitational collapse and become bound. The outermost boundary of the mass shell that experiences collapse at the cluster redshift is defined as the infall radius.
 
Followup observations with {\em HST} were possible due to the extensive spectroscopic and photometric coverage of the EDisCS sample in \cite{Just_2019} and the proven abilities of the grism with 3DHST \citep{mom16} and GLASS in dense cluster environments \citep{treu15}. From the full EDisCS sample of 17 EDisCS clusters we selected four clusters for follow-up with the {\em HST}/WFC3 G102 IR grism to produce high-spatial resolution emission line maps for individual galaxies. These clusters were chosen according to the following criteria: 1) The ability of HST to observe H$\alpha$ with the G102 at the redshift of the cluster and 2) the degree to which the infall region of the cluster was populated with groups at a range of cluster-centric radii and with enough galaxies in each group so as to maximize the multiplexing efficiency for the grism observations. 

\begin{figure}

\centering
\includegraphics[scale=0.18]{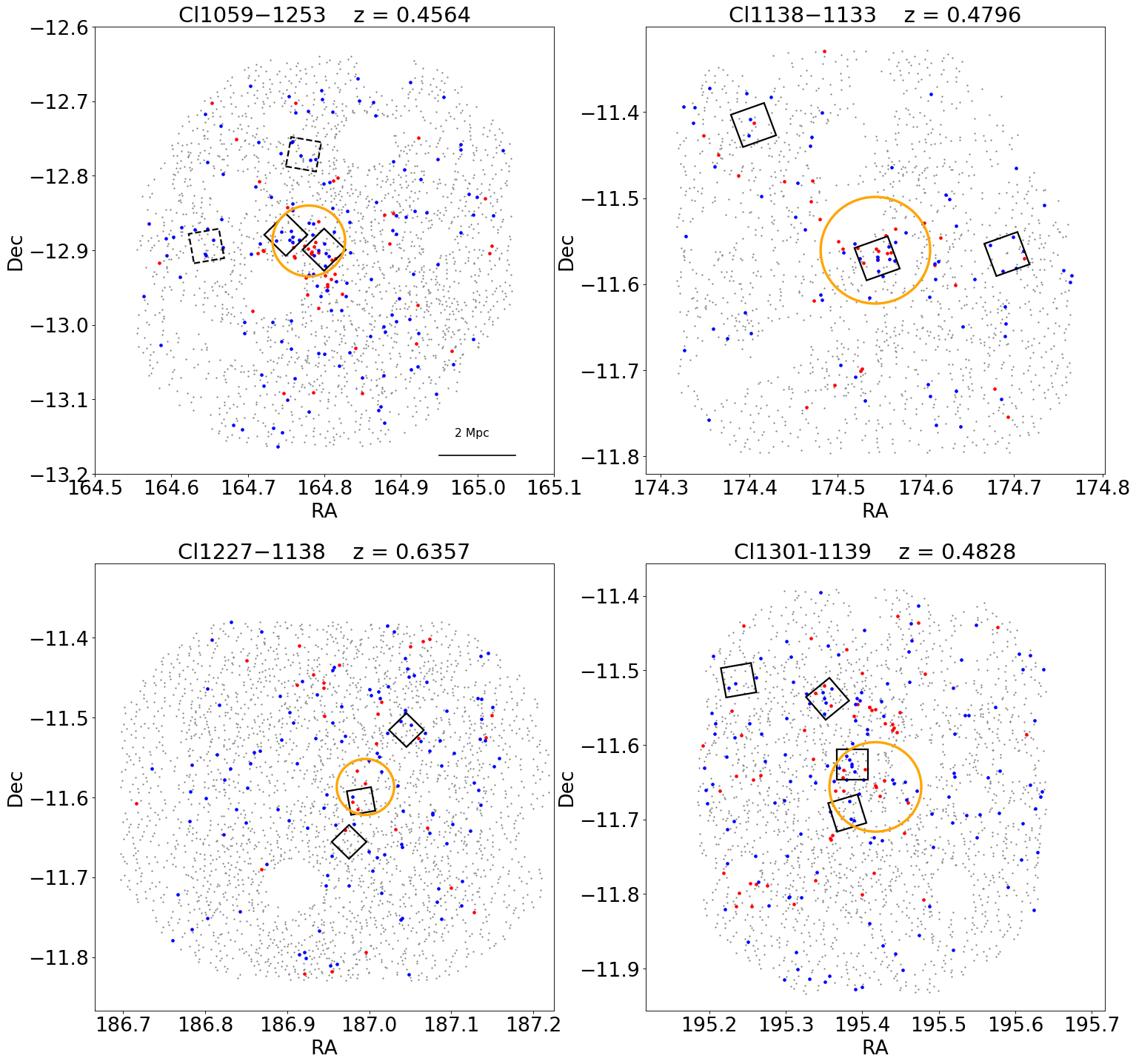}
\caption{The RA and Dec spatial distribution of galaxies in each cluster. Grey dots represent all galaxies in the FOV that have an LDP redshift, red/blue points signify {\em UVJ}-identified quiescent/star-forming cluster member sources, and the virial radius is indicated by the orange circle. No magnitude or mass limits are taken into account.} {\em HST}/WFC3 G102 observations are represented by the black squares, where the two unused infall pointings in Cl1059 are dashed. The distributed sampling among the core and infall region allows for a direct comparison of SFRs by environment.
\label{fig:prop}
\end{figure}

These clusters have a velocity dispersion ranging from $500-800$~km s$^{-1}$, which is squarely in the middle of the velocity dispersion range of EDisCS clusters \citep{halliday04, Milvang-Jensen08}. The limited range in velocity dispersion of our target clusters will help to minimize and halo-mass dependent cluster-to-cluster variations \citep{pogg06,moran07}. 
\cite{Just_2019} performed a characterization of the infall region of the EDisCS clusters using the LDP spectroscopy, which showed that red galaxies are more clustered than blue galaxies. Because of the magnitude limited density-dependent sparse spectroscopic target sampling, the limited number of galaxies in each cluster's infall region, and because of the aggressive masking of bright stars in the targeting (Fig.~\ref{fig:prop}), it is significantly more difficult to characterize local densities in the infall region and to identify other structures, like filaments. 
Details for each pointing including cluster membership and location are listed in Table~\ref{tab:pointings}. 

\subsection{Field Sample}
 In order to form a comparison set of galaxies in an effort to constrain environmental effects from the cosmic web, we establish a field sample that is assumed to occupy a less dense and interactive region of the universe. Nearly $\frac{3}{4}$ of all galaxies in the universe reside in the field and have been the subject of many surveys such as 3DHST \citep{mom16} and CANDELS \citep{grogin11}. However, no large H$\alpha$ field sample with significant ancillary data exists at the redshift of our clusters, so we therefore construct a field sample from our own data. In our study, we construct the field using {\em HST}-observed galaxies within each pointing FOV in the range of 0.4 $<$ z $<$ 0.7 that lie outside $\pm$ 0.02 of each cluster redshift. This span in redshift is dictated by the range of our four target clusters. 

\begin{figure*}
\centering
\includegraphics[scale=0.2175]{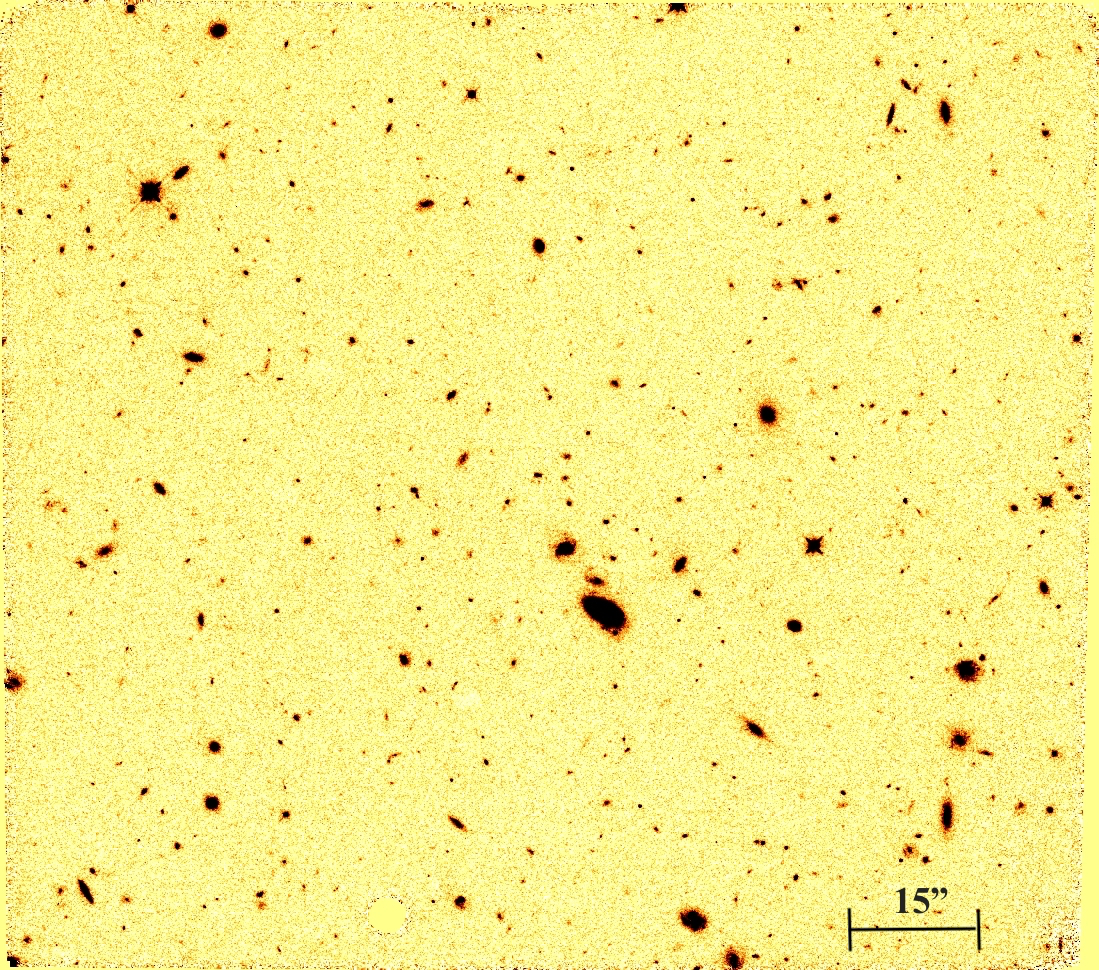}
\includegraphics[scale=0.2175]{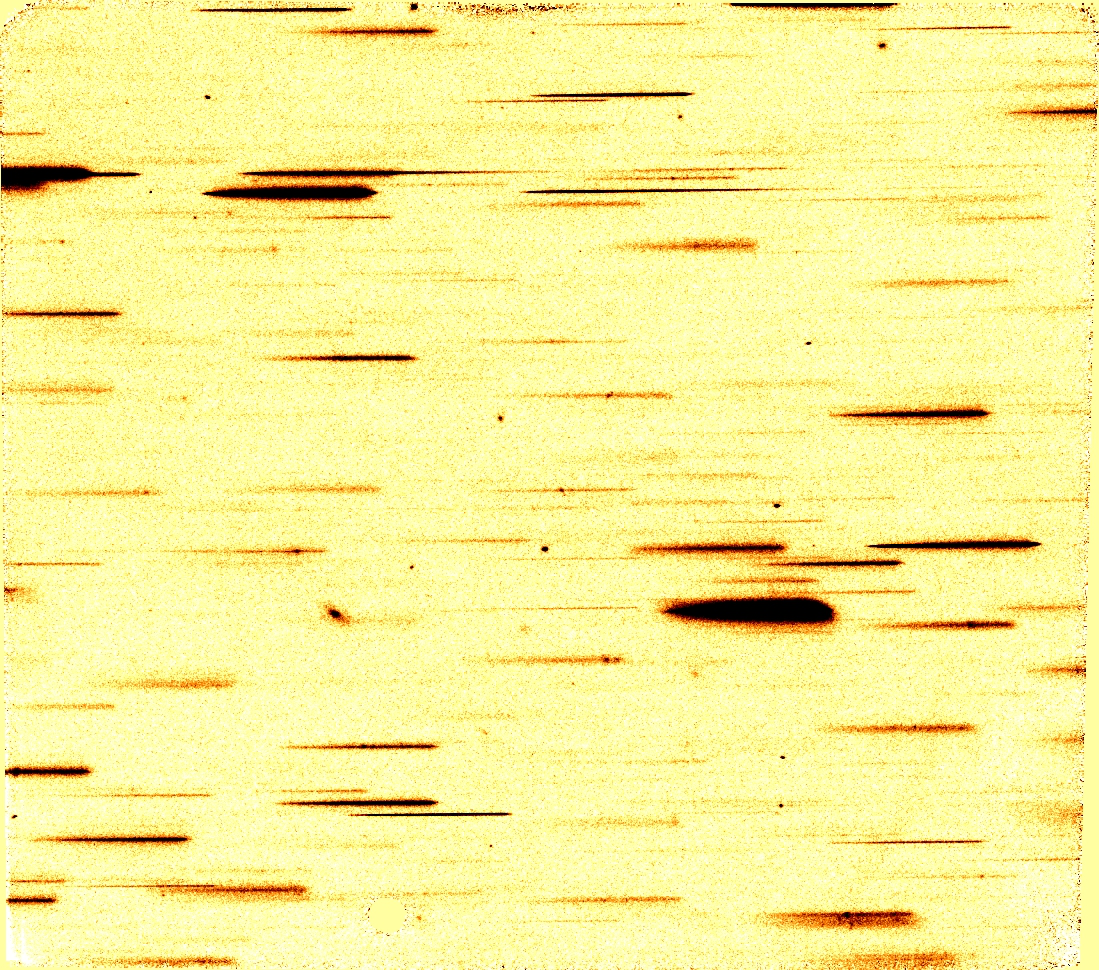}
\caption{(Left) {\em HST} WFC3 F105W (1.05$\mu$m) infrared direct image of the CL1301-11.2 pointing. (Right) {\em HST} WFC3 G102 grism data, which provides a spatially resolved spectrum for every object in the FOV of the F105W image. The spectrum for each object is extracted within the grism wavelength range after a polynomial fit to the continuum is subtracted off.}
\label{fig:products}

\end{figure*}
\subsection{{\em HST}/WFC3 observations}
\label{sec:hst} 
We obtained {\em HST}/Wide Field Camera 3 F105W imaging and G102 grism spectroscopy in a Cycle 20 program (GO-12945: PI Rudnick) for four EDisCS clusters at z $\sim$ 0.5 to target star-forming H${\alpha}$ emitters. Details for each cluster in this study are listed in Table~\ref{tab:clusters}.  

\par
There are 14 pointings consisting of two orbits each (2800 seconds) that are distributed over the four clusters, where $\sim$ 15\% of the time is devoted to F105W (rest-frame R-band) direct imaging and the remaining 85\% used for G102 grism spectroscopy. This is a similar split between modes as in 3DHST \citep{nelson12,mom16}. The distribution of the pointings aims to equally cover the cluster core and infalling region in each cluster in order to sample a range of environments, as shown in Figure~\ref{fig:prop}. The infall region pointings were chosen based on a preliminary LDP catalog to contain projected overdensities of blue galaxies spectroscopically confirmed to lie at the cluster redshift. The LDP catalog underwent significant revisions following the original targeting and some of the originally chosen projected groups reduced their density contrast. Of the 14 pointings, only 12 are utilized due to unreliable photometry in Cl1059; this results in the loss of two infall pointings, which are designated as dashes in Figure~\ref{fig:prop}. There are a total of 581 galaxies with LDP redshifts in these 12 pointings, which will be further reduced based on H$\alpha$ detection.

\par 
The G102 grism spans a wavelength range of 0.7 -- 1.1$\mu$m, which contains the H$\alpha$ emission for 0.4 $<$ z $<$ 0.7. As the brightest Balmer series emission line, the H${\alpha}$ flux can straightforwardly be transformed into a SFR (see \S~\ref{sec:SFR} for a further explanation) and is an excellent tracer of nearly instantaneous star-formation on $\sim$10 million year timescales, despite having typical attenuation of $1-2$~magnitudes. 

The G102 grism resolution of 700 km s$^{-1}$ is much higher than the typical internal galaxy velocity dispersion, which results in a resolved H$\alpha$ map of the galaxy. The emission line map is produced by subtracting a polynomial fit to the background from the 2D spectrum, where the emission line is initially masked. The residual provides an image of the galaxy at a given wavelength within the grism range for the masked emission line. An example of z $\sim$ 1 H$\alpha$ emission line maps are available from 3DHST observations in \cite{nelson12}.
Additionally, as a robust optical tracer, H$\alpha$ can detect SFR to low surface brightness levels, which is crucial for creating a sample that encompasses galaxies as they are shutting off star formation. 
The SFR detection limit is variable depending on the extent and morphology of the galaxy, which makes defining a detection limit nontrivial. The lowest log(SFR) derived in this study are $\sim -0.5~$M\textsubscript{\(\odot\)} yr$^{-1}$, which is considered a typical value for a regular star forming galaxy at masses similar to our target galaxies.

\begin{table*}

 \begin{tabular}[t]{ccccccc }
  \hline
 Pointing ID & RA & Dec  &  Location  & N$_{galaxies}$& N$_{cluster}$ & N$_{field}$ \\ 
  & (hours) & (degrees) & (cluster/infall) & & (H$\alpha$) & (H$\alpha$) \\ 
  (1)       &   (2)     & (3)       & (4)       & (5)             & (6)     & (7)   \\ 
  \hline
Cl1059-12.0 & 10:59:08.16 & -12:45:05.04 & I & 161 & x & x \\ 
Cl1059-12.1 & 10:59:03.36 & -12:51:59.04 & I & 152 & x &  x \\ 
Cl1059-12.2& 10:59:14.16  & -12:53:11.04 & C & 152 & 8(4) & 15 \\ 
Cl1059-12.3 & 10:59:32.16 & -12:54:12.64 &  C & 179 & 17(2) & 12 \\ 
Cl1138-11.0 & 11:38:16.56 & -11:33:23.04  & C & 80 & 15(6) & 13(24) \\ 
Cl1138-11.1 & 11:38:51.60 & -11:33:30.24  & I & 108 & 4 & 5(4) \\ 
Cl1138-11.2 & 11:37:54.48 & -11:30:23.04  & I & 179 & 3(1) & 7 \\ 
Cl1227-11.0 & 12:28:02.40 & -11:35:11.04  & C & 117 & 3(4) & 3(2)  \\ 
Cl1227-11.1 & 12:28:08.16 & -11:31:02.64  & I & 167 & 3 & 5(3) \\ 
Cl1227-11.2 & 12:28:20.64 & -11:30:59.04  & I & 129 & 3 & 9(3) \\ 
Cl1301-11.0 & 13:01:35.76 & -11:36:59.04  & C & 167 & 4(3) & 3(6) \\ 
Cl1301-11.1 & 13:01:25.44 & -11:31:42.24  & I & 143 & 6(3) & 6(1) \\ 
Cl1301-11.2 & 13:01:33.36 & -11:40:27.84  & C & 184 & 5(7) & 1(5) \\ 
Cl1301-11.3 & 13:01:02.88 & -11:30:15.84  & I & 145 & 4 & 7(5)\\ 

  \hline

 \end{tabular} \par
 \bigskip
\caption{ Information for each of the 14 pointings observed with {\em HST}/WFC3. In column 1, the prefix of the Pointing ID relates to the Cluster ID from Table~\ref{tab:clusters} column 1. Columns 2 and 3 contain the RA/Dec information. For the Location column, I and C refer to {\em infall} and {\em core}, respectively, where  {\em infall} is outside of R$_{200}$ as specified in Table~\ref{tab:clusters} column 6. The number of sources extracted with GRIZLI in each pointing are listed in column 5. The number of galaxies with H${\alpha}$ S/N $>$ 3 and without  contamination in the cluster (6) and field (7) for each pointing, where H${\alpha}$ S/N $<$ 3 are designated within parentheses. An x signifies that the pointing was not utilized. All of these sources have a wide-field catalog counterpart with rest-frame colors and stellar mass calculations. For Cl1059, the two pointings (12.2 \& 12.3) in the core do not have well-calibrated photometric wide-field data and thus replacement observations and redshifts are utilized from previous VLT/FORS observations \protect\citep{white05}. The two infall pointings (Cl1059-12.0 \& Cl1059-12.1) do not have substitute coverage and are not included in analysis.}
 \label{tab:pointings}
\end{table*}

\subsection{Data Reduction}
\label{sec:red}
GRIZLI (grism redshift \& line analysis software for space-based slitless spectroscopy)\footnote{GRIZLI is written and developed by Gabriel Brammer and is publicly available as open-source software \citep{grizli}. github.com/gbrammer/grizli} is a reduction and extraction pipeline in Python that allows for end-to-end processing of WFC3 data, starting from a query of the ESA Hubble Science archive to download all of the data associated with an observation ID. It then performs a routine calibration of the data, including image background sky subtraction, alignment and flat-fielding, resulting in the two drizzled mosaic data products shown in Figure~\ref{fig:products}. The WFC3 camera captures both an infrared 1.05$\mu$m direct image (F105W) and the spectrum as a dispersed image for each object in the FOV (G102 grism). The 2D spectra are the streaks, which represent the flux of each object as it is spread out over the range (0.7 -- 1.1$\mu$m) of the grism. Several conditions may make a grism spectrum unusable, including contamination from a bright source, low signal-to-noise, or FOV restrictions. All sources included in our analysis  are visually inspected for artifacts or poor modeling. While the analysis focuses on galaxies with S/N H$\alpha >$ 3, those with $<$ 3 are presented as down arrows in several figures.

\subsection{H$\alpha$ Line Extraction \& Redshift Prior}
\label{sec:SFR}
 The redshifts in GRIZLI are fit using a coarse grid (resolution $\sim$0.005) with three line complex templates composed of 1) [OII]+[NeIII], 2) [OIII]+H$\beta$, and 3) H$\alpha$+[SII] + weaker red lines. Each of the line complexes has fixed line ratios in order to reduce line misidentification and break redshift degeneracies. A minima in the $\chi$-squared fit on the redshift grid allows for the best fit determination of the redshift. 
 
\par To reduce the misidentification of other emission lines as H$\alpha$, a redshift prior is utilized during extraction within GRIZLI. Priors are derived from the LDP spectroscopic or the wide-field photometric redshift surveys. To determine the probability distribution (P(z)) in Equation~\ref{eq:1},
\par
\begin{equation}
P(z) = (\sigma\sqrt{2\pi})^{-1}{e^{\frac{-(z - z_{prior})^{2}}{2\sigma_z^{2}}}}
\label{eq:1}
\end{equation}
\par
\vspace{.6cm}

\begin{figure*}
\centering
\includegraphics[scale=0.52]{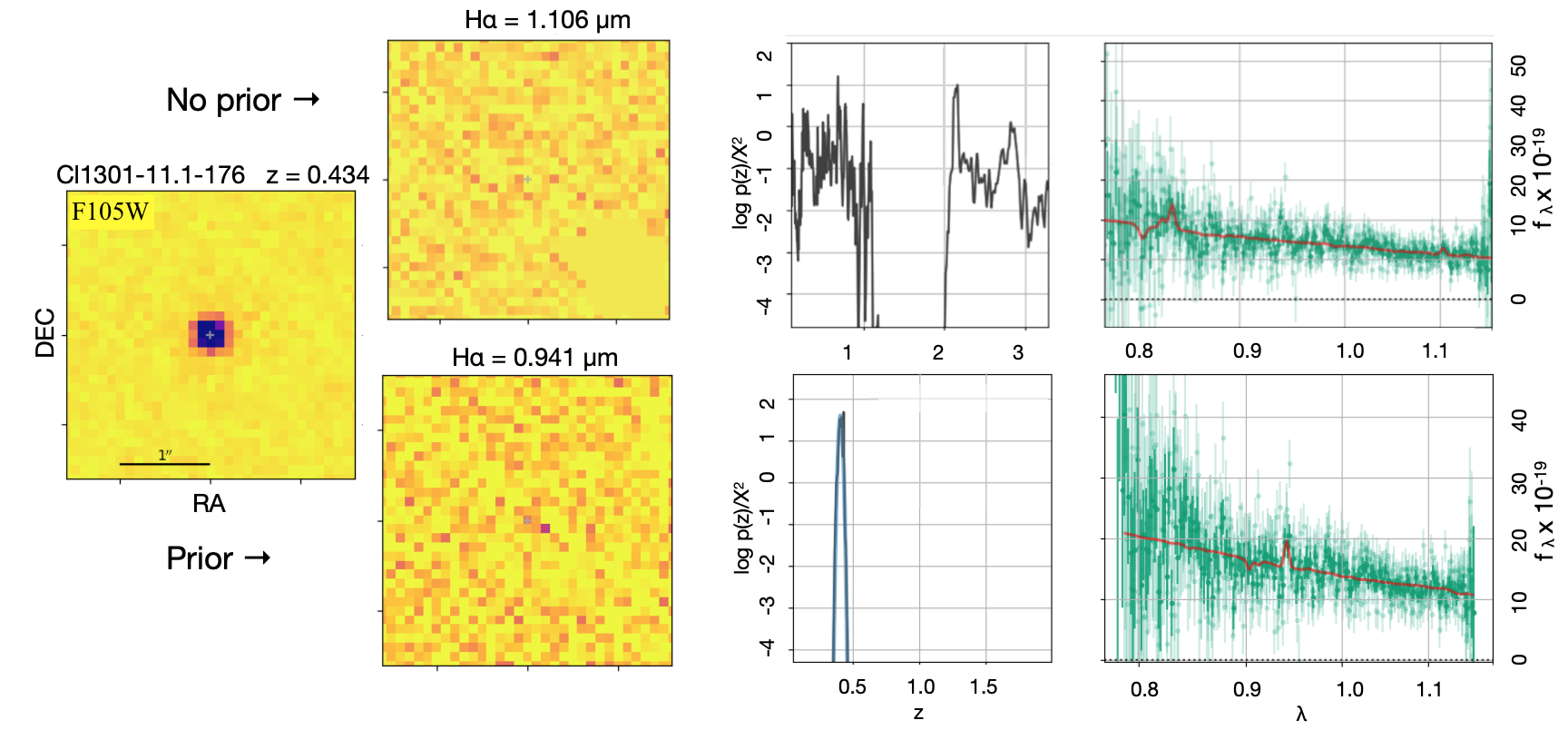}
\caption{ This collection of data products represents a comparison between the same galaxy with a blind GRIZLI extraction (top row) and with an LDP prior (bottom row). Initially, GRIZLI measured the H$\alpha$ flux with $S/N=2.99$ at $z = 0.685$. With the redshift prior, the redshift changed by 0.25 to $z=0.434$ with the resultant the H$\alpha$ flux having $S/N=3.99$. The F105W direct image of the stellar content is on the far left, followed by the detected H$\alpha$ emission line map. In the third panel is the p(z) from the redshift fitting algorithm (black line), with a blue line indicating the applied Gaussian redshift-prior in the bottom panel. The p(z) after the prior is applied (black line - bottom row, middle panel) is much more constrained than the blind p(z). Note the redshift scale differences between the extractions. The rightmost panel shows the 1D spectra in green with a fit (red). The blind extraction for the p(z) is very uncertain and could easily be a high or low-z galaxy. The application of the prior dramatically alters the results of the redshift determination. The ability of the prior to be successfully applied to a low S/N H$\alpha$ galaxy is important towards creating a sample that is not biased towards strong H${\alpha}$ line galaxies. }
\label{fig:nop}
\end{figure*}

\hspace{-.66cm}the prior is multiplied by the GRILZI redshift fit, using either a Gaussian probability \citep{Just_2019} with a $\sigma$ = 0.007 or the average of the 68\% photometric redshift confidence levels, respectively. Figure~\ref{fig:nop} is a demonstration of applying the prior to a low S/N H$\alpha$ galaxy that changes the determined redshift by $>$0.2, which is significant when cluster membership is determined within a 0.02 range in z.

\begin{figure*}
\centering
\includegraphics[scale=0.58]{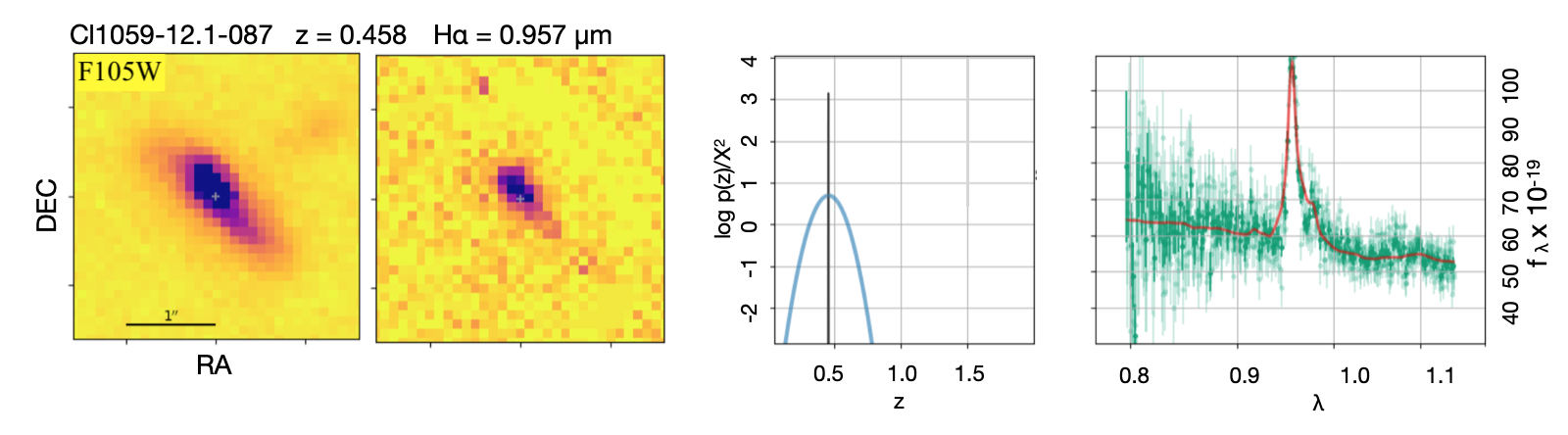}

\caption{(2 left panels): RA and Dec postage stamps show the stellar continuum from the F105W 1.05$\mu$m direct image and the H$\alpha$ emission extraction at 0.957$\mu$m. (2 right panels): The fitted redshift, shown as the black line, is fully consistent with the photometric redshift prior probability distribution in blue. The 1D spectrum data are shown in green, with a best fit template in red. Note the prominent H$\alpha$ emission line at 0.95$\mu$m with a S/N of 36.6. Note that is one of the brightest H$\alpha$ emission lines from our sample}.
\label{fig:extract}
\end{figure*}

\begin{figure}
\centering
\includegraphics[scale=0.25]{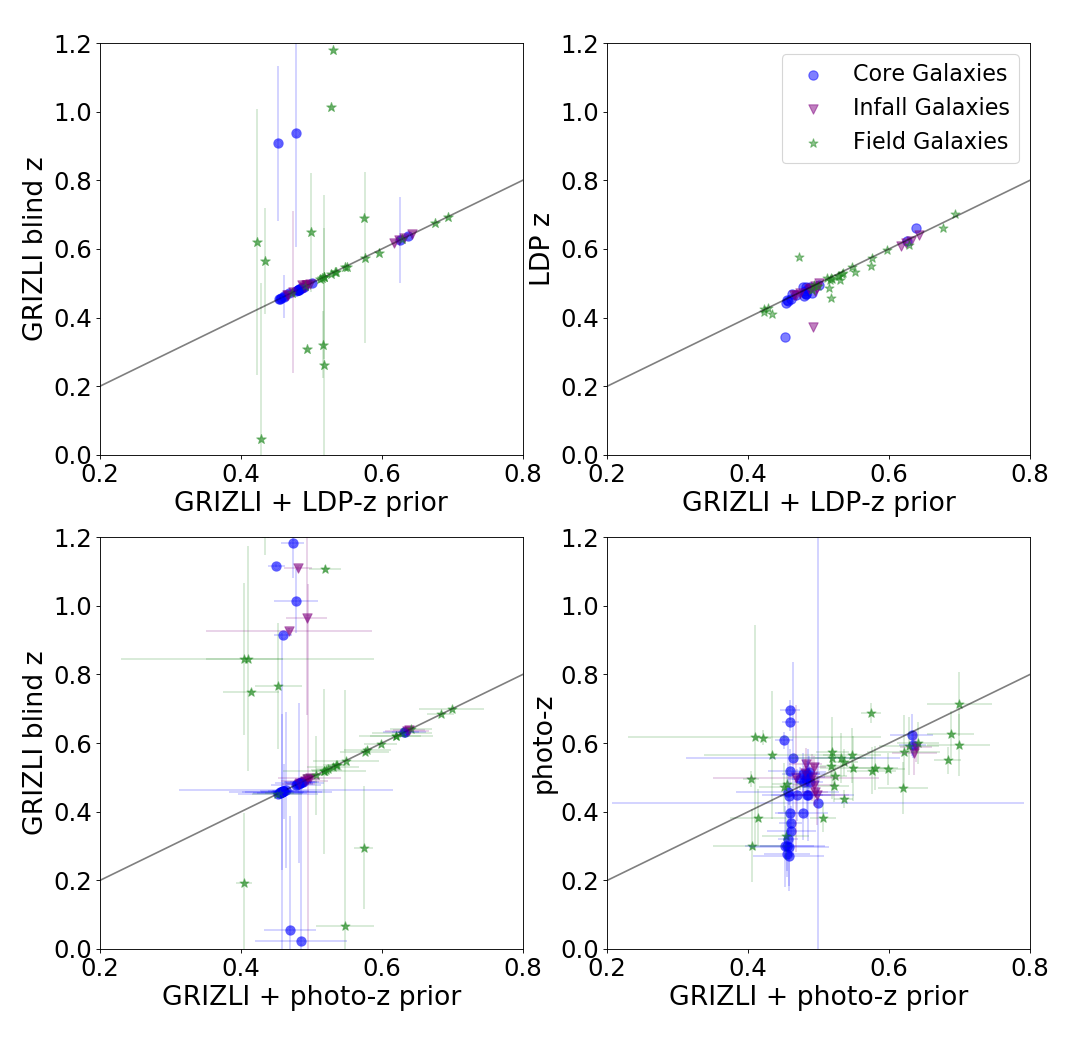}
\caption{The GRIZLI-extracted redshift (no prior) vs the GRIZLI-extracted redshift with an applied spectroscopic (top left) or photometric (bottom left) prior for cluster core (blue circle), infall (purple triangle), and field (green star) galaxies. A majority of objects fall along the 1-1 line in black, indicating that the GRIZLI extractions without a prior can be reliable. The galaxies significantly above 1-to-1 line in the top and bottom left originate from a single pointing with a significantly higher background, and hence poor blind GRIZLI redshifts. These galaxies are mostly corrected with the LDP spectroscopic or photometric redshift prior. Several low S/N H$\alpha$ galaxies are also corrected through the prior. The 68\% confidence levels for the photometric, GRIZLI blind and GRIZLI + photometric redshifts are shown with error bars, while the errorbars on the LDP or GRIZLI + LDP redshifts are too small to be visible. This photometric relation has noticeably more scatter around the 1-to-1 line, which is a reflection of the reduced accuracy of photo-z measurements. The GRIZLI-extracted redshifts with a spectroscopic (top right) or photometric (bottom right) prior are shown in comparison to their blind redshift. }
\label{fig:zcomp}
\end{figure}

A full set of data products for a strong H$\alpha$ emission line Cl1059 cluster member is shown in Figure~\ref{fig:extract}. This galaxy has a spectroscopic prior applied, but it also had a well-determined redshift based solely on the blind GRIZLI extraction. A comparison between the available redshifts for each galaxy in this sample with and without priors is shown in Figure~\ref{fig:zcomp}. The general agreement of GRIZLI z extractions without a prior to the wide-field catalog of spectroscopic and photometric redshifts supports the usage of this software in H$\alpha$ line identification without previous information, but it is most important for low S/N emission lines or quiescent galaxies where a prominent emission line may not exist. These lower S/N sources are critical for encompassing a range of SFRs in a main sequence analysis and exclusion of these galaxies would introduce a bias towards strong emission line galaxies. 
\par 
This dataset has three types of redshifts available: GRIZLI, GRIZLI + Gaussian prior from a spectroscopic LDP, and GRIZLI + Gaussian prior from a photometric wide-field, where the prior is described in Equation~\ref{eq:1}. GRIZLI is first run without any priors, and is then rerun to include a prior with either a spectroscopic LDP or photometric wide-field redshift for each galaxy. When compared for sources with H$\alpha$ S/N $>$ 3, the blind GRIZLI redshifts do remarkably well, with $\sim$85\% matching the extracted redshift with an LDP prior and $\sim$62\% for the photometric prior as shown in Figure~\ref{fig:zcomp}.

\subsection{Stellar Masses and Star-formation Rate Corrections}
\label{sec:corrs}
As described in \S\ref{sec:wff} and Appendix~\ref{sec:app}, due to the calibration issues with our photometry we cannot derive stellar masses for our sample using SED fitting techniques. However, we have calibrated our rest-frame $UVJ$ colors by comparing them to the valid colors from the EDisCS survey. We therefore determine our stellar masses using the relation between rest-frame $U-V$ color and $V$-band stellar mass-to-light ratio that is derived from continuous star formation history (SFH) models with a variety of \cite{calzetti00} attenuation laws. This approach is similar to that used by various authors \citep[e.g.][]{Bell01, vul10,taylor11} and is primarily valid for galaxies with continuous SFHs. The derivation of our specific relation is provided in Mann et al. (in prep.) The relation we adopt is:
\begin{equation}
\label{eq:3}
 \ {\rm log}_{10} M_* = 0.997\times ( U - V) - 1.272 + {\rm log}_{10} L_V
\end{equation}

where
\begin{equation}
    \frac{L_V}{{\rm L_{\odot,V}}} = \frac{4\pi f_V D_{L}^2}{(1 + z )}\frac{1}{{\rm L}_{\odot,V}}
\end{equation} 
and $f_V$ is the flux through the redshifted rest-frame {\em V}-band filter, D$_L$ is the distance luminosity, and L$\textsubscript{\(\odot\)V}$ is the rest-frame luminosity of the sun in V-band. Stellar masses determined from a single color can have systematic errors on the order of 0.3~dex stemming from variations in the dust attenuation, metallicity, and SFH. In addition, we applied systematic corrections of less than 0.2 mag to our $U-V$ colors that may also be uncertain and can result in additional stellar mass errors up to 0.2~dex. We therefore conservatively assume stellar mass uncertainties up to 0.5~dex, with the consideration that the statistical uncertainties on photometric measurements are significantly less than this, especially for the bright galaxies with LDP redshifts, and can be ignored. 

The mass-completeness level is independently identified at this value through 1) a comparison of the stellar masses of the 163 {\em HST}-observed galaxies to the ULTRAVISTA star-forming subset at a similar redshift range and 2) by the 2$\sigma$ distribution of all star-forming galaxies in the EDisCS sample from \cite{Just_2019} within the photometric completeness limit and redshift range. This results in a mass-complete value of 10$^{9.75}M_*$/$M\textsubscript{\(\odot\)}$.

\par
GRIZLI outputs a line flux, but there are several intrinsic properties that need to be accounted for while calculating a SFR. Following the prescription in \cite{carl20}, a series of corrections are applied to achieve a correct H$\alpha$-based SFR. The resolution of the grism is not fine enough to distinguish between the H$\alpha$ and [NII] line doublet emission, indicating that measured line fluxes include the contribution of [NII] and therefore need to be reduced to account for the additional flux. \cite{strom17} find that the [NII] contribution is uniform across SFR per given stellar mass, so \cite{carl20} calculates this reduction through a mass-dependent metallicity relation. The mass-metallicity relation is derived from \cite{zahid14}, which is then transformed into an H$\alpha$/[NII] ratio \citep{kewley08}, resulting in a flux reduction of $\sim 33\%$ for our sample. \cite{carl20} required a $\sim 25\%$ correction for z $\sim$ 1 galaxies, while 3D-HST \citep{wuyts11} found $\sim 20\%$. There is a secondary dependence of the metallicity on the SFR at a fixed stellar mass known as the Fundamental Metallicity Relation \citep[FMR;][]{Mannucci10}. We used the FMR to determine how much the metallicity correction changes over the range of SFRs in our sample. As we discuss in \S\ref{sec:sfr-m}, our mass-complete galaxies range in $\log_{10}$(SFR) from -0.5 to 0.5. At $\log_{10}(M_{*}$) = 10.0, the FMR predicts that log(O/H) changes from 8.82 to 8.98. In comparison, at $\log_{10}(SFR)$ = 0, the FMR predicts that log(O/H) changes from 8.8 to 9.07 over the full mass range of our mass-complete sample $\log_{10}(M_{*}$) = 9.7 to 11. While the mass dependence of the metallicity dominates over the residual dependence on the SFR, the dependence on the SFR is not negligible. This implies that we may be underestimating the uncertainty in this correction. However, [\ion{N}{ii}]/H$\alpha$ saturates at high metallicity, which should minimize the effect of this residual SFR dependence on our results.

\par The H$\alpha$ line is also contaminated with emission from post-AGB stars and this is remedied by subtracting $f_{AGB} = 2\times 1.37 \times 10^{29}$ erg s$^{-1}$ M$_{\odot}^{-1}$ from the line luminosity \citep{carl20}, where the factor of two comes from the 1:1 ratio of [NII]/H$\alpha$ lines \citep{bel16} and the $1.37\times 10^{29}$ factor comes from the expected contribution of ionization by the post-AGB stars. When compared to the H$\alpha$ line luminosity ($\sim 10^{40}$ -- $10^{42}$), the post-AGB emission is negligible. This correction is equivalent to a reduction in the specific SFR of $1.2 \times 10^{-12}$ yr$^{-1}$. 

Dust within each galaxy is responsible for the extinction and scattering of  light and thus, contributes towards suppressed H$\alpha$ emission lines and SFRs. We correct for H$\alpha$ attenuation following the approach from \cite{wutys13}, who relate the attentuation at H$\alpha$ (A$_{H\alpha}$) to that in the continuum at 6563\AA\ ($A_{6563\text{\AA}}$) using the relation $A_{H\alpha} = 1.9A_{6563\text{\AA}} - 0.15 A_{6563\text{\AA}}^{2}$. As demonstrated in \cite{carl20} using Balmer decrements from the LEGA-C survey \citep{vanderwel16}, the \cite{wutys13} approach yields line attenuations within 0.32~mag of the Balmer-decrement approach, with a 1.2 mag scatter and no dependence of the disagreement on the measured extinction. 

Given the limited spectral coverage of the G102 grism and the issues with our SED calibration discussed earlier in the text, we cannot compute A$_{6563\text{\AA}}$ directly for each galaxy. Instead, we develop a  method to derive a statistical attenuation correction in which we determine the dependence of A$_{6563\text{\AA}}$ on $UVJ$ color for a sample of galaxies from UltraVISTA which have extremely well characterized SEDs and redshifts \citep{muzzin13}. We use a subset of ULTRAVISTA galaxies with a similar redshift and stellar mass distribution to our own and fit their SEDs using MAGPHYS \citep{dacun08}, which is a Bayesian SED fitting code that deals with attenuation both from birth clouds and from diffuse dust. Neither of these is a good approximation for the total attenuation used in the \cite{wutys13} formula used above. We therefore derive the effective total continuum extinction by taking the ratio of the attenuated and unattenuated model for each galaxy at 6563\AA, where the attenuated model folds in the distinct attenuation sources for the different stellar populations. We then derive the optical depth at 6563\AA\ $\tau_{6563\AA} = -{\rm ln}(10^{{\rm log} I_{att}}/10^{{\rm log} I_{o}}$), where ${\rm log}_{att}$ is the flux of the attenuated SED and {\rm log}$_{o}$ is the unattenuated SED curve. We compute the attenuation in magnitudes  as $A_{6563\text{\AA}} = 1.086 \times \tau_{6563\text{\AA}}$.

We then compute the median ULTRAVISTA attenuation in 0.2~mag bins of $UVJ$ color-space and apply to each EDisCS galaxy the attenuation corresponding to the appropriate $UVJ$ color cell (Figure~\ref{fig:uvjtau}). This final correction for A$_{H\alpha}$ is folded into the H$\alpha$ SFR, which is calculated from \cite{kenn12} as 
\begin{center}
\begin{equation}
{\rm log}_{10} (SFR_{H\alpha} / M\textsubscript{\(\odot\)} yr^{-1}) = {\rm log}_{10}(L \textsubscript{H${\alpha}$}/L\textsubscript{\(\odot\)})  - 41.27 + 0.4A_{H\alpha}    
\label{eq:sfr}
\end{equation}
\end{center}
where
\begin{center}
\begin{equation}
L \textsubscript{H${\alpha}$}(L\textsubscript{\(\odot\)}) = 4\pi f \textsubscript{H${\alpha}$}D_{\rm L}^{2} - 2 \times 1.37 \times 10^{29} {\rm erg~s^{-1}} \times {\rm log}_{10} (M_{*}/M\textsubscript{\(\odot\)}).
\label{eq:2b }
\end{equation}
\end{center}
The median H$\alpha$ attenuation correction is 0.48 for the mass-complete sample, and is 0.79 and 0.26 for our UVJ star-forming and quiescent galaxies respectively.

\section{Results}
\label{sec:results}

\subsection{Galaxy Sample Properties}
\label{sec:props}
\begin{figure}
\centering
\includegraphics[scale=0.40]{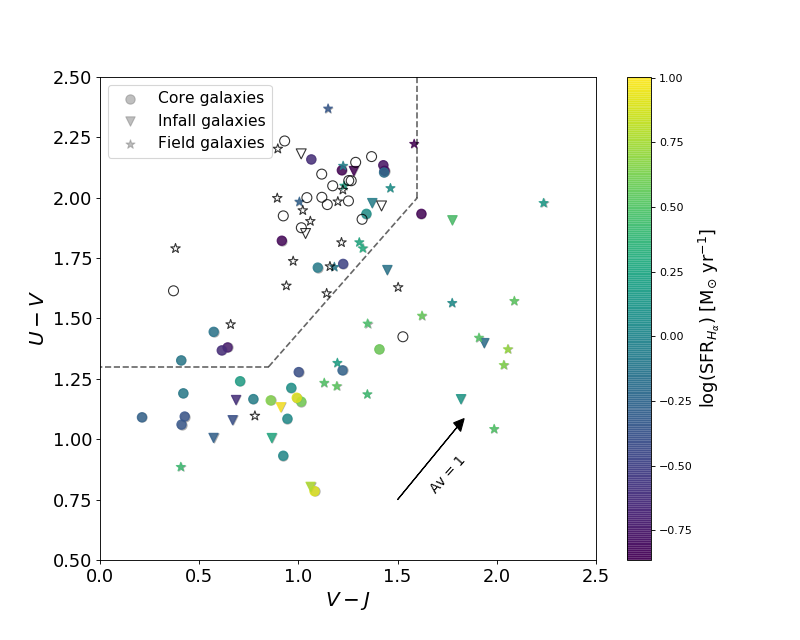}
\caption{{\em U -- V} vs. {\em V -- J} rest-frame colors color-coded by H$\alpha$-based SFR for the mass-complete sample of {\em HST}-observed galaxies. Galaxies with S/N $<3$ in H$\alpha$ are grey open symbols, which are predominantly located in the quiescent clump. The SFR color-coded points are the final sample of 67 galaxies selected for S/N $>$ 3 and emission line extraction quality. 51 of the 67 of the sources with F(H$\alpha$) S/N $>$ 3 lie in the star-forming region, with 21 residing in the quiescent region. The stellar continuum and H$\alpha$ emission line maps for the 21 passively-classified galaxies are shown in Appendix~\ref{fig:app2}. Black dashed lines follow the quiescent and star-forming definition of \protect\cite{will09}.  }
\label{fig:UVJ}
\end{figure}

From the \cite{Just_2019} catalog, there are 581 EDisCS galaxies in the 12 {\em HST} pointings FOV. This is further reduced to 326 after limiting the redshift range to 0.4 $<$ z $<$ 0.7. Adopting a S/N in H$\alpha$ cut $>$ 3 results in 190 sources in the sample. Finally, removing extractions that are unsatisfactory due to poor contamination modeling, artifacts, or being on the edge of the chip result in a sample of 163 galaxies, of which 67 (30 core, 13 infall, 24 field) are above the mass-complete limit of log$_{10}$($M_*$/M\textsubscript{\(\odot\)}) = 9.75. This mass-complete sample of H$\alpha$-emitter galaxies is dominated by blue, star-forming objects as shown in the {\em UVJ} diagram in Figure~\ref{fig:UVJ}. We discuss corrections to the wide-field photometry in more detail in Section~\ref{sec:app}.
\par
\begin{figure}
\centering
\includegraphics[scale=0.33]{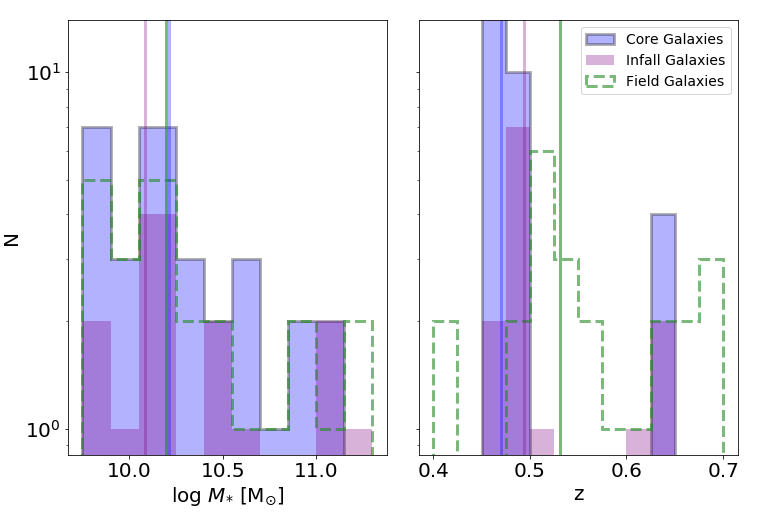}
\caption{(Left) The distribution of the stellar masses for each of the core, infall, and field mass-complete samples is shown in the blue, purple, and green-dashed histograms. The median value for each sample is the vertical line. (Right) The same three samples are shown with their distributions in redshift space, with the median shown again as the vertical lines. A more significant difference in z is apparent between the field and cluster samples and is corrected following the equation in \S~\ref{sec:corrs}. The elevated median for the infall sample is due to the lack of two samples for Cl1059 at z = 0.4564, which is the lowest z cluster. Thus, the median z is offset to a higher value than the core which includes galaxies from this cluster.}
\label{fig:mz}
\end{figure}

In Figure~\ref{fig:mz}, we present the distributions of the stellar masses and redshift for each environment in the mass-complete sample. While the core and infall have similar median values for {M$_*$} (K-S statistic of 0.12), the field masses average slightly higher. However, they follow a similar distribution with a 2-sample K-S statistic of 0.32 and 0.29 with the core and infall regions, respectively. 

In contrast, the redshift distributions have significant differences (K-S statistics: core-infall (0.60), core-field (0.65), infall-field (0.46)), with the field having the highest median value of the three environments. We therefore correct the SFRs of field galaxies to the median redshift of the cluster sample (0.48) using the following relation from \cite{schr15}. 

\begin{center}
\begin{equation}
log_{10}(SFR_{MS}[M\odot/yr]) = m - m_0 + a_0r - a_1[max(0,m - m_1 - a_2r)]^2.
\label{eq:sch}
\end{equation}
\end{center}

Here, r = log$_{10}$(1 + z), where z is difference between the median and individual redshift, m$_0$ = 0.5, a$_0$ = 0.15, a$_1$ = 0.3, m$_1$ = 0.6, a$_2$ = 2.5 and {\em m = log$_{10}$(M$_*$/10$^9$M$_{\odot}$}). For each galaxy in the field we compute the difference in SFR that would be expected from Eq.~\ref{eq:sch} between the galaxy and median redshift.  We apply that difference to correct the SFRs to the median redshift. This allows us to correct for any variation in the SFR that comes from redshift evolution. The lack of an infall sample in Cl1059 at z = 0.4564 is likely driving the variation in the median z's between the core and infall distributions.

\subsection{Stellar Mass -- SFR Relations}
\label{sec:sfr-m}

\begin{figure*}
\centering
\includegraphics[scale=0.475]{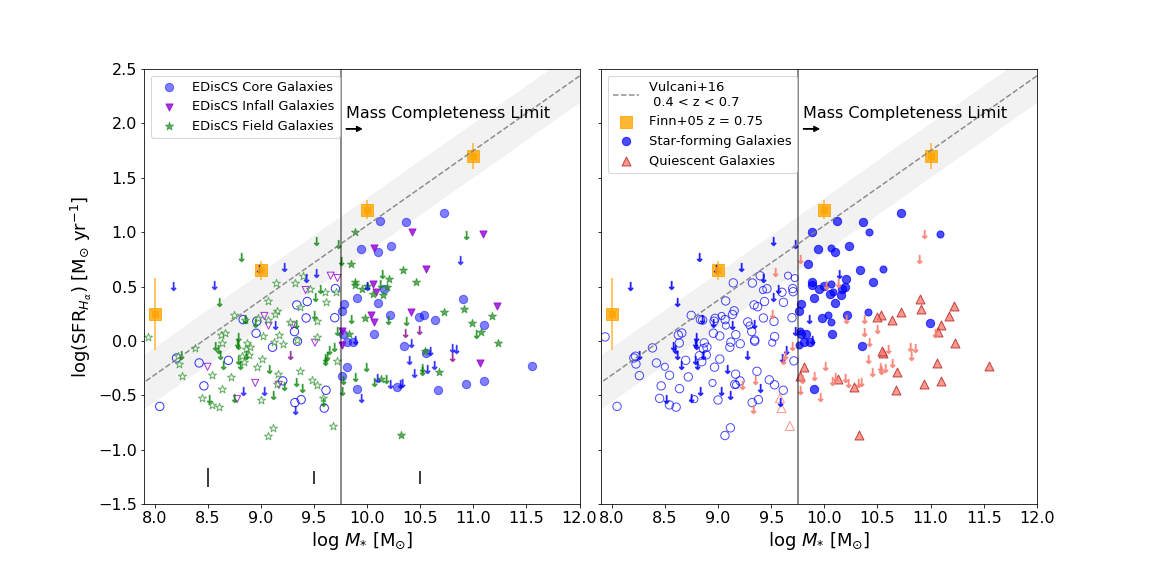}
\caption{The H$\alpha$-based SFR -- M$_*$ main sequence relation for 163 S/N >3 and 82 S/N <3 (down arrow) galaxies. (Left) The locations of these galaxies are distributed among the cluster core (blue circles), infall region (purple triangles), and field samples (green stars), where the mass-completeness line is denoted by the vertical black line and filled in symbols and with median errors shown as black lines in 3 mass bins for all environments at the bottom of the plot. The <3 S/N galaxies are plotted at their 3$\sigma$ limit, which occupy the lowest SFRs of the main sequence here. These data are systematically above the main sequence relations defined by \protect\cite{whitaker12} and \protect\cite{schr15}, which is shown in Appendix~\ref{sec:app3}. The scatter is larger than the literature, but is still expected due to varying star-formation histories and other disturbances throughout a galaxy lifetime. A comparison is shown to the EDisCS narrow band H${\alpha}$ SFRs in \protect\cite{finn05} (orange squares) at z = 0.75 and H${\alpha}$ SFRs \protect\cite{vul16} from the GLASS clusters at 0.4 $<$ z $<$ 0.7 (grey dashed line + 1$\sigma$ scatter). The median statistical error in the SFR in 3 mass bins is shown at the bottom of the figure. (Right) This same sample is now color-coded by location in Figure~\ref{fig:UVJ}, where blue circles are star-forming and salmon triangles are quiescent. The quiescent galaxies mostly occupy the lower portion of the main sequence, which also have suppressed SFRs.}
\label{fig:MS}
\end{figure*}

\begin{figure}
\centering
\includegraphics[scale=0.46]{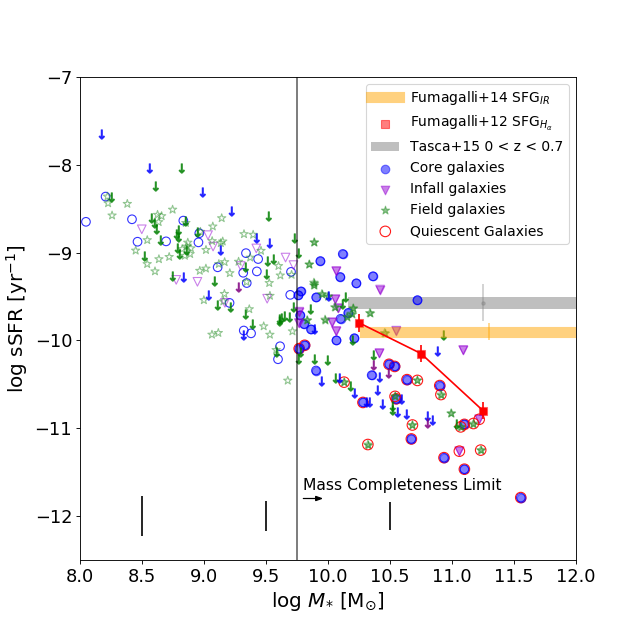}
\caption{The H$\alpha$-based sSFR -- M$_*$ relation for 163 S/N >3 and 82 S/N <3 (down arrow) galaxies, with median errors shown as black lines in 3 mass bins for all environments at the bottom of the plot. The environmental locations of these galaxies are distributed among the cluster core (blue circles), infall region (purple triangles), and field samples (green stars), where the mass-completeness line is denoted by the vertical black line and filled in symbols. The <3 S/N galaxies are plotted at their 3$\sigma$ limit, which occupy the lowest SFRs of the main sequence here. Comparisons to \protect\cite{tasca15} at 0 $<$ z $<$ 0,7, 3DHST \protect\citep{fum12} at z $\sim$0.9 and \protect\cite{fum14} at 0.3 $<$ z $<$ 0.7 are shown as a grey bar, red square and orange bar, respectively. Note that we have very few star-forming galaxies in the mass range as the comparison sets, but the \protect\cite{fum12} H$\alpha$ field relation aligns well with our sample, though at slightly higher sSFR. The {\em UVJ}-quiescent galaxies, which are designated by red circles, mostly occupy the lower right portion of the relation. The statistical error of the sSFR in 3 mass bins is shown at the bottom of the figure for each environment. }
\label{fig:ssfr}
\end{figure}

\begin{figure*}
\centering
\includegraphics[scale=0.475]{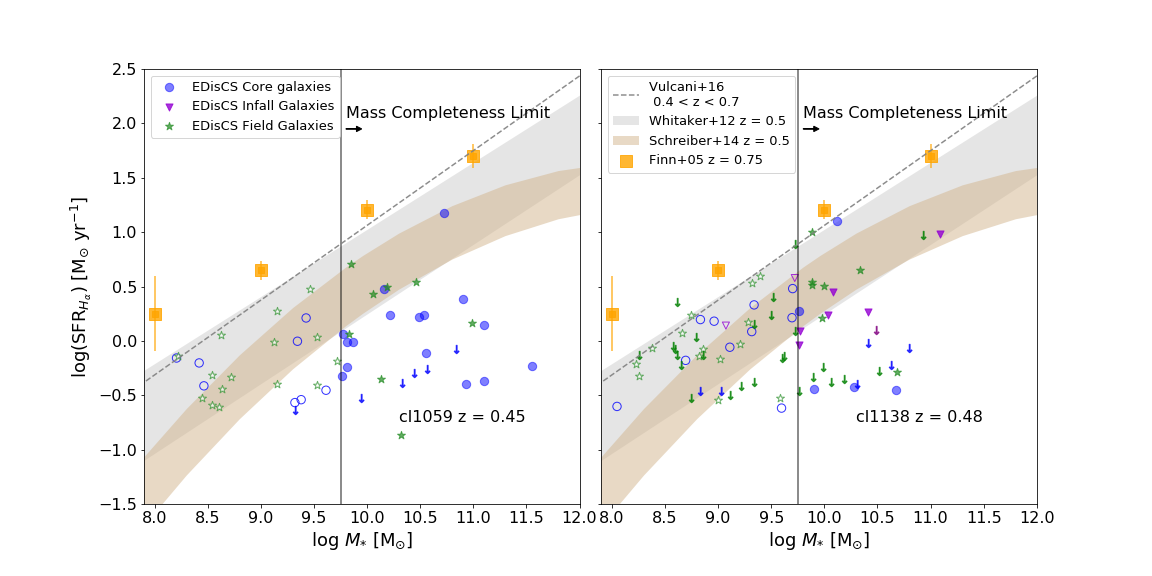}
\includegraphics[scale=0.475]{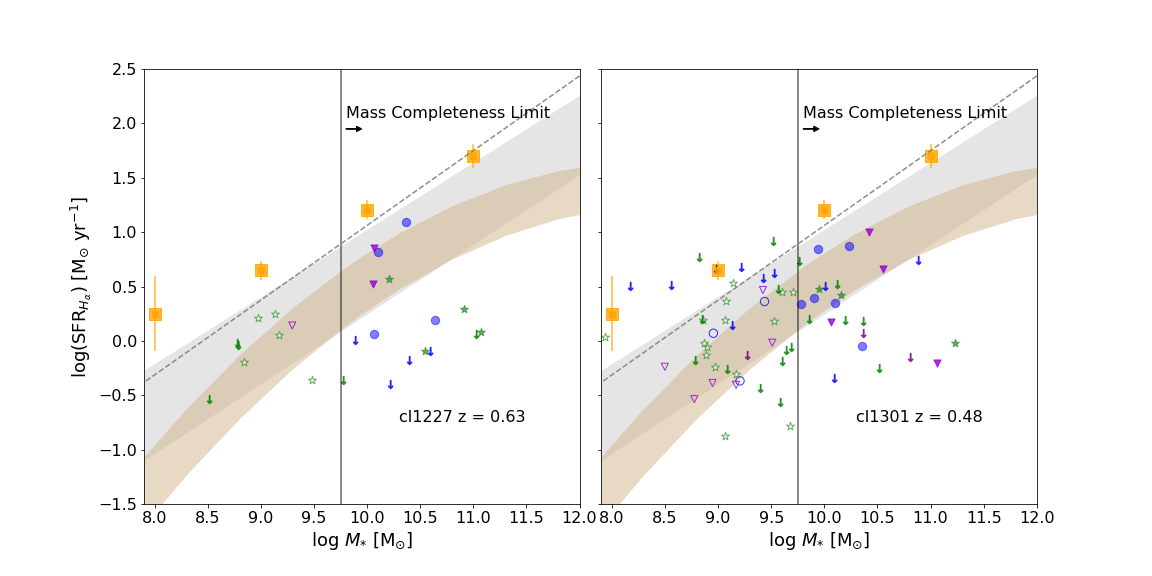}
\caption{Each panel shows the SFR vs. stellar mass as in the left panel of Figure~\ref{fig:MS}, but separated by cluster. Each cluster reveals a similar distribution of galaxies below the \protect\cite{vul16} mean distribution. This is confirmed with a 2-sample K-S test for each individual cluster compared to all cluster galaxies as follows (p-value, statistic): 1059: (0.31, 0.27) 1138: (0.82, 0.20) 1227: (0.3. 0.34) 1301: (0.32, 0.31). This indicates that no cluster is offset with respect to the others and influencing the combined relation. The highest-z cluster in the bottom right, Cl1227, has noticeably fewer galaxies, which is also evident in Figure~\ref{fig:prop}. Comparison lines to \protect\cite{whitaker12} and \protect\cite{schr15} at z = 0.5 are sown as the grey and tan curves, respectively. The full sample in a single panel compared to \protect\cite{vul16}, \protect\cite{whitaker12} and \protect\cite{schr15} is available in Appendix~\ref{sec:app3}.}
\label{fig:4MS}
\end{figure*}

In Figures~\ref{fig:MS} and ~\ref{fig:ssfr}, we present the H$\alpha$-derived SFR -- M$_*$ main sequence relation and specific SFR for four EDisCS clusters separated into three environments: core (blue circles), infall (purple triangles) and field (green stars) for 163 galaxies in the left panel. In the right panel, galaxies are divided by their classification from Figure~\ref{fig:UVJ}, where red triangles are defined as {\em UVJ}-quiescent. Both panels include galaxies with S/N in H$\alpha$ $<$ 3 as down arrows at their 3$\sigma$ upper limit SFR. A scatter of $\sim$1 dex is observed across all masses with a lack of flattening of the SFR relation for more massive galaxies in the cluster core as shown with \cite{schr15}. The median $\log_{10}$(SFR) for the mass-complete sample with S/N $>$ 3 in H$\alpha$ is 0.25 with a standard deviation of 0.43, while the $UVJ$ star-forming sample has a median of 0.46 and standard deviation of 0.37. The infall times of galaxies into the cluster environment can vary and contribute towards this large scatter, which is double the 1$\sigma$ value of 0.25 dex in GLASS clusters from \cite{vul16}. The mean SFRs for the three EDisCS clusters in \cite{finn05} at z = 0.75 are shown as orange squares. 2D image cutouts of the stellar and H${\alpha}$ maps for the {\em UVJ} quiescent are available in Figure~\ref{fig:app2}. We also show the SFR -- M$_*$ distribution on a cluster-by-cluster basis in Figure~\ref{fig:4MS}. The apparent distribution of galaxies seen in Figure~\ref{fig:MS} is not dominated by any individual cluster, but rather contains small contributions from each cluster. The distribution of SFRs appears to be similar between the clusters.

\begin{figure}
\centering
\includegraphics[scale=0.43]{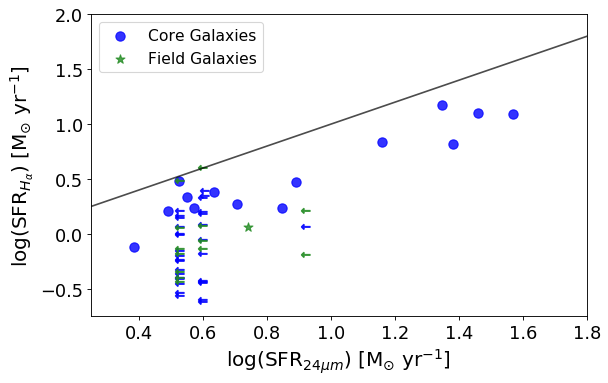}
\caption{The MIPS 24$\mu$m SFRs derived from the EDisCS core pointings of this sample from \protect\cite{finn10} compared to the H${\alpha}$ SFRs in this study. There are only matches between the core (blue circle) and field (green star) because these {\em Spitzer} pointings did not extend to the infall region. Left arrows are non-detections in 24$\mu$m at the 80\% completeness limit. The $\sim$0.2 -- 0.4 dex offset from the 24$\mu$ SFRs is similar to the one seen with GLASS. }
\label{fig:MSa}
\end{figure}

\begin{figure}
\centering
\includegraphics[scale=0.43]{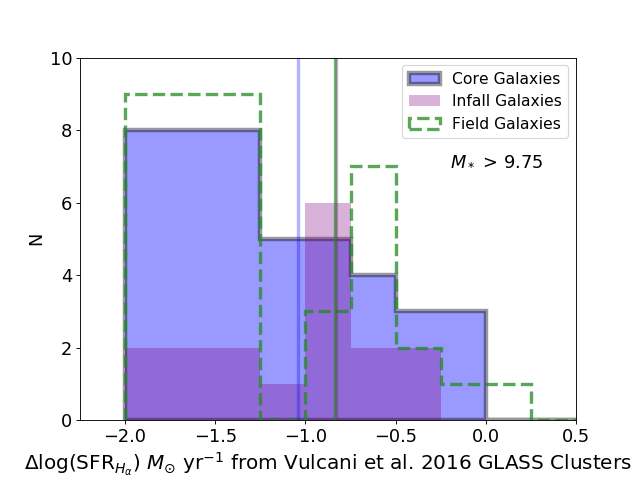}
\caption{The difference in SFR for each of the three environments (blue = core, purple = infall, field = green) from the \protect\cite{vul16} GLASS clusters H$\alpha$ main sequence. The median values for each environment are signified by the corresponding vertical line, which are $\sim$1 dex below the GLASS relation. 
}
\label{fig:hist-sfr}
\end{figure}

There are 21 galaxies identified as quiescent based upon the \cite{Just_2019} {\em UVJ} rest-frame colors, with five of them being in the core, two in the infall, and nine in the field. These are identified as red triangles in the right panel of Figure~\ref{fig:MS}. As seen in Figure~\ref{fig:UVJ}, there are galaxies that are quiescent based on their {\em UVJ} colors, but which have significant H$\alpha$ emission. We will discuss these galaxies in \S\ref{sec:dis}. GRIZLI produces a stellar continuum and emission line map for each observed galaxy, which is shown in Figure~\ref{fig:app2} for select galaxies. 

In Figure~\ref{fig:MSa}, we compare our H$\alpha$ SFRs to those derived from {\em Spitzer} MIPS 24$\mu$m emission \citep{finn10}. The observations from \cite{finn10} only covered the central regions of the EDisCS clusters. Therefore, our comparison only involves galaxies that are either in the core or are field galaxies in the central projected area of the cluster. Many of the galaxies in our H$\alpha$ sample are not detected in 24\micron. 15 galaxies (14 core and one field) in our H$\alpha$ sample are detected at 24\micron\ and these have SFR(H$\alpha$) that are very well correlated with SFR(24\micron) but are lower by 0.2 -- 0.3 dex. Given the median attenuation at H$\alpha$ of 0.46~mag, it is reasonable to assume that we might have slightly underestimated our attenuation values towards H$\alpha$. It might also be that the 24\micron\ detections are biased towards galaxies with a higher-than-average amount of obscured star formation. To test the robustness of our attenuation correction, we computed an alternate correction from \cite{kenn12} in which observed H$\alpha$ is corrected for attenuation using a scale applied to the total IR luminosity ($L({\rm TIR})$) such that $L({\rm H}\alpha)_{\rm corr} = L({\rm H}\alpha)_{\rm obs} + 0.0024L(TIR)$. $L({\rm TIR})$ was determined in \cite{finn10} using \cite{CharyElbaz01} to scale the observed 24\micron\ flux to $L({\rm TIR})$. The SFR derived from this alternatively corrected H$\alpha$ luminosity is very close to our default value, with a median difference of only 0.13~dex and a scatter of 0.17~dex. These comparisons give us confidence that our $UVJ$-derived attenuation is comparable to that derived using IR estimates. As a final note, in Appendix~\ref{fig:dist1}, we show a comparison of the main sequence from different authors at the same redshift. These estimates use both H$\alpha$ and UV$+$IR SFR indicators and also differ by $\sim 0.3$~dex. This underscores the systematic uncertainty inherent when comparing different SFR indicators. 
As an alternative way of comparing the SFRs across environment, in Figure~\ref{fig:hist-sfr} we show the distribution of the SFR with respect to the cluster-based main sequence from \cite{vul16} for each of the three environments in the mass-complete sample. 
The median SFR for each environment is $\sim$1 dex below the relation from \cite{vul16}.  While this offset may indicate problems with our extinction correction, we showed above that our H$\alpha$ based SFR measurements are within 0.2 -- 0.3 dex of those based on the IR.  We also demonstrate the systematic offsets in SFR estimates among different authors and attribute part of our disagreement with \cite{vul16} to this difference.  We performed 2-sample K-S tests on the mass-complete sample comparing core, infall, and field galaxies for all galaxies, $UVJ$ star-forming galaxies, and quiescent galaxies.  In all cases the K-S probabilities were significantly larger than 0.05.  
Therefore, we cannot rule out the null hypothesis that the core, infall, and field galaxies are drawn from the same SFR distribution.

\par
\begin{figure}
\centering
\includegraphics[scale=0.43]{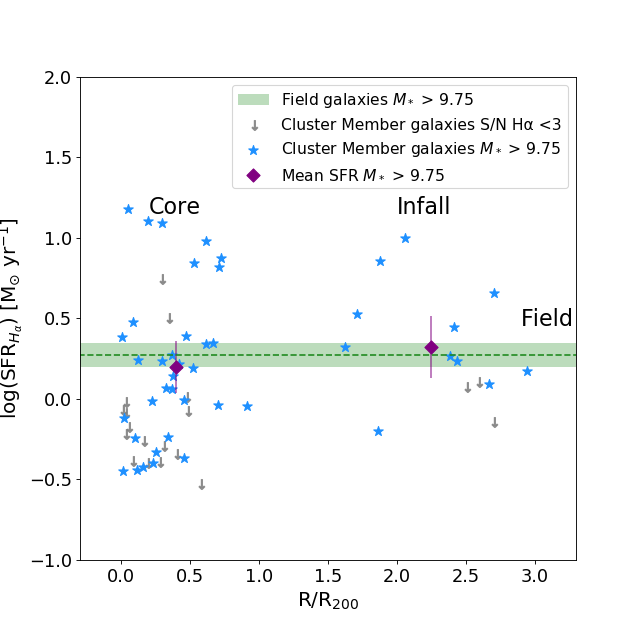}
\caption{The distance from the center of the cluster (defined as the BCG in \protect\cite{white05} is computed for each core and infall galaxy and compared to the SFR. The mass-complete sample is shown as blue stars for cluster members and the average for the field is the dashed green line, where error bars on the median are the 1$\sigma$ confidence level from a bootstrap resampling. Galaxies with S/N in H$\alpha$ $<$ 3 are the grey arrows and are not accounted for in the median values. All but three of these upper limits are for $UVJ$ quiescent galaxies and so the apparent radial dependence in the fraction of galaxies with upper limits just reflects the well-known radial dependence in the quiescent fraction in clusters. For the robust H$\alpha$ detections, no significant radial trend exists for the mass-complete sample.}
\label{fig:dist}
\end{figure}

In Figure~\ref{fig:dist}, we plot the SFR result as a function of distance from the cluster center in relation to R$_{200}$, where the mass-complete cluster sample ($>$ 10$^{9.75}$M\textsubscript{\(\odot\)}) is represented as the blue stars and the field galaxies are shown in green as a median SFR. Cluster galaxies below the mass-complete limit are plotted as grey down arrows at the 3$\sigma$ limit. The median for the mass complete cluster member sample in the core and infall region is shown as a purple triangle with 1$\sigma$ bootstrap resampling error bars. There is no observable difference in the SFRs between the three environments as in Figures~\ref{fig:MS}. 
\par

\par

\section{Discussion}
\label{sec:dis}
\par In this study, there is no significant difference in the distribution of SFRs between environments. The EDisCS cluster galaxies are roughly aligned with the GLASS clusters H$\alpha$-SFRs main sequence relation, which also does not reveal a variation from the field SFRs \citep{vul16}. \cite{koyama13} finds a similar result with H$\alpha$ observations of clusters, but the SFR limits are not deep enough to detect significantly suppressed galaxies. However, the lack of an environmental dependence on the SFR that is normally seen across all masses contradicts the notion that dense environments are contributing or directly responsible for gas quenching as evidenced by the buildup of quiescent galaxies \citep{patel09,vul10,pacc16}. Indeed, the cores \citep{pogg06} and infall regions (Cooper et al. in prep.) of the EDisCS clusters have a higher quenched fraction than the co-eval field, indicating that our clusters and their environments host processes that suppress galaxy star formation. 

The lack of a dependence of the SFR on environment that we find here can potentially be explained in the following ways. First, our sensitivity limits are not low enough to detect galaxies with significantly suppressed SFRs. This is evident in Figure~\ref{fig:MS} in that the majority of the $<$3 S/N galaxies appear to populate the bottom of the main sequence. Thus, there may be a tail of galaxies to lower SFRs, but we would be unable to detect this population with our data. The importance of highly sensitive SFR limits to interpret the distribution of SFRs in dense environments is illustrated in \cite{vul10}, in which they do find an excess of galaxies in EDisCS clusters with low SFRs compared to those in the field, but only because they probe well below the main sequence. 
\par 
It is also possible that the galaxies within our sample have not experienced significant quenching and the reasons for this vary by environment. Within the core, galaxies may have been recently accreted and are still within the `delay' period of the quenching process. In the infall region, the local density is lower than that of the core and may not create conditions capable of quenching.
\par 
Thirdly, the lack of a difference in field vs. cluster SFRs could mean that the timescale for the truncation of SF is rapid. In this case, galaxies that are undergoing external quenching will fall below our detection limits before we can observe them in their reduced SFR state. Such a rapid decline in SFRs caused by dense environments is consistent with the excess of post-starburst galaxies in dense environments as seen by \cite{pogg08}, \cite{muzzin12} and \cite{wild16}. It is not immediately clear why there is a diversity in the distribution of SFRs in different environments among different published works. It may be that much of the `action' is in the tails of the distribution, which requires not only deep observations, but also large sample sizes to characterize the distribution shapes well away from the median. Observations with the {\em James Webb Space Telescope} or deep UV+IR observations with WISE in the local universe may satisfy this criteria.
\par 
We should also consider that our cluster core and infall samples could likely contain interlopers, which has been estimated to be 15\% or more for clusters with historical datasets \citep{duarte15,wot18}. These galaxies may appear to be spectroscopic members by superposition or our redshift determination is incorrect.

Finally, some of our clusters have significant substructure, e.g. Cl1138.2-1133 \citep{DeLucia09}, which may make clustercentric radius a poor proxy for environment.  As a result, a given range in clustercentric radius could contain a large variation in local density, and in fact does for some of our clusters \citep{Just_2019}.  As locally dense regions host higher quenched fractions \citep{patel09, Just_2019}, this variation in local density could translate directly to a variation in the quenching efficiency at a fixed clustercentric radius.

\subsection{{\em UVJ}-quiescent galaxies with H$\alpha$ emission}

There are 21 galaxies in the $UVJ$ quiescent region that have H$\alpha$ emission that is detected with S/N $>3$ (Figure~\ref{fig:UVJ}). These galaxies lie systematically closer to the dividing line between quiescent and star-forming galaxies than the rest of $UVJ$ quiescent galaxies, however they still exist at red colors consistent with the larger passive population. As can be seen in Figure~\ref{fig:MS}, these $UVJ$-quiescent H$\alpha$ emitters also have systematically lower SFRs than $UVJ$-SF galaxies of the same stellar mass, and none have SFR greater than 3~M$_\odot$~yr$^{-1}$. The extend to SFR as low as 0.3~M$_\odot$~yr$^{-1}$. Continuum and emission-line postage stamps for all these galaxies are shown in Figure~\ref{fig:app2}. The emission is faint but visible in all 2D stamps and in the 1D spectrum and the spectra are free of artifacts. We entertain four possibilities to explain these sources.

First, we must explore the possibility that our rest-frame colors are uncertain and that these nominally $UVJ$-quiescent galaxies with H$\alpha$ emission actually lie in the SF region but were moved into the quiescent $UVJ$ region by random and systematic rest-frame color errors. This is a potential concern, especially given the calibration challenges that we experienced with the wide-field data and the additional rest-frame color corrections described in Appendix~\ref{sec:app}. We test for this possibility by comparing the $UVJ$ colors as derived from the photometry in this paper with the $UVJ$ colors derived from the original EDisCS photometry in the cluster cores. The original EDisCS photometry is well calibrated and results in a very well defined passive clump at the correct color location. We verified that the $UVJ$ colors derived from the wide-field data are slightly different from the EDisCS $UVJ$ colors on a galaxy-by-galaxy basis, but that the differences are not significant enough to move galaxies in and out of the passive region. Therefore, we conclude that these galaxies are indeed in the $UVJ$-quiescent region and that we should discuss the implication of them having significant amounts of H$\alpha$ emission.

\par 
Second, it is possible that weak AGN may be contributing to some of the emission. With our data we cannot explicitly rule out the role of an AGN. \cite{martini09} found only two X-ray AGN in 17 clusters at $z<0.4$. There are some objects that have spatially compact and linearly extended residuals in the emission line maps in Figure~\ref{fig:app2}, e.g. Cl1059-12.2-447, Cl1227-11.2-259. This could be an indicator that the continuum shape is not well modeled by the GRIZLI continuum subtraction, which could occur of the continuum has significant non-stellar contributions from an AGN as such templates are not included in the GRIZLI continuum models.  While this is a possibility, the emission lines for these galaxies, which admittedly have low signal-to-noise, do not look broad in the 1D spectra. We examine the position in the SFR -- M$_*$ plane of the 11 objects with such linear residuals and find that they do not occupy any favored place in either stellar mass or SFR, being sparsely spread in both quantities and not preferentially biasing the main sequence in any parameter. If these linear features do indeed correspond to AGN, the lack of bias with respect to the SFR -- M$_*$ plan would indicate that contamination by AGN is a minor contributor to the H$\alpha$ flux in this population.   

Third, it is possible that the H$\alpha$ emission comes from a `LIER'-like phenomena \citep{sarzi06,singh13,bel16,Rudnick17} in which gas from mass loss and accretion in quiescent galaxies is being heated by preexisting stellar populations, mostly post-AGB stars. `LIER' stands for `low-ionization emission-line region', which occurs in passive galaxies that have an emission line, much like the subset of {\em UVJ}-passive galaxies with H$\alpha$ emission. In a similar emission line study, \cite{Rudnick17} showed that [OII] emission in EDisCS quiescent galaxies was less common in galaxies in the EDisCS clusters and groups than in the field, where quiescent [OII] emitters comprised $\sim 5\%$ of the quiescent population with M$_*>10.4$ in clusters and groups, and $30\%$ in the field. Those authors attributed this suppression of [OII] in clusters to a combination of hydrodynamic stripping and a cutoff of gas accretion in dense environments. We do not have enough galaxies in this EDisCS subsample to make the same comparison but this could be a similar population of red emission line galaxies.
\par The fourth possibility is that we are catching galaxies as they are in the process of quenching their star formation and moving from the star-forming to quiescent region. In this case the low SFRs and position closer to the boundary of the $UVJ$-quiescent region could indicate that these galaxies are leaving the main sequence and joining the population with much lower SFRs \citep{cant16,foltz18,belli19,carnall20}. Such red emission line galaxies may be similar to those seen in other works \citep{Wolf09,vul10,koyama11} and could represent a distinct phase in the quenching of galaxies.
These results imply that caution must be taken in interpreting the true quiescent nature of galaxies classified by $UVJ$ techniques as truly quiescent. To assess if these $UVJ$-quiescent H$\alpha$ emitters are truly quenching, it would be beneficial to obtain high signal-to-noise spectra at medium resolution to model the spectra and search for evidence of young stellar populations \citep{webb20}. We could also obtain deep molecular gas observations to probe the cold gas reservoirs that would be needed to power the observed star formation.

\par

\section{Conclusions \& Future Work}
\label{sec:fut}
\par 
In this paper, we explore the environmental dependence of spectroscopically-derived H$\alpha$ star-formation in three distinct regimes in the vicinity of four galaxy clusters at $0.4<z<0.7$: cluster cores, infall regions, and the field. We combine {\em HST}/WFC3 G102 grism observations at 1$\mu$m with photometric and spectroscopic redshift priors to obtain a sample of 67 galaxies with secure redshifts, S/N in H$\alpha$ $>$ 3 and which are above our mass completeness limit for star-forming galaxies of M$_{*} $>$ 10^{9.75}$. 
\par
Our main findings are summarized as the following points: 
\par
\begin{enumerate}
\item With the combination of grism and redshift priors, we can obtain precise and accurate redshifts for galaxies with a range of stellar masses and intracluster locations.
\item We find no difference in the distribution of SFRs for galaxies in the three environments or as a function of radius from the cluster out to 3R$_{200}$.
\item We find 21 galaxies that are identified as $UVJ$-quiescent galaxies, but which have significant amounts of H$\alpha$ emission. We explore possible explanations for this emission that include star formation in quenching galaxies, AGN, and excitation of the gas by post-AGB stars. We conclude that there may be contributions from all of these scenarios.
\item The similarity of the SFR distributions for our core, infall, and field samples may be attributed to the delayed-then-rapid quenching scenario, where galaxies are unaffected for the first two -- four Gyr that they reside in the cluster environment, followed by a rapid quenching event that leaves the distribution of SFRs for star forming galaxies unaffected. We cannot conclusively test this scenario without significantly more galaxies measured to lower SFR sensitivity limits. However, it is possible that our H$\alpha$-detected galaxies have not experienced significant quenching processes. For the infall galaxies, this can be because of the relatively low densities that they inhabit while for the core galaxies it may be that they have recently been accreted by the cluster and are still in the ``delay" phase of their eventual quenching. Whichever intrinsic and extrinsic processes that do affect star formation in the infall regions and cores of our clusters must do so in a way that preserves the indistinguishable distribution of SFRs in the different environments, at least at the level constrained by our data. 
\end{enumerate}
\par
One possibility for using this dataset to explore the effect of environment on the star formation properties of galaxies would be to analyze the relative size of the stellar (traced by F105W) and H$\alpha$ disks. As different processes may result in a different ratio of these sizes, this may provide a new constraint on the quenching process. We will explore this in a future work.

\section*{Acknowledgements}

GR acknowledges support from the National Science Foundation grants AST-1517815, AST-1716690 and NASA HST grant HST-GO-12945.001. GR also acknowledges the support of an ESO visiting science fellowship. The work in this paper benefited significantly from interaction conducted as part of a team at the International Space Sciences Institute in Bern, Switzerland. JC thanks the Madison \& Lila Self Graduate Fellowship at the University of Kansas for financial and professional support. YJ acknowledges financial support from CONICYT PAI (Concurso Nacional de Inserci\'on en la Academia 2017) No. 79170132 and FONDECYT Iniciaci\'on 2018 No. 11180558.
BV acknowledges financial contribution from the contract ASI-INAF n.2017-14-H.0, from the grant PRIN MIUR 2017 n.20173ML3WW\_001 (PI Cimatti) and from the INAF main-stream funding programme (PI Vulcani). We thank Matthew Kirby for his assistance with producing the LDP redshifts that were used in this work.

\section*{Data Availability}
The raw HST data are available through MAST (Program ID 12945). The ground-based redshift catalog is available in \citealt{Just_2019} for clusters Cl11138, Cl1227, and Cl1301, while Cl1059 is available in \cite{white05}; DOI - 10.3847/1538-4357/ab44a0. Additional data on derived physical parameters are available in this paper.



\bibliographystyle{mnras}
\bibliography{allcites1} 

\begin{thebibliography}{}
\makeatletter
\relax
\def\mn@urlcharsother{\let\do\@makeother \do\$\do\&\do\#\do\^\do\_\do\%\do\~}
\def\mn@doi{\begingroup\mn@urlcharsother \@ifnextchar [ {\mn@doi@}
  {\mn@doi@[]}}
\def\mn@doi@[#1]#2{\def\@tempa{#1}\ifx\@tempa\@empty \href
  {http://dx.doi.org/#2} {doi:#2}\else \href {http://dx.doi.org/#2} {#1}\fi
  \endgroup}
\def\mn@eprint#1#2{\mn@eprint@#1:#2::\@nil}
\def\mn@eprint@arXiv#1{\href {http://arxiv.org/abs/#1} {{\tt arXiv:#1}}}
\def\mn@eprint@dblp#1{\href {http://dblp.uni-trier.de/rec/bibtex/#1.xml}
  {dblp:#1}}
\def\mn@eprint@#1:#2:#3:#4\@nil{\def\@tempa {#1}\def\@tempb {#2}\def\@tempc
  {#3}\ifx \@tempc \@empty \let \@tempc \@tempb \let \@tempb \@tempa \fi \ifx
  \@tempb \@empty \def\@tempb {arXiv}\fi \@ifundefined
  {mn@eprint@\@tempb}{\@tempb:\@tempc}{\expandafter \expandafter \csname
  mn@eprint@\@tempb\endcsname \expandafter{\@tempc}}}

\bibitem[\protect\citeauthoryear{{Abramson} et~al.,}{{Abramson}
  et~al.}{2018}]{abra18}
{Abramson} L.~E.,  et~al., 2018, \mn@doi [\aj] {10.3847/1538-3881/aac822},
  \href {https://ui.adsabs.harvard.edu/abs/2018AJ....156...29A} {156, 29}

\bibitem[\protect\citeauthoryear{{Baade} et~al.,}{{Baade}
  et~al.}{1999}]{baade99}
{Baade} D.,  et~al., 1999, The Messenger, \href
  {https://ui.adsabs.harvard.edu/abs/1999Msngr..95...15B} {95, 15}

\bibitem[\protect\citeauthoryear{{Balogh}, {Morris}, {Yee}, {Carlberg}  \&
  {Ellingson}}{{Balogh} et~al.}{1997}]{bal97}
{Balogh} M.~L.,  {Morris} S.~L.,  {Yee} H.~K.~C.,  {Carlberg} R.~G.,
  {Ellingson} E.,  1997, \mn@doi [\apjl] {10.1086/310927}, \href
  {https://ui.adsabs.harvard.edu/abs/1997ApJ...488L..75B} {488, L75}

\bibitem[\protect\citeauthoryear{{Balogh} et~al.,}{{Balogh}
  et~al.}{2016}]{Balogh16}
{Balogh} M.~L.,  et~al., 2016, \mn@doi [\mnras] {10.1093/mnras/stv2949}, \href
  {https://ui.adsabs.harvard.edu/abs/2016MNRAS.456.4364B} {456, 4364}

\bibitem[\protect\citeauthoryear{{Barro} et~al.,}{{Barro}
  et~al.}{2014}]{barro14}
{Barro} G.,  et~al., 2014, \mn@doi [\apj] {10.1088/0004-637X/791/1/52}, \href
  {https://ui.adsabs.harvard.edu/abs/2014ApJ...791...52B} {791, 52}

\bibitem[\protect\citeauthoryear{{Belfiore} et~al.,}{{Belfiore}
  et~al.}{2016}]{bel16}
{Belfiore} F.,  et~al., 2016, \mn@doi [\mnras] {10.1093/mnras/stw1234}, \href
  {https://ui.adsabs.harvard.edu/abs/2016MNRAS.461.3111B} {461, 3111}

\bibitem[\protect\citeauthoryear{{Bell} \& {de Jong}}{{Bell} \& {de
  Jong}}{2001}]{Bell01}
{Bell} E.~F.,  {de Jong} R.~S.,  2001, \mn@doi [\apj] {10.1086/319728}, \href
  {https://ui.adsabs.harvard.edu/abs/2001ApJ...550..212B} {550, 212}

\bibitem[\protect\citeauthoryear{{Bell} et~al.,}{{Bell} et~al.}{2004}]{bell04}
{Bell} E.~F.,  et~al., 2004, \mn@doi [\apj] {10.1086/420778}, \href
  {https://ui.adsabs.harvard.edu/abs/2004ApJ...608..752B} {608, 752}

\bibitem[\protect\citeauthoryear{{Belli}, {Newman}  \& {Ellis}}{{Belli}
  et~al.}{2019}]{belli19}
{Belli} S.,  {Newman} A.~B.,   {Ellis} R.~S.,  2019, \mn@doi [\apj]
  {10.3847/1538-4357/ab07af}, \href
  {https://ui.adsabs.harvard.edu/abs/2019ApJ...874...17B} {874, 17}

\bibitem[\protect\citeauthoryear{{Bouch{\'e}} et~al.,}{{Bouch{\'e}}
  et~al.}{2010}]{bouche}
{Bouch{\'e}} N.,  et~al., 2010, \mn@doi [\apj] {10.1088/0004-637X/718/2/1001},
  \href {https://ui.adsabs.harvard.edu/abs/2010ApJ...718.1001B} {718, 1001}

\bibitem[\protect\citeauthoryear{{Bouwens}, {Illingworth}, {Oesch}, {Caruana},
  {Holwerda}, {Smit}  \& {Wilkins}}{{Bouwens} et~al.}{2015}]{bouwens15}
{Bouwens} R.~J.,  {Illingworth} G.~D.,  {Oesch} P.~A.,  {Caruana} J.,
  {Holwerda} B.,  {Smit} R.,   {Wilkins} S.,  2015, \mn@doi [\apj]
  {10.1088/0004-637X/811/2/140}, \href
  {https://ui.adsabs.harvard.edu/abs/2015ApJ...811..140B} {811, 140}

\bibitem[\protect\citeauthoryear{{Brammer}, {Lundgren}, {Marchesini},
  {Momcheva}, {Pirzkal}, {Ryan}, {Vang}  \& {Wake}}{{Brammer}
  et~al.}{2016}]{grizli}
{Brammer} G.,  {Lundgren} B.~F.,  {Marchesini} D.,  {Momcheva} I.~G.,
  {Pirzkal} N.,  {Ryan} R.~E.,  {Vang} A.,   {Wake} D.~A.,  2016, {Grizli: The
  Grism redshift \& Line Database for HST WFC3/IR Spectroscopy}, HST Proposal

\bibitem[\protect\citeauthoryear{{Brinchmann}, {Charlot}, {White}, {Tremonti},
  {Kauffmann}, {Heckman}  \& {Brinkmann}}{{Brinchmann} et~al.}{2004}]{brinch07}
{Brinchmann} J.,  {Charlot} S.,  {White} S.~D.~M.,  {Tremonti} C.,  {Kauffmann}
  G.,  {Heckman} T.,   {Brinkmann} J.,  2004, \mn@doi [\mnras]
  {10.1111/j.1365-2966.2004.07881.x}, \href
  {https://ui.adsabs.harvard.edu/abs/2004MNRAS.351.1151B} {351, 1151}

\bibitem[\protect\citeauthoryear{{Calzetti}, {Armus}, {Bohlin}, {Kinney},
  {Koornneef}  \& {Storchi-Bergmann}}{{Calzetti} et~al.}{2000}]{calzetti00}
{Calzetti} D.,  {Armus} L.,  {Bohlin} R.~C.,  {Kinney} A.~L.,  {Koornneef} J.,
   {Storchi-Bergmann} T.,  2000, \mn@doi [\apj] {10.1086/308692}, \href
  {https://ui.adsabs.harvard.edu/abs/2000ApJ...533..682C} {533, 682}

\bibitem[\protect\citeauthoryear{{Cantale} et~al.,}{{Cantale}
  et~al.}{2016}]{cant16}
{Cantale} N.,  et~al., 2016, \mn@doi [\aap] {10.1051/0004-6361/201525801},
  \href {https://ui.adsabs.harvard.edu/abs/2016A&A...589A..82C} {589, A82}

\bibitem[\protect\citeauthoryear{{Carleton}, {Guo}, {Nayyeri}, {Cooper},
  {Rudnick}  \& {Whitaker}}{{Carleton} et~al.}{2020}]{carl20}
{Carleton} T.,  {Guo} Y.,  {Nayyeri} H.,  {Cooper} M.,  {Rudnick} G.,
  {Whitaker} K.,  2020, \mn@doi [\mnras] {10.1093/mnras/stz3216}, \href
  {https://ui.adsabs.harvard.edu/abs/2020MNRAS.491.2822C} {491, 2822}

\bibitem[\protect\citeauthoryear{{Carnall} et~al.,}{{Carnall}
  et~al.}{2020}]{carnall20}
{Carnall} A.~C.,  et~al., 2020, \mn@doi [\mnras] {10.1093/mnras/staa1535},
  \href {https://ui.adsabs.harvard.edu/abs/2020MNRAS.496..695C} {496, 695}

\bibitem[\protect\citeauthoryear{{Chabrier}}{{Chabrier}}{2003}]{chabimf}
{Chabrier} G.,  2003, \mn@doi [\pasp] {10.1086/376392}, \href
  {https://ui.adsabs.harvard.edu/abs/2003PASP..115..763C} {115, 763}

\bibitem[\protect\citeauthoryear{Chary \& Elbaz}{Chary \&
  Elbaz}{2001}]{CharyElbaz01}
Chary R.,  Elbaz D.,  2001, The Astrophysical Journal, 556, 562

\bibitem[\protect\citeauthoryear{{Clowe} et~al.,}{{Clowe}
  et~al.}{2006}]{clowe06}
{Clowe} D.,  et~al., 2006, \mn@doi [\aap] {10.1051/0004-6361:20041787}, \href
  {https://ui.adsabs.harvard.edu/abs/2006A&A...451..395C} {451, 395}

\bibitem[\protect\citeauthoryear{{Cucciati} et~al.,}{{Cucciati}
  et~al.}{2012}]{cucc12}
{Cucciati} O.,  et~al., 2012, \mn@doi [\aap] {10.1051/0004-6361/201118010},
  \href {https://ui.adsabs.harvard.edu/abs/2012A&A...539A..31C} {539, A31}

\bibitem[\protect\citeauthoryear{{Daddi} et~al.,}{{Daddi}
  et~al.}{2007}]{daddi07}
{Daddi} E.,  et~al., 2007, \mn@doi [\apj] {10.1086/521818}, \href
  {https://ui.adsabs.harvard.edu/abs/2007ApJ...670..156D} {670, 156}

\bibitem[\protect\citeauthoryear{{De Lucia} et~al.,}{{De Lucia}
  et~al.}{2004}]{delucia04}
{De Lucia} G.,  et~al., 2004, \mn@doi [\apjl] {10.1086/423373}, \href
  {https://ui.adsabs.harvard.edu/abs/2004ApJ...610L..77D} {610, L77}

\bibitem[\protect\citeauthoryear{{De Lucia}, {Poggianti}, {Halliday},
  {Milvang-Jensen}, {Noll}, {Smail}  \& {Zaritsky}}{{De Lucia}
  et~al.}{2009}]{DeLucia09}
{De Lucia} G.,  {Poggianti} B.~M.,  {Halliday} C.,  {Milvang-Jensen} B.,
  {Noll} S.,  {Smail} I.,   {Zaritsky} D.,  2009, \mn@doi [\mnras]
  {10.1111/j.1365-2966.2009.15435.x}, \href
  {https://ui.adsabs.harvard.edu/abs/2009MNRAS.400...68D} {400, 68}

\bibitem[\protect\citeauthoryear{{De Lucia}, {Weinmann}, {Poggianti},
  {Arag{\'o}n-Salamanca}  \& {Zaritsky}}{{De Lucia} et~al.}{2012}]{DeLucia12}
{De Lucia} G.,  {Weinmann} S.,  {Poggianti} B.~M.,  {Arag{\'o}n-Salamanca} A.,
   {Zaritsky} D.,  2012, \mn@doi [\mnras] {10.1111/j.1365-2966.2012.20983.x},
  \href {https://ui.adsabs.harvard.edu/abs/2012MNRAS.423.1277D} {423, 1277}

\bibitem[\protect\citeauthoryear{{Desai} et~al.,}{{Desai}
  et~al.}{2007}]{desai07}
{Desai} V.,  et~al., 2007, \mn@doi [\apj] {10.1086/513310}, \href
  {https://ui.adsabs.harvard.edu/abs/2007ApJ...660.1151D} {660, 1151}

\bibitem[\protect\citeauthoryear{{Dom{\'\i}nguez S{\'a}nchez}
  et~al.,}{{Dom{\'\i}nguez S{\'a}nchez} et~al.}{2014}]{dom14}
{Dom{\'\i}nguez S{\'a}nchez} H.,  et~al., 2014, \mn@doi [\mnras]
  {10.1093/mnras/stu503}, \href
  {https://ui.adsabs.harvard.edu/abs/2014MNRAS.441....2D} {441, 2}

\bibitem[\protect\citeauthoryear{{Dressler}}{{Dressler}}{1980}]{dressler80}
{Dressler} A.,  1980, \mn@doi [\apj] {10.1086/157753}, \href
  {https://ui.adsabs.harvard.edu/abs/1980ApJ...236..351D} {236, 351}

\bibitem[\protect\citeauthoryear{{Duarte} \& {Mamon}}{{Duarte} \&
  {Mamon}}{2015}]{duarte15}
{Duarte} M.,  {Mamon} G.~A.,  2015, \mn@doi [\mnras] {10.1093/mnras/stv1799},
  \href {https://ui.adsabs.harvard.edu/abs/2015MNRAS.453.3848D} {453, 3848}

\bibitem[\protect\citeauthoryear{{Dutton}, {Conroy}, {van den Bosch}, {Prada}
  \& {More}}{{Dutton} et~al.}{2010}]{dutton}
{Dutton} A.~A.,  {Conroy} C.,  {van den Bosch} F.~C.,  {Prada} F.,   {More} S.,
   2010, \mn@doi [\mnras] {10.1111/j.1365-2966.2010.16911.x}, \href
  {https://ui.adsabs.harvard.edu/abs/2010MNRAS.407....2D} {407, 2}

\bibitem[\protect\citeauthoryear{{Elbaz} et~al.,}{{Elbaz}
  et~al.}{2007}]{elbaz07}
{Elbaz} D.,  et~al., 2007, \mn@doi [\aap] {10.1051/0004-6361:20077525}, \href
  {https://ui.adsabs.harvard.edu/abs/2007A&A...468...33E} {468, 33}

\bibitem[\protect\citeauthoryear{{Erfanianfar} et~al.,}{{Erfanianfar}
  et~al.}{2016}]{erf16}
{Erfanianfar} G.,  et~al., 2016, \mn@doi [\mnras] {10.1093/mnras/stv2485},
  \href {https://ui.adsabs.harvard.edu/abs/2016MNRAS.455.2839E} {455, 2839}

\bibitem[\protect\citeauthoryear{{Finn} et~al.,}{{Finn} et~al.}{2005}]{finn05}
{Finn} R.~A.,  et~al., 2005, \mn@doi [\apj] {10.1086/431642}, \href
  {https://ui.adsabs.harvard.edu/abs/2005ApJ...630..206F} {630, 206}

\bibitem[\protect\citeauthoryear{{Finn} et~al.,}{{Finn} et~al.}{2010}]{finn10}
{Finn} R.~A.,  et~al., 2010, \mn@doi [\apj] {10.1088/0004-637X/720/1/87}, \href
  {https://ui.adsabs.harvard.edu/abs/2010ApJ...720...87F} {720, 87}

\bibitem[\protect\citeauthoryear{{Finn} et~al.,}{{Finn} et~al.}{2018}]{Finn18}
{Finn} R.~A.,  et~al., 2018, \mn@doi [\apj] {10.3847/1538-4357/aac32a}, \href
  {https://ui.adsabs.harvard.edu/abs/2018ApJ...862..149F} {862, 149}

\bibitem[\protect\citeauthoryear{{Foltz} et~al.,}{{Foltz}
  et~al.}{2018}]{foltz18}
{Foltz} R.,  et~al., 2018, \mn@doi [\apj] {10.3847/1538-4357/aad80d}, \href
  {https://ui.adsabs.harvard.edu/abs/2018ApJ...866..136F} {866, 136}

\bibitem[\protect\citeauthoryear{{Fossati} et~al.,}{{Fossati}
  et~al.}{2017}]{foss17}
{Fossati} M.,  et~al., 2017, \mn@doi [\apj] {10.3847/1538-4357/835/2/153},
  \href {https://ui.adsabs.harvard.edu/abs/2017ApJ...835..153F} {835, 153}

\bibitem[\protect\citeauthoryear{{Fumagalli} et~al.,}{{Fumagalli}
  et~al.}{2012}]{fum12}
{Fumagalli} M.,  et~al., 2012, \mn@doi [\apjl] {10.1088/2041-8205/757/2/L22},
  \href {https://ui.adsabs.harvard.edu/abs/2012ApJ...757L..22F} {757, L22}

\bibitem[\protect\citeauthoryear{{Fumagalli} et~al.,}{{Fumagalli}
  et~al.}{2014}]{fum14}
{Fumagalli} M.,  et~al., 2014, \mn@doi [\apj] {10.1088/0004-637X/796/1/35},
  \href {https://ui.adsabs.harvard.edu/abs/2014ApJ...796...35F} {796, 35}

\bibitem[\protect\citeauthoryear{{G{\'o}mez} et~al.,}{{G{\'o}mez}
  et~al.}{2003}]{gomez03}
{G{\'o}mez} P.~L.,  et~al., 2003, \mn@doi [\apj] {10.1086/345593}, \href
  {https://ui.adsabs.harvard.edu/abs/2003ApJ...584..210G} {584, 210}

\bibitem[\protect\citeauthoryear{{Gonzalez}, {Zaritsky}, {Dalcanton}  \&
  {Nelson}}{{Gonzalez} et~al.}{2001}]{gonz01}
{Gonzalez} A.~H.,  {Zaritsky} D.,  {Dalcanton} J.~J.,   {Nelson} A.,  2001,
  \mn@doi [\apjs] {10.1086/322541}, \href
  {https://ui.adsabs.harvard.edu/abs/2001ApJS..137..117G} {137, 117}

\bibitem[\protect\citeauthoryear{{Grogin} et~al.,}{{Grogin}
  et~al.}{2011}]{grogin11}
{Grogin} N.~A.,  et~al., 2011, \mn@doi [\apjs] {10.1088/0067-0049/197/2/35},
  \href {https://ui.adsabs.harvard.edu/abs/2011ApJS..197...35G} {197, 35}

\bibitem[\protect\citeauthoryear{{Haines} et~al.,}{{Haines}
  et~al.}{2015}]{haines15}
{Haines} C.~P.,  et~al., 2015, \mn@doi [\apj] {10.1088/0004-637X/806/1/101},
  \href {https://ui.adsabs.harvard.edu/abs/2015ApJ...806..101H} {806, 101}

\bibitem[\protect\citeauthoryear{{Halliday} et~al.,}{{Halliday}
  et~al.}{2004}]{halliday04}
{Halliday} C.,  et~al., 2004, \mn@doi [\aap] {10.1051/0004-6361:20041304},
  \href {https://ui.adsabs.harvard.edu/abs/2004A&A...427..397H} {427, 397}

\bibitem[\protect\citeauthoryear{{Hopkins}, {Kere{\v{s}}}, {O{\~n}orbe},
  {Faucher-Gigu{\`e}re}, {Quataert}, {Murray}  \& {Bullock}}{{Hopkins}
  et~al.}{2014}]{hopkins14}
{Hopkins} P.~F.,  {Kere{\v{s}}} D.,  {O{\~n}orbe} J.,  {Faucher-Gigu{\`e}re}
  C.-A.,  {Quataert} E.,  {Murray} N.,   {Bullock} J.~S.,  2014, \mn@doi
  [\mnras] {10.1093/mnras/stu1738}, \href
  {https://ui.adsabs.harvard.edu/abs/2014MNRAS.445..581H} {445, 581}

\bibitem[\protect\citeauthoryear{{Jaff{\'e}} et~al.,}{{Jaff{\'e}}
  et~al.}{2016}]{jaffe16}
{Jaff{\'e}} Y.~L.,  et~al., 2016, \mn@doi [\mnras] {10.1093/mnras/stw984},
  \href {https://ui.adsabs.harvard.edu/abs/2016MNRAS.461.1202J} {461, 1202}

\bibitem[\protect\citeauthoryear{Just et~al.,}{Just et~al.}{2019}]{Just_2019}
Just D.~W.,  et~al., 2019, \mn@doi [The Astrophysical Journal]
  {10.3847/1538-4357/ab44a0}, 885, 6

\bibitem[\protect\citeauthoryear{{Karim} et~al.,}{{Karim}
  et~al.}{2011}]{karim11}
{Karim} A.,  et~al., 2011, \mn@doi [\apj] {10.1088/0004-637X/730/2/61}, \href
  {https://ui.adsabs.harvard.edu/abs/2011ApJ...730...61K} {730, 61}

\bibitem[\protect\citeauthoryear{{Kauffmann}, {White}, {Heckman}, {M{\'e}nard},
  {Brinchmann}, {Charlot}, {Tremonti}  \& {Brinkmann}}{{Kauffmann}
  et~al.}{2004}]{kauff04}
{Kauffmann} G.,  {White} S. D.~M.,  {Heckman} T.~M.,  {M{\'e}nard} B.,
  {Brinchmann} J.,  {Charlot} S.,  {Tremonti} C.,   {Brinkmann} J.,  2004,
  \mn@doi [\mnras] {10.1111/j.1365-2966.2004.08117.x}, \href
  {https://ui.adsabs.harvard.edu/abs/2004MNRAS.353..713K} {353, 713}

\bibitem[\protect\citeauthoryear{{Kennicutt}}{{Kennicutt}}{1998}]{Kennicutt98}
{Kennicutt} Robert~C. J.,  1998, \mn@doi [\araa]
  {10.1146/annurev.astro.36.1.189}, \href
  {https://ui.adsabs.harvard.edu/abs/1998ARA&A..36..189K} {36, 189}

\bibitem[\protect\citeauthoryear{{Kennicutt} \& {Evans}}{{Kennicutt} \&
  {Evans}}{2012}]{kenn12}
{Kennicutt} R.~C.,  {Evans} N.~J.,  2012, \mn@doi [\araa]
  {10.1146/annurev-astro-081811-125610}, \href
  {https://ui.adsabs.harvard.edu/abs/2012ARA&A..50..531K} {50, 531}

\bibitem[\protect\citeauthoryear{{Kewley} \& {Ellison}}{{Kewley} \&
  {Ellison}}{2008}]{kewley08}
{Kewley} L.~J.,  {Ellison} S.~L.,  2008, \mn@doi [\apj] {10.1086/587500}, \href
  {https://ui.adsabs.harvard.edu/abs/2008ApJ...681.1183K} {681, 1183}

\bibitem[\protect\citeauthoryear{{Kodama}, {Balogh}, {Smail}, {Bower}  \&
  {Nakata}}{{Kodama} et~al.}{2004}]{kodama04}
{Kodama} T.,  {Balogh} M.~L.,  {Smail} I.,  {Bower} R.~G.,   {Nakata} F.,
  2004, \mn@doi [\mnras] {10.1111/j.1365-2966.2004.08271.x}, \href
  {https://ui.adsabs.harvard.edu/abs/2004MNRAS.354.1103K} {354, 1103}

\bibitem[\protect\citeauthoryear{{Koyama}, {Kodama}, {Nakata}, {Shimasaku}  \&
  {Okamura}}{{Koyama} et~al.}{2011}]{koyama11}
{Koyama} Y.,  {Kodama} T.,  {Nakata} F.,  {Shimasaku} K.,   {Okamura} S.,
  2011, \mn@doi [\apj] {10.1088/0004-637X/734/1/66}, \href
  {https://ui.adsabs.harvard.edu/abs/2011ApJ...734...66K} {734, 66}

\bibitem[\protect\citeauthoryear{{Koyama} et~al.,}{{Koyama}
  et~al.}{2013}]{koyama13}
{Koyama} Y.,  et~al., 2013, \mn@doi [\mnras] {10.1093/mnras/stt1035}, \href
  {https://ui.adsabs.harvard.edu/abs/2013MNRAS.434..423K} {434, 423}

\bibitem[\protect\citeauthoryear{{Lang} et~al.,}{{Lang} et~al.}{2014}]{lang14}
{Lang} P.,  et~al., 2014, \mn@doi [\apj] {10.1088/0004-637X/788/1/11}, \href
  {https://ui.adsabs.harvard.edu/abs/2014ApJ...788...11L} {788, 11}

\bibitem[\protect\citeauthoryear{{Larson}, {Tinsley}  \& {Caldwell}}{{Larson}
  et~al.}{1980}]{larson80}
{Larson} R.~B.,  {Tinsley} B.~M.,   {Caldwell} C.~N.,  1980, \mn@doi [\apj]
  {10.1086/157917}, \href
  {https://ui.adsabs.harvard.edu/abs/1980ApJ...237..692L} {237, 692}

\bibitem[\protect\citeauthoryear{{Lee-Brown} et~al.,}{{Lee-Brown}
  et~al.}{2017}]{LeeBrown17}
{Lee-Brown} D.~B.,  et~al., 2017, \mn@doi [\apj] {10.3847/1538-4357/aa7948},
  \href {https://ui.adsabs.harvard.edu/abs/2017ApJ...844...43L} {844, 43}

\bibitem[\protect\citeauthoryear{{Lewis} et~al.,}{{Lewis}
  et~al.}{2002}]{lewis02}
{Lewis} I.,  et~al., 2002, \mn@doi [\mnras] {10.1046/j.1365-8711.2002.05558.x},
  \href {https://ui.adsabs.harvard.edu/abs/2002MNRAS.334..673L} {334, 673}

\bibitem[\protect\citeauthoryear{{Lotz} et~al.,}{{Lotz} et~al.}{2013}]{Lotz13}
{Lotz} J.~M.,  et~al., 2013, \mn@doi [\apj] {10.1088/0004-637X/773/2/154},
  \href {https://ui.adsabs.harvard.edu/abs/2013ApJ...773..154L} {773, 154}

\bibitem[\protect\citeauthoryear{{Madau} \& {Dickinson}}{{Madau} \&
  {Dickinson}}{2014}]{md}
{Madau} P.,  {Dickinson} M.,  2014, \mn@doi [\araa]
  {10.1146/annurev-astro-081811-125615}, \href
  {https://ui.adsabs.harvard.edu/abs/2014ARA&A..52..415M} {52, 415}

\bibitem[\protect\citeauthoryear{{Mannucci}, {Cresci}, {Maiolino}, {Marconi}
  \& {Gnerucci}}{{Mannucci} et~al.}{2010}]{Mannucci10}
{Mannucci} F.,  {Cresci} G.,  {Maiolino} R.,  {Marconi} A.,   {Gnerucci} A.,
  2010, \mn@doi [\mnras] {10.1111/j.1365-2966.2010.17291.x}, \href
  {https://ui.adsabs.harvard.edu/abs/2010MNRAS.408.2115M} {408, 2115}

\bibitem[\protect\citeauthoryear{{Martig}, {Bournaud}, {Teyssier}  \&
  {Dekel}}{{Martig} et~al.}{2009}]{martig09}
{Martig} M.,  {Bournaud} F.,  {Teyssier} R.,   {Dekel} A.,  2009, \mn@doi
  [\apj] {10.1088/0004-637X/707/1/250}, \href
  {https://ui.adsabs.harvard.edu/abs/2009ApJ...707..250M} {707, 250}

\bibitem[\protect\citeauthoryear{{Martini}, {Sivakoff}  \&
  {Mulchaey}}{{Martini} et~al.}{2009}]{martini09}
{Martini} P.,  {Sivakoff} G.~R.,   {Mulchaey} J.~S.,  2009, \mn@doi [\apj]
  {10.1088/0004-637X/701/1/66}, \href
  {https://ui.adsabs.harvard.edu/abs/2009ApJ...701...66M} {701, 66}

\bibitem[\protect\citeauthoryear{{McGee}, {Balogh}, {Wilman}, {Bower},
  {Mulchaey}, {Parker}  \& {Oemler}}{{McGee} et~al.}{2011}]{mcgee11}
{McGee} S.~L.,  {Balogh} M.~L.,  {Wilman} D.~J.,  {Bower} R.~G.,  {Mulchaey}
  J.~S.,  {Parker} L.~C.,   {Oemler} A.,  2011, \mn@doi [\mnras]
  {10.1111/j.1365-2966.2010.18189.x}, \href
  {https://ui.adsabs.harvard.edu/abs/2011MNRAS.413..996M} {413, 996}

\bibitem[\protect\citeauthoryear{{Milvang-Jensen} et~al.,}{{Milvang-Jensen}
  et~al.}{2008a}]{mj08}
{Milvang-Jensen} B.,  et~al., 2008a, \mn@doi [\aap]
  {10.1051/0004-6361:20079148}, \href
  {https://ui.adsabs.harvard.edu/abs/2008A&A...482..419M} {482, 419}

\bibitem[\protect\citeauthoryear{{Milvang-Jensen} et~al.,}{{Milvang-Jensen}
  et~al.}{2008b}]{Milvang-Jensen08}
{Milvang-Jensen} B.,  et~al., 2008b, \mn@doi [\aap]
  {10.1051/0004-6361:20079148}, \href
  {https://ui.adsabs.harvard.edu/abs/2008A&A...482..419M} {482, 419}

\bibitem[\protect\citeauthoryear{{Momcheva} et~al.,}{{Momcheva}
  et~al.}{2016}]{mom16}
{Momcheva} I.~G.,  et~al., 2016, \mn@doi [\apjs] {10.3847/0067-0049/225/2/27},
  \href {https://ui.adsabs.harvard.edu/abs/2016ApJS..225...27M} {225, 27}

\bibitem[\protect\citeauthoryear{{Moran}, {Ellis}, {Treu}, {Smith}, {Rich}  \&
  {Smail}}{{Moran} et~al.}{2007}]{moran07}
{Moran} S.~M.,  {Ellis} R.~S.,  {Treu} T.,  {Smith} G.~P.,  {Rich} R.~M.,
  {Smail} I.,  2007, \mn@doi [\apj] {10.1086/522303}, \href
  {https://ui.adsabs.harvard.edu/abs/2007ApJ...671.1503M} {671, 1503}

\bibitem[\protect\citeauthoryear{{Moustakas}, {Kennicutt}  \&
  {Tremonti}}{{Moustakas} et~al.}{2006}]{Moustakas06}
{Moustakas} J.,  {Kennicutt} Robert~C. J.,   {Tremonti} C.~A.,  2006, \mn@doi
  [\apj] {10.1086/500964}, \href
  {https://ui.adsabs.harvard.edu/abs/2006ApJ...642..775M} {642, 775}

\bibitem[\protect\citeauthoryear{{Muzzin} et~al.,}{{Muzzin}
  et~al.}{2012}]{muzzin12}
{Muzzin} A.,  et~al., 2012, \mn@doi [\apj] {10.1088/0004-637X/746/2/188}, \href
  {https://ui.adsabs.harvard.edu/abs/2012ApJ...746..188M} {746, 188}

\bibitem[\protect\citeauthoryear{{Muzzin} et~al.,}{{Muzzin}
  et~al.}{2013}]{muzzin13}
{Muzzin} A.,  et~al., 2013, \mn@doi [\apjs] {10.1088/0067-0049/206/1/8}, \href
  {https://ui.adsabs.harvard.edu/abs/2013ApJS..206....8M} {206, 8}

\bibitem[\protect\citeauthoryear{{Muzzin} et~al.,}{{Muzzin}
  et~al.}{2014}]{muzzin14}
{Muzzin} A.,  et~al., 2014, \mn@doi [\apj] {10.1088/0004-637X/796/1/65}, \href
  {https://ui.adsabs.harvard.edu/abs/2014ApJ...796...65M} {796, 65}

\bibitem[\protect\citeauthoryear{{Nelson} et~al.,}{{Nelson}
  et~al.}{2012}]{nelson12}
{Nelson} E.~J.,  et~al., 2012, \mn@doi [\apjl] {10.1088/2041-8205/747/2/L28},
  \href {https://ui.adsabs.harvard.edu/abs/2012ApJ...747L..28N} {747, L28}

\bibitem[\protect\citeauthoryear{Newman, Treu, Ellis, Sand, Nipoti, Richard  \&
  Jullo}{Newman et~al.}{2013}]{Newman13}
Newman A.~B.,  Treu T.,  Ellis R.~S.,  Sand D.~J.,  Nipoti C.,  Richard J.,
  Jullo E.,  2013, \mn@doi [The Astrophysical Journal]
  {10.1088/0004-637x/765/1/24}, 765, 24

\bibitem[\protect\citeauthoryear{{Noeske} et~al.,}{{Noeske}
  et~al.}{2007}]{noeske}
{Noeske} K.~G.,  et~al., 2007, \mn@doi [\apjl] {10.1086/517926}, \href
  {https://ui.adsabs.harvard.edu/abs/2007ApJ...660L..43N} {660, L43}

\bibitem[\protect\citeauthoryear{{Oemler}, {Dressler}, {Gladders}, {Fritz},
  {Poggianti}, {Vulcani}  \& {Abramson}}{{Oemler} et~al.}{2013}]{oemler13}
{Oemler} Augustus J.,  {Dressler} A.,  {Gladders} M.~G.,  {Fritz} J.,
  {Poggianti} B.~M.,  {Vulcani} B.,   {Abramson} L.,  2013, \mn@doi [\apj]
  {10.1088/0004-637X/770/1/63}, \href
  {https://ui.adsabs.harvard.edu/abs/2013ApJ...770...63O} {770, 63}

\bibitem[\protect\citeauthoryear{{Old} et~al.,}{{Old} et~al.}{2020}]{old19}
{Old} L.~J.,  et~al., 2020, \mn@doi [\mnras] {10.1093/mnras/staa579}, \href
  {https://ui.adsabs.harvard.edu/abs/2020MNRAS.493.5987O} {493, 5987}

\bibitem[\protect\citeauthoryear{{Paccagnella} et~al.,}{{Paccagnella}
  et~al.}{2016}]{pacc16}
{Paccagnella} A.,  et~al., 2016, \mn@doi [\apjl] {10.3847/2041-8205/816/2/L25},
  \href {https://ui.adsabs.harvard.edu/abs/2016ApJ...816L..25P} {816, L25}

\bibitem[\protect\citeauthoryear{{Patel}, {Holden}, {Kelson}, {Illingworth}  \&
  {Franx}}{{Patel} et~al.}{2009}]{patel09}
{Patel} S.~G.,  {Holden} B.~P.,  {Kelson} D.~D.,  {Illingworth} G.~D.,
  {Franx} M.,  2009, \mn@doi [\apjl] {10.1088/0004-637X/705/1/L67}, \href
  {https://ui.adsabs.harvard.edu/abs/2009ApJ...705L..67P} {705, L67}

\bibitem[\protect\citeauthoryear{{Patel}, {Kelson}, {Holden}, {Franx}  \&
  {Illingworth}}{{Patel} et~al.}{2011}]{patel11}
{Patel} S.~G.,  {Kelson} D.~D.,  {Holden} B.~P.,  {Franx} M.,   {Illingworth}
  G.~D.,  2011, \mn@doi [\apj] {10.1088/0004-637X/735/1/53}, \href
  {https://ui.adsabs.harvard.edu/abs/2011ApJ...735...53P} {735, 53}

\bibitem[\protect\citeauthoryear{{Peng} et~al.,}{{Peng} et~al.}{2010}]{peng10}
{Peng} Y.-j.,  et~al., 2010, \mn@doi [\apj] {10.1088/0004-637X/721/1/193},
  \href {https://ui.adsabs.harvard.edu/abs/2010ApJ...721..193P} {721, 193}

\bibitem[\protect\citeauthoryear{{Planck Collaboration} et~al.,}{{Planck
  Collaboration} et~al.}{2016}]{planck}
{Planck Collaboration} et~al., 2016, \mn@doi [\aap]
  {10.1051/0004-6361/201525830}, \href
  {https://ui.adsabs.harvard.edu/abs/2016A&A...594A..13P} {594, A13}

\bibitem[\protect\citeauthoryear{{Poggianti} et~al.,}{{Poggianti}
  et~al.}{2006}]{pogg06}
{Poggianti} B.~M.,  et~al., 2006, \mn@doi [\apj] {10.1086/500666}, \href
  {https://ui.adsabs.harvard.edu/abs/2006ApJ...642..188P} {642, 188}

\bibitem[\protect\citeauthoryear{{Poggianti} et~al.,}{{Poggianti}
  et~al.}{2008}]{pogg08}
{Poggianti} B.~M.,  et~al., 2008, \mn@doi [\apj] {10.1086/589936}, \href
  {https://ui.adsabs.harvard.edu/abs/2008ApJ...684..888P} {684, 888}

\bibitem[\protect\citeauthoryear{{Poggianti} et~al.,}{{Poggianti}
  et~al.}{2009}]{pogg09}
{Poggianti} B.~M.,  et~al., 2009, \mn@doi [\apj] {10.1088/0004-637X/693/1/112},
  \href {https://ui.adsabs.harvard.edu/abs/2009ApJ...693..112P} {693, 112}

\bibitem[\protect\citeauthoryear{{Poggianti} et~al.,}{{Poggianti}
  et~al.}{2016}]{pogg16}
{Poggianti} B.~M.,  et~al., 2016, \mn@doi [\aj] {10.3847/0004-6256/151/3/78},
  \href {https://ui.adsabs.harvard.edu/abs/2016AJ....151...78P} {151, 78}

\bibitem[\protect\citeauthoryear{{Quilis}, {Moore}  \& {Bower}}{{Quilis}
  et~al.}{2000}]{quilis2000}
{Quilis} V.,  {Moore} B.,   {Bower} R.,  2000, \mn@doi [Science]
  {10.1126/science.288.5471.1617}, \href
  {https://ui.adsabs.harvard.edu/abs/2000Sci...288.1617Q} {288, 1617}

\bibitem[\protect\citeauthoryear{{Rudnick} et~al.,}{{Rudnick}
  et~al.}{2009}]{rudnick09}
{Rudnick} G.,  et~al., 2009, \mn@doi [\apj] {10.1088/0004-637X/700/2/1559},
  \href {https://ui.adsabs.harvard.edu/abs/2009ApJ...700.1559R} {700, 1559}

\bibitem[\protect\citeauthoryear{{Rudnick} et~al.,}{{Rudnick}
  et~al.}{2017}]{Rudnick17}
{Rudnick} G.,  et~al., 2017, \mn@doi [\apj] {10.3847/1538-4357/aa866c}, \href
  {https://ui.adsabs.harvard.edu/abs/2017ApJ...850..181R} {850, 181}

\bibitem[\protect\citeauthoryear{{Saglia} et~al.,}{{Saglia}
  et~al.}{2010}]{saglia10}
{Saglia} R.~P.,  et~al., 2010, \mn@doi [\aap] {10.1051/0004-6361/201014703},
  \href {https://ui.adsabs.harvard.edu/abs/2010A&A...524A...6S} {524, A6}

\bibitem[\protect\citeauthoryear{{Sarzi} et~al.,}{{Sarzi}
  et~al.}{2006}]{sarzi06}
{Sarzi} M.,  et~al., 2006, \mn@doi [\mnras] {10.1111/j.1365-2966.2005.09839.x},
  \href {https://ui.adsabs.harvard.edu/abs/2006MNRAS.366.1151S} {366, 1151}

\bibitem[\protect\citeauthoryear{{Schreiber} et~al.,}{{Schreiber}
  et~al.}{2015}]{schr15}
{Schreiber} C.,  et~al., 2015, \mn@doi [\aap] {10.1051/0004-6361/201425017},
  \href {https://ui.adsabs.harvard.edu/abs/2015A&A...575A..74S} {575, A74}

\bibitem[\protect\citeauthoryear{{Simard} et~al.,}{{Simard}
  et~al.}{2009}]{simard09}
{Simard} L.,  et~al., 2009, \mn@doi [\aap] {10.1051/0004-6361/20078872}, \href
  {https://ui.adsabs.harvard.edu/abs/2009A&A...508.1141S} {508, 1141}

\bibitem[\protect\citeauthoryear{{Singh} et~al.,}{{Singh}
  et~al.}{2013}]{singh13}
{Singh} R.,  et~al., 2013, \mn@doi [\aap] {10.1051/0004-6361/201322062}, \href
  {https://ui.adsabs.harvard.edu/abs/2013A&A...558A..43S} {558, A43}

\bibitem[\protect\citeauthoryear{{Sobral}, {Best}, {Smail}, {Geach},
  {Cirasuolo}, {Garn}  \& {Dalton}}{{Sobral} et~al.}{2011}]{sobral11}
{Sobral} D.,  {Best} P.~N.,  {Smail} I.,  {Geach} J.~E.,  {Cirasuolo} M.,
  {Garn} T.,   {Dalton} G.~B.,  2011, \mn@doi [\mnras]
  {10.1111/j.1365-2966.2010.17707.x}, \href
  {https://ui.adsabs.harvard.edu/abs/2011MNRAS.411..675S} {411, 675}

\bibitem[\protect\citeauthoryear{{Speagle}, {Steinhardt}, {Capak}  \&
  {Silverman}}{{Speagle} et~al.}{2014}]{speagle14}
{Speagle} J.~S.,  {Steinhardt} C.~L.,  {Capak} P.~L.,   {Silverman} J.~D.,
  2014, \mn@doi [\apjs] {10.1088/0067-0049/214/2/15}, \href
  {https://ui.adsabs.harvard.edu/abs/2014ApJS..214...15S} {214, 15}

\bibitem[\protect\citeauthoryear{{Strom}, {Steidel}, {Rudie}, {Trainor},
  {Pettini}  \& {Reddy}}{{Strom} et~al.}{2017}]{strom17}
{Strom} A.~L.,  {Steidel} C.~C.,  {Rudie} G.~C.,  {Trainor} R.~F.,  {Pettini}
  M.,   {Reddy} N.~A.,  2017, \mn@doi [\apj] {10.3847/1538-4357/836/2/164},
  \href {https://ui.adsabs.harvard.edu/abs/2017ApJ...836..164S} {836, 164}

\bibitem[\protect\citeauthoryear{{Tacchella}, {Dekel}, {Carollo}, {Ceverino},
  {DeGraf}, {Lapiner}, {Mand elker}  \& {Primack Joel}}{{Tacchella}
  et~al.}{2016}]{tacch16}
{Tacchella} S.,  {Dekel} A.,  {Carollo} C.~M.,  {Ceverino} D.,  {DeGraf} C.,
  {Lapiner} S.,  {Mand elker} N.,   {Primack Joel} R.,  2016, \mn@doi [\mnras]
  {10.1093/mnras/stw131}, \href
  {https://ui.adsabs.harvard.edu/abs/2016MNRAS.457.2790T} {457, 2790}

\bibitem[\protect\citeauthoryear{{Taranu}, {Hudson}, {Balogh}, {Smith},
  {Power}, {Oman}  \& {Krane}}{{Taranu} et~al.}{2014}]{taranu14}
{Taranu} D.~S.,  {Hudson} M.~J.,  {Balogh} M.~L.,  {Smith} R.~J.,  {Power} C.,
  {Oman} K.~A.,   {Krane} B.,  2014, \mn@doi [\mnras] {10.1093/mnras/stu389},
  \href {https://ui.adsabs.harvard.edu/abs/2014MNRAS.440.1934T} {440, 1934}

\bibitem[\protect\citeauthoryear{{Tasca} et~al.,}{{Tasca}
  et~al.}{2015}]{tasca15}
{Tasca} L.~A.~M.,  et~al., 2015, \mn@doi [\aap] {10.1051/0004-6361/201425379},
  \href {https://ui.adsabs.harvard.edu/abs/2015A&A...581A..54T} {581, A54}

\bibitem[\protect\citeauthoryear{{Taylor} et~al.,}{{Taylor}
  et~al.}{2011}]{taylor11}
{Taylor} E.~N.,  et~al., 2011, \mn@doi [\mnras]
  {10.1111/j.1365-2966.2011.19536.x}, \href
  {https://ui.adsabs.harvard.edu/abs/2011MNRAS.418.1587T} {418, 1587}

\bibitem[\protect\citeauthoryear{{Tiley} et~al.,}{{Tiley}
  et~al.}{2020}]{tiley20}
{Tiley} A.~L.,  et~al., 2020, \mn@doi [\mnras] {10.1093/mnras/staa1418}, \href
  {https://ui.adsabs.harvard.edu/abs/2020MNRAS.496..649T} {496, 649}

\bibitem[\protect\citeauthoryear{{Treu} et~al.,}{{Treu} et~al.}{2015}]{treu15}
{Treu} T.,  et~al., 2015, \mn@doi [\apj] {10.1088/0004-637X/812/2/114}, \href
  {https://ui.adsabs.harvard.edu/abs/2015ApJ...812..114T} {812, 114}

\bibitem[\protect\citeauthoryear{{Vulcani}, {Poggianti}, {Finn}, {Rudnick},
  {Desai}  \& {Bamford}}{{Vulcani} et~al.}{2010}]{vul10}
{Vulcani} B.,  {Poggianti} B.~M.,  {Finn} R.~A.,  {Rudnick} G.,  {Desai} V.,
  {Bamford} S.,  2010, \mn@doi [\apjl] {10.1088/2041-8205/710/1/L1}, \href
  {https://ui.adsabs.harvard.edu/abs/2010ApJ...710L...1V} {710, L1}

\bibitem[\protect\citeauthoryear{{Vulcani} et~al.,}{{Vulcani}
  et~al.}{2011a}]{vul11b}
{Vulcani} B.,  et~al., 2011a, \mn@doi [\mnras]
  {10.1111/j.1365-2966.2010.17904.x}, \href
  {https://ui.adsabs.harvard.edu/abs/2011MNRAS.412..246V} {412, 246}

\bibitem[\protect\citeauthoryear{{Vulcani} et~al.,}{{Vulcani}
  et~al.}{2011b}]{vul11a}
{Vulcani} B.,  et~al., 2011b, \mn@doi [\mnras]
  {10.1111/j.1365-2966.2010.18182.x}, \href
  {https://ui.adsabs.harvard.edu/abs/2011MNRAS.413..921V} {413, 921}

\bibitem[\protect\citeauthoryear{{Vulcani} et~al.,}{{Vulcani}
  et~al.}{2012}]{vul12}
{Vulcani} B.,  et~al., 2012, \mn@doi [\aap] {10.1051/0004-6361/201219397},
  \href {https://ui.adsabs.harvard.edu/abs/2012A&A...544A.104V} {544, A104}

\bibitem[\protect\citeauthoryear{{Vulcani} et~al.,}{{Vulcani}
  et~al.}{2015}]{vul15}
{Vulcani} B.,  et~al., 2015, \mn@doi [\apj] {10.1088/0004-637X/814/2/161},
  \href {https://ui.adsabs.harvard.edu/abs/2015ApJ...814..161V} {814, 161}

\bibitem[\protect\citeauthoryear{{Vulcani} et~al.,}{{Vulcani}
  et~al.}{2016}]{vul16}
{Vulcani} B.,  et~al., 2016, \mn@doi [\apj] {10.3847/1538-4357/833/2/178},
  \href {https://ui.adsabs.harvard.edu/abs/2016ApJ...833..178V} {833, 178}

\bibitem[\protect\citeauthoryear{{Vulcani} et~al.,}{{Vulcani}
  et~al.}{2017}]{vul17}
{Vulcani} B.,  et~al., 2017, \mn@doi [\apj] {10.3847/1538-4357/aa618b}, \href
  {https://ui.adsabs.harvard.edu/abs/2017ApJ...837..126V} {837, 126}

\bibitem[\protect\citeauthoryear{{Vulcani} et~al.,}{{Vulcani}
  et~al.}{2018}]{vul18}
{Vulcani} B.,  et~al., 2018, \mn@doi [\apjl] {10.3847/2041-8213/aae68b}, \href
  {https://ui.adsabs.harvard.edu/abs/2018ApJ...866L..25V} {866, L25}

\bibitem[\protect\citeauthoryear{{Webb} et~al.,}{{Webb} et~al.}{2020}]{webb20}
{Webb} K.,  et~al., 2020, \mn@doi [\mnras] {10.1093/mnras/staa2752}, \href
  {https://ui.adsabs.harvard.edu/abs/2020MNRAS.498.5317W} {498, 5317}

\bibitem[\protect\citeauthoryear{{Wetzel}, {Tinker}, {Conroy}  \& {van den
  Bosch}}{{Wetzel} et~al.}{2013}]{wetzel13}
{Wetzel} A.~R.,  {Tinker} J.~L.,  {Conroy} C.,   {van den Bosch} F.~C.,  2013,
  \mn@doi [\mnras] {10.1093/mnras/stt469}, \href
  {https://ui.adsabs.harvard.edu/abs/2013MNRAS.432..336W} {432, 336}

\bibitem[\protect\citeauthoryear{{Whiley} et~al.,}{{Whiley}
  et~al.}{2008}]{whiley08}
{Whiley} I.~M.,  et~al., 2008, \mn@doi [\mnras]
  {10.1111/j.1365-2966.2008.13324.x}, \href
  {https://ui.adsabs.harvard.edu/abs/2008MNRAS.387.1253W} {387, 1253}

\bibitem[\protect\citeauthoryear{{Whitaker}, {van Dokkum}, {Brammer}  \&
  {Franx}}{{Whitaker} et~al.}{2012}]{whitaker12}
{Whitaker} K.~E.,  {van Dokkum} P.~G.,  {Brammer} G.,   {Franx} M.,  2012,
  \mn@doi [\apjl] {10.1088/2041-8205/754/2/L29}, \href
  {https://ui.adsabs.harvard.edu/abs/2012ApJ...754L..29W} {754, L29}

\bibitem[\protect\citeauthoryear{{Whitaker} et~al.,}{{Whitaker}
  et~al.}{2014}]{whitaker14}
{Whitaker} K.~E.,  et~al., 2014, \mn@doi [\apj] {10.1088/0004-637X/795/2/104},
  \href {https://ui.adsabs.harvard.edu/abs/2014ApJ...795..104W} {795, 104}

\bibitem[\protect\citeauthoryear{{White} \& {Zaritsky}}{{White} \&
  {Zaritsky}}{1992}]{white92}
{White} S. D.~M.,  {Zaritsky} D.,  1992, \mn@doi [\apj] {10.1086/171552}, \href
  {https://ui.adsabs.harvard.edu/abs/1992ApJ...394....1W} {394, 1}

\bibitem[\protect\citeauthoryear{{White} et~al.,}{{White}
  et~al.}{2005}]{white05}
{White} S.~D.~M.,  et~al., 2005, A\&A, \href
  {https://ui.adsabs.harvard.edu/abs/2005A&A...444..365W} {444, 365}

\bibitem[\protect\citeauthoryear{{Wild}, {Almaini}, {Dunlop}, {Simpson},
  {Rowlands}, {Bowler}, {Maltby}  \& {McLure}}{{Wild} et~al.}{2016}]{wild16}
{Wild} V.,  {Almaini} O.,  {Dunlop} J.,  {Simpson} C.,  {Rowlands} K.,
  {Bowler} R.,  {Maltby} D.,   {McLure} R.,  2016, \mn@doi [\mnras]
  {10.1093/mnras/stw1996}, \href
  {https://ui.adsabs.harvard.edu/abs/2016MNRAS.463..832W} {463, 832}

\bibitem[\protect\citeauthoryear{{Williams}, {Quadri}, {Franx}, {van Dokkum}
  \& {Labb{\'e}}}{{Williams} et~al.}{2009}]{will09}
{Williams} R.~J.,  {Quadri} R.~F.,  {Franx} M.,  {van Dokkum} P.,   {Labb{\'e}}
  I.,  2009, \mn@doi [\apj] {10.1088/0004-637X/691/2/1879}, \href
  {https://ui.adsabs.harvard.edu/abs/2009ApJ...691.1879W} {691, 1879}

\bibitem[\protect\citeauthoryear{{Wojtak} et~al.,}{{Wojtak}
  et~al.}{2018}]{wot18}
{Wojtak} R.,  et~al., 2018, \mn@doi [\mnras] {10.1093/mnras/sty2257}, \href
  {https://ui.adsabs.harvard.edu/abs/2018MNRAS.481..324W} {481, 324}

\bibitem[\protect\citeauthoryear{Wolf et~al.,}{Wolf et~al.}{2009}]{Wolf09}
Wolf C.,  et~al., 2009, Monthly Notices of the Royal Astronomical Society, 393,
  1302

\bibitem[\protect\citeauthoryear{{Wuyts} et~al.,}{{Wuyts}
  et~al.}{2007}]{Wuyts07}
{Wuyts} S.,  et~al., 2007, \mn@doi [\apj] {10.1086/509708}, \href
  {https://ui.adsabs.harvard.edu/abs/2007ApJ...655...51W} {655, 51}

\bibitem[\protect\citeauthoryear{{Wuyts} et~al.,}{{Wuyts}
  et~al.}{2011}]{wuyts11}
{Wuyts} S.,  et~al., 2011, \mn@doi [\apj] {10.1088/0004-637X/738/1/106}, \href
  {https://ui.adsabs.harvard.edu/abs/2011ApJ...738..106W} {738, 106}

\bibitem[\protect\citeauthoryear{{Wuyts} et~al.,}{{Wuyts}
  et~al.}{2013}]{wutys13}
{Wuyts} S.,  et~al., 2013, \mn@doi [\apj] {10.1088/0004-637X/779/2/135}, \href
  {https://ui.adsabs.harvard.edu/abs/2013ApJ...779..135W} {779, 135}

\bibitem[\protect\citeauthoryear{{Zahid}, {Dima}, {Kudritzki}, {Kewley},
  {Geller}, {Hwang}, {Silverman}  \& {Kashino}}{{Zahid} et~al.}{2014}]{zahid14}
{Zahid} H.~J.,  {Dima} G.~I.,  {Kudritzki} R.-P.,  {Kewley} L.~J.,  {Geller}
  M.~J.,  {Hwang} H.~S.,  {Silverman} J.~D.,   {Kashino} D.,  2014, \mn@doi
  [\apj] {10.1088/0004-637X/791/2/130}, \href
  {https://ui.adsabs.harvard.edu/abs/2014ApJ...791..130Z} {791, 130}

\bibitem[\protect\citeauthoryear{{da Cunha}, {Charlot}  \& {Elbaz}}{{da Cunha}
  et~al.}{2008}]{dacun08}
{da Cunha} E.,  {Charlot} S.,   {Elbaz} D.,  2008, \mn@doi [\mnras]
  {10.1111/j.1365-2966.2008.13535.x}, \href
  {https://ui.adsabs.harvard.edu/abs/2008MNRAS.388.1595D} {388, 1595}

\bibitem[\protect\citeauthoryear{{van Dokkum} et~al.,}{{van Dokkum}
  et~al.}{2011}]{vandokkum11}
{van Dokkum} P.~G.,  et~al., 2011, \mn@doi [\apjl]
  {10.1088/2041-8205/743/1/L15}, \href
  {https://ui.adsabs.harvard.edu/abs/2011ApJ...743L..15V} {743, L15}

\bibitem[\protect\citeauthoryear{{van Dokkum} et~al.,}{{van Dokkum}
  et~al.}{2013}]{van13a}
{van Dokkum} P.~G.,  et~al., 2013, \mn@doi [\apjl]
  {10.1088/2041-8205/771/2/L35}, \href
  {https://ui.adsabs.harvard.edu/abs/2013ApJ...771L..35V} {771, L35}

\bibitem[\protect\citeauthoryear{{van der Wel} et~al.,}{{van der Wel}
  et~al.}{2014}]{van14}
{van der Wel} A.,  et~al., 2014, \mn@doi [\apj] {10.1088/0004-637X/788/1/28},
  \href {https://ui.adsabs.harvard.edu/abs/2014ApJ...788...28V} {788, 28}

\bibitem[\protect\citeauthoryear{{van der Wel} et~al.,}{{van der Wel}
  et~al.}{2016}]{vanderwel16}
{van der Wel} A.,  et~al., 2016, \mn@doi [\apjs] {10.3847/0067-0049/223/2/29},
  \href {https://ui.adsabs.harvard.edu/abs/2016ApJS..223...29V} {223, 29}

\makeatother
\end{thebibliography}




\appendix

\section{Color Corrections}
\label{sec:app}
As described in \cite{Just_2019}, there were multiple calibration challenges with the wide-field photometry that we use in this work, which was not taken under photometric conditions. The WFI photometry in \cite{Just_2019} was calibrated to the EDisCS core photometry and a subsequent calibration step was applied to minimize the residuals of photometric vs. spectroscopic redshifts. Despite this, the calibration has produced reasonably good photometric redshifts and it later became apparent that the rest-frame $UVJ$ colors had additional calibration issues, likely resulting from a non-trivial by-product of the multiple zeropoint calibration steps that we undertook in \cite{Just_2019}. While the $UVJ$ colors for each cluster had a clear quiescent clump and SF sequence, they were each systematically shifted with respect to each other, and to the quiescent clump as defined from the well-calibrated EDisCS photometry on the cluster cores.

We therefore undertook an additional calibration step in which we used the median colors of quiescent galaxies in the wide-field sample and shifted the $U-V$ and $V-J$ colors such that this clump matched the median $UVJ$ colors of the spectroscopically confirmed galaxies from EDisCS that had no emission lines in their spectra. Although calculated for just the quiescent galaxies, These shifts were applied to all galaxies on a cluster-by-cluster basis. These shifts were $<0.2$ in color but resulted in all of our fields having well matched $UVJ$ sequences. This gives us the ability to robustly separate galaxies in different regions of $UVJ$ space. The adjustment to the colors was also important for our use of the $U-V$ color to compute stellar mass to light ratios and stellar masses (\S\ref{sec:props}). In practice, the correction was mostly applied to the rest-frame $V$-band magnitude. The rest-frame $U$-band magnitude was derived from well-calibrated $B$-band observations and the $J$-band magnitude was calibrated well to the 2MASS photometry. The $V$ magnitude was more tied to the problematic WFI photometry.

\section{Visual Inspection of Galaxies with GRIZLI}
\label{sec:app2}
\begin{figure*}
\centering

\includegraphics[scale=0.25]{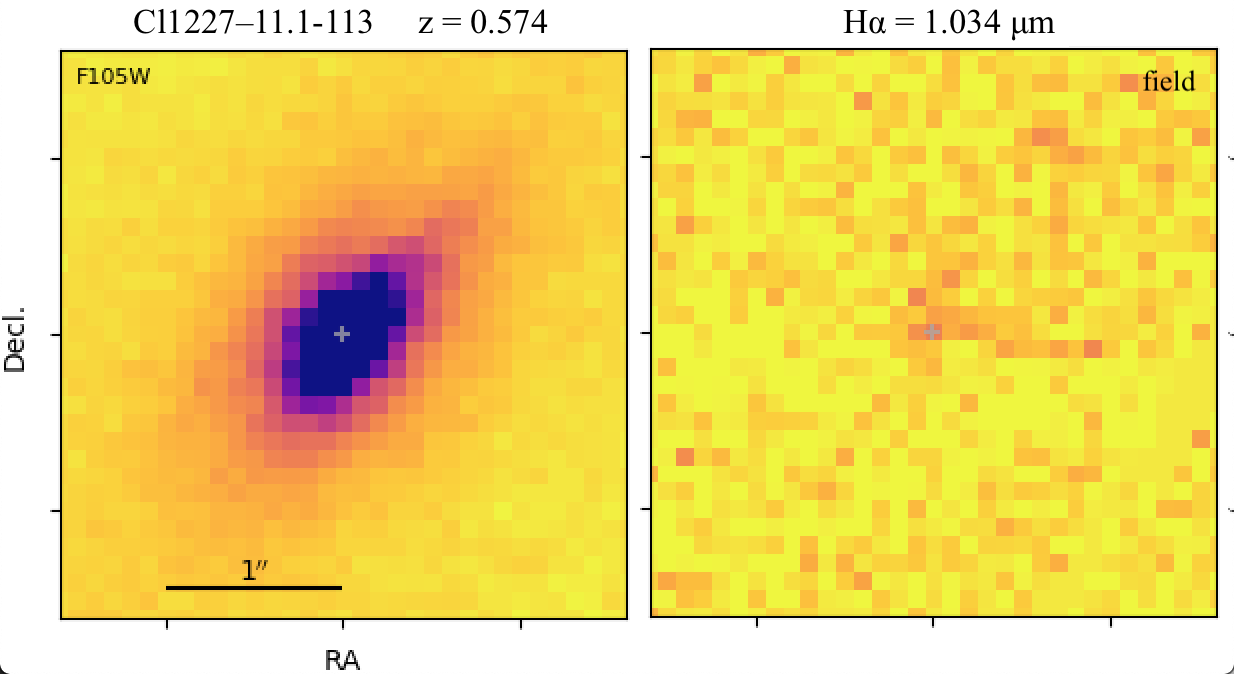}
\includegraphics[scale=0.25]{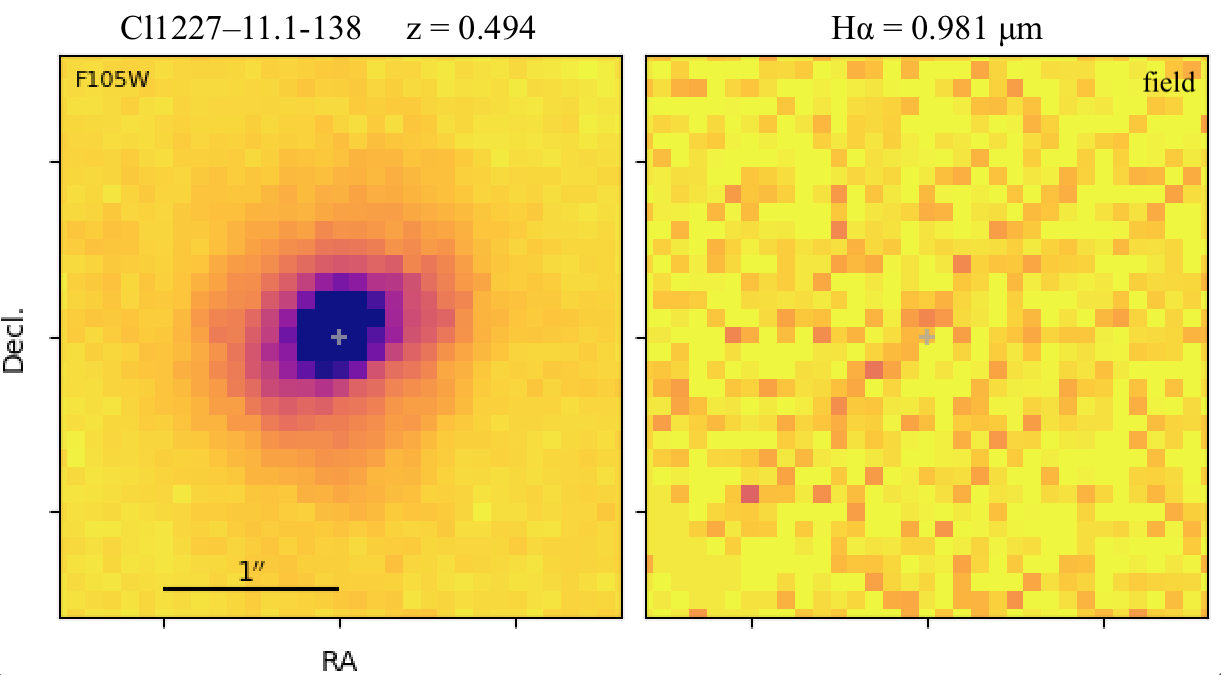}
\includegraphics[scale=0.25]{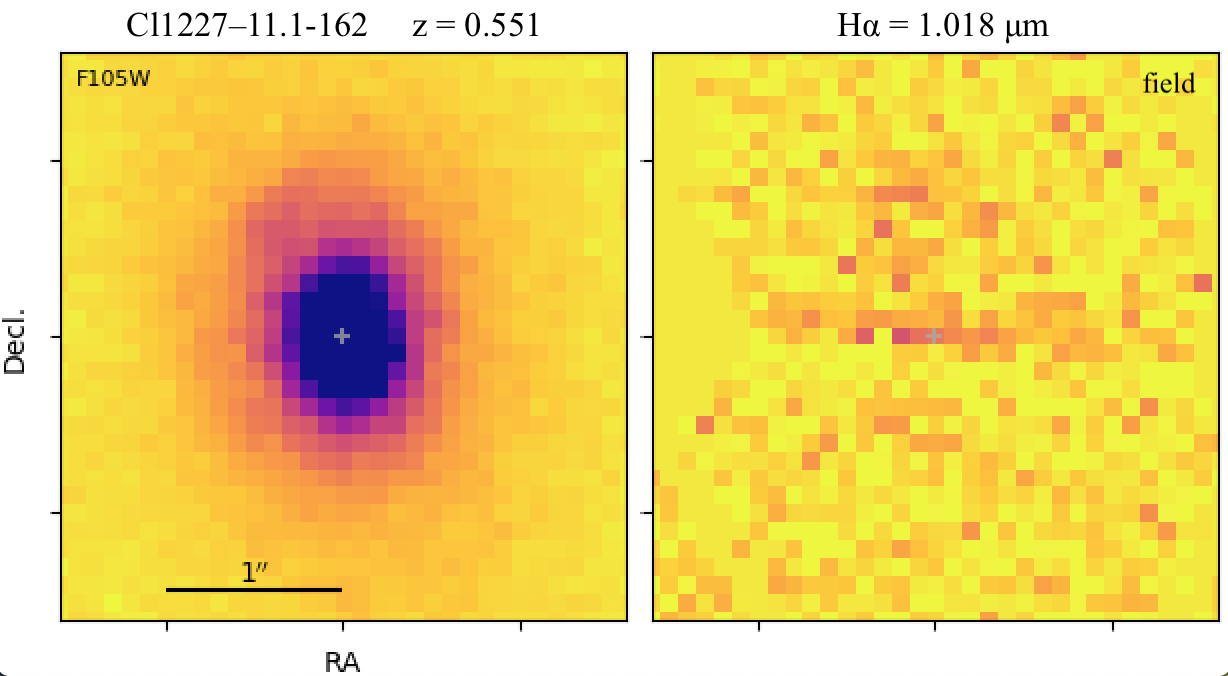}
\includegraphics[scale=0.25]{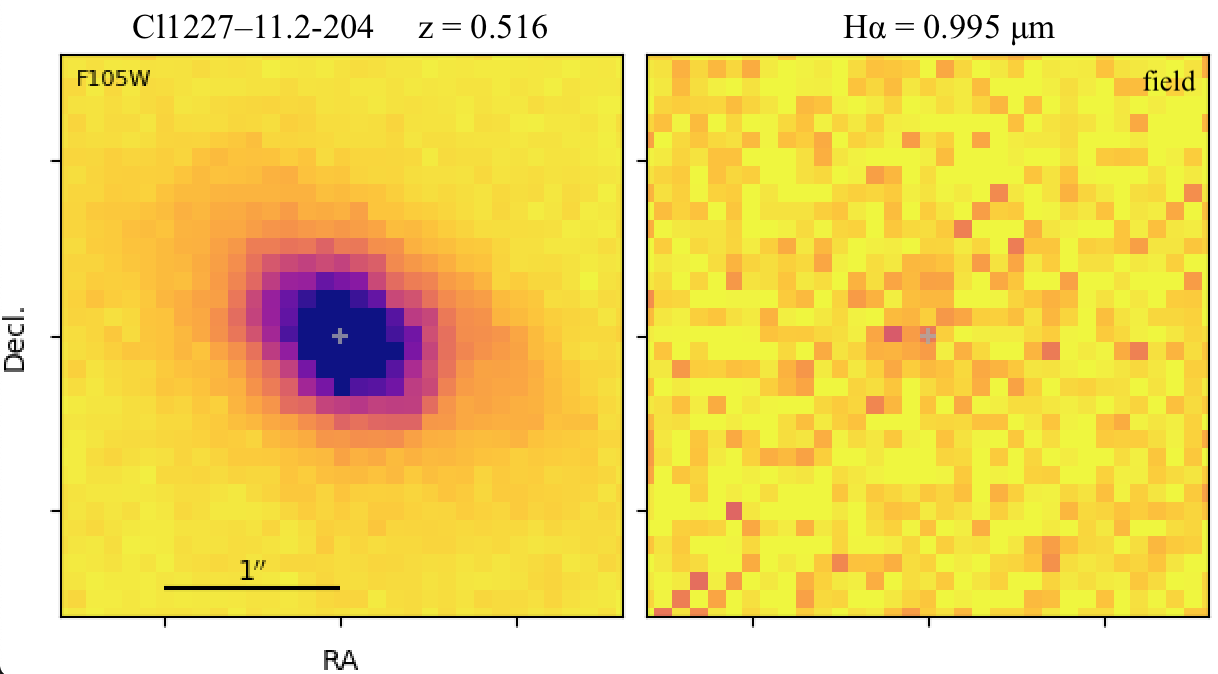}
\includegraphics[scale=0.25]{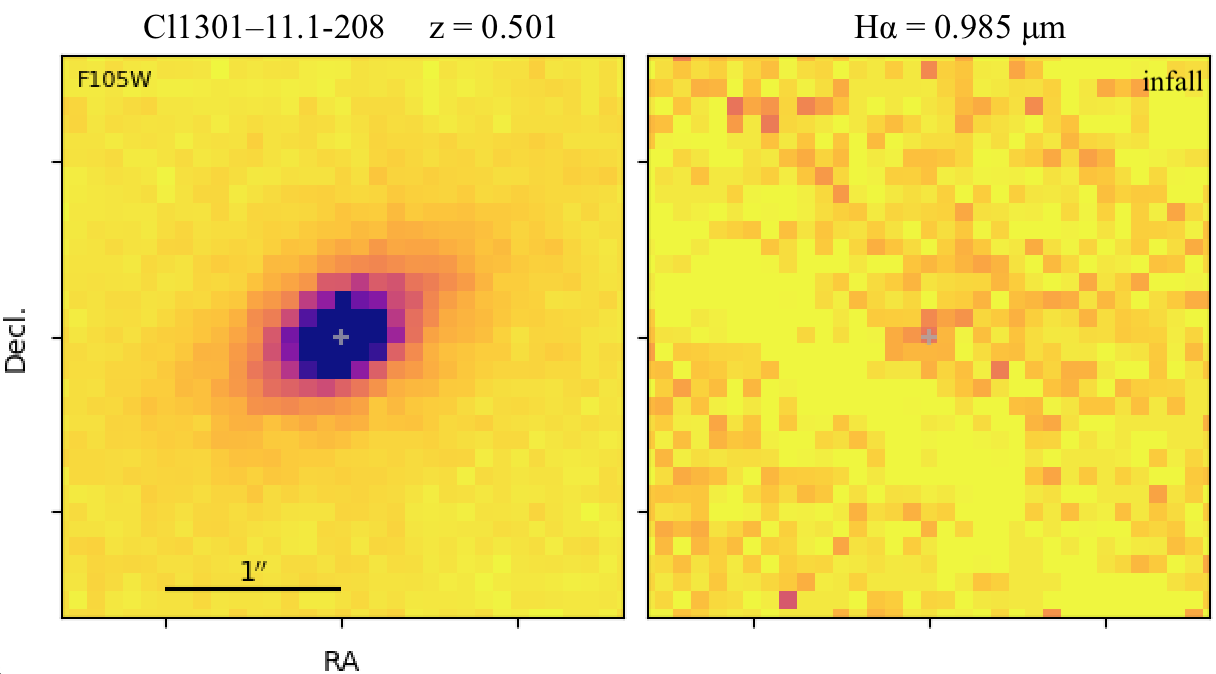}
\includegraphics[scale=0.25]{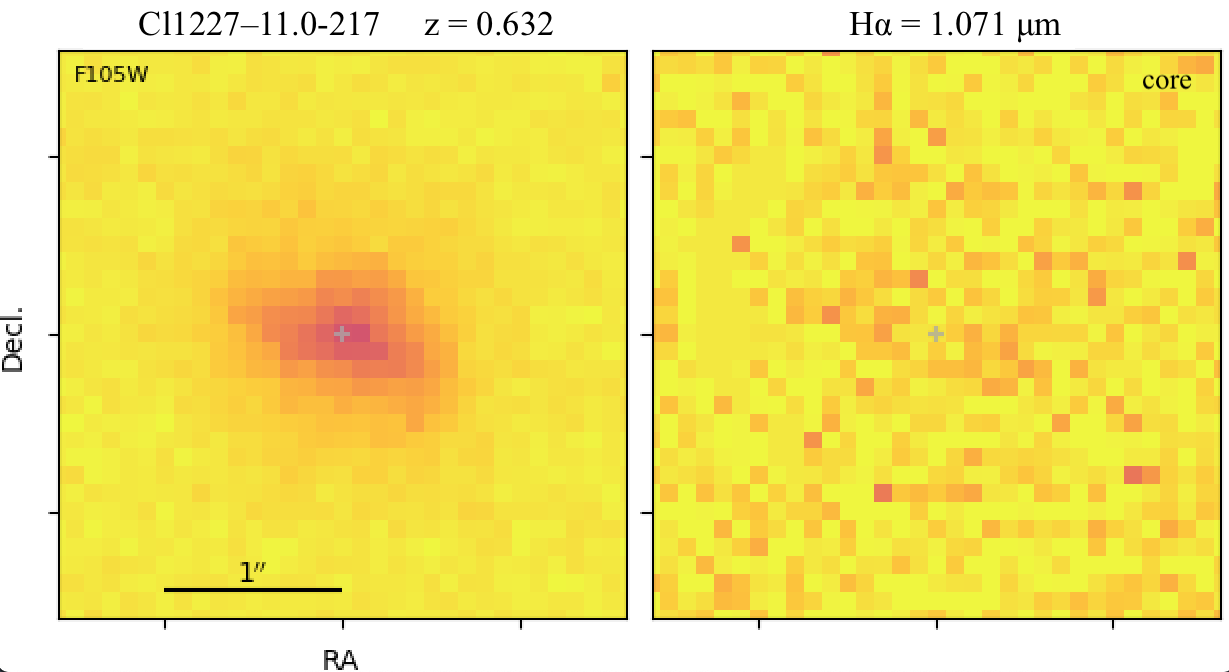}
\includegraphics[scale=0.25]{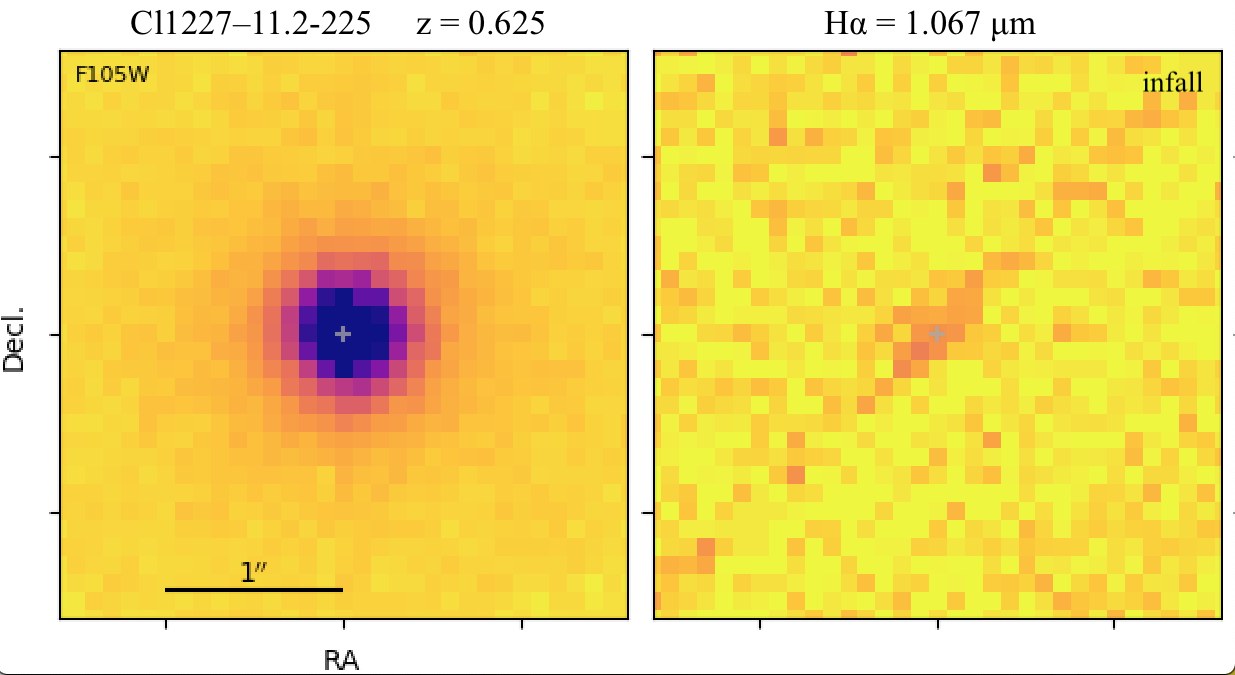}
\includegraphics[scale=0.25]{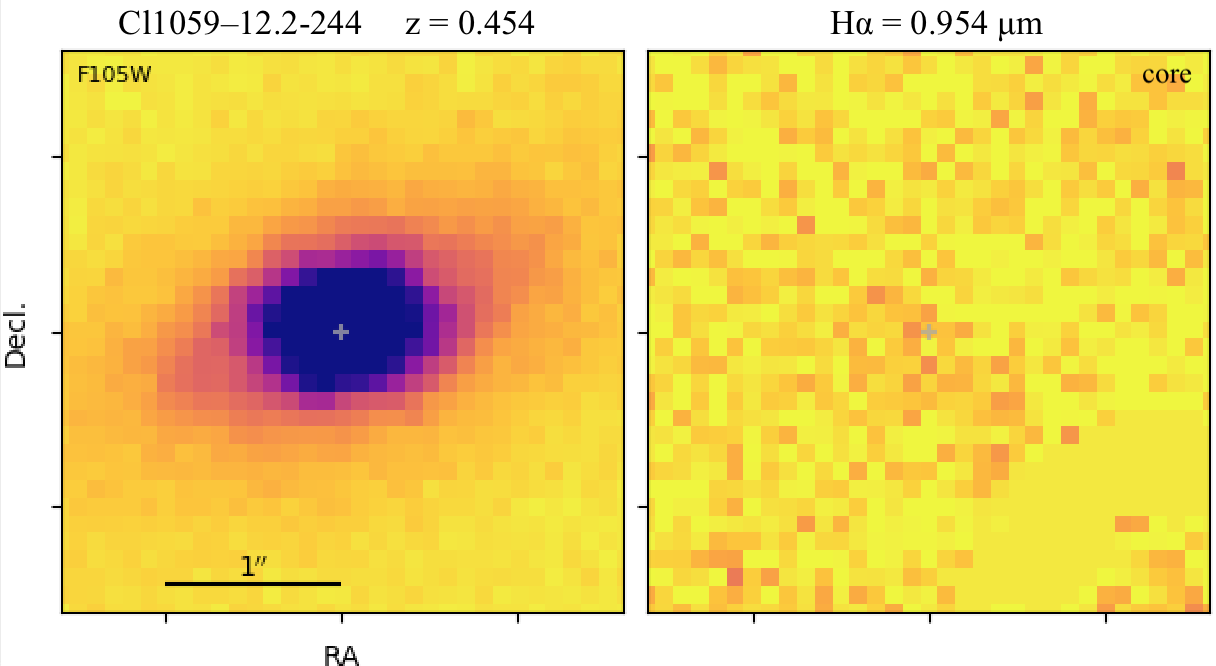}
\includegraphics[scale=0.25]{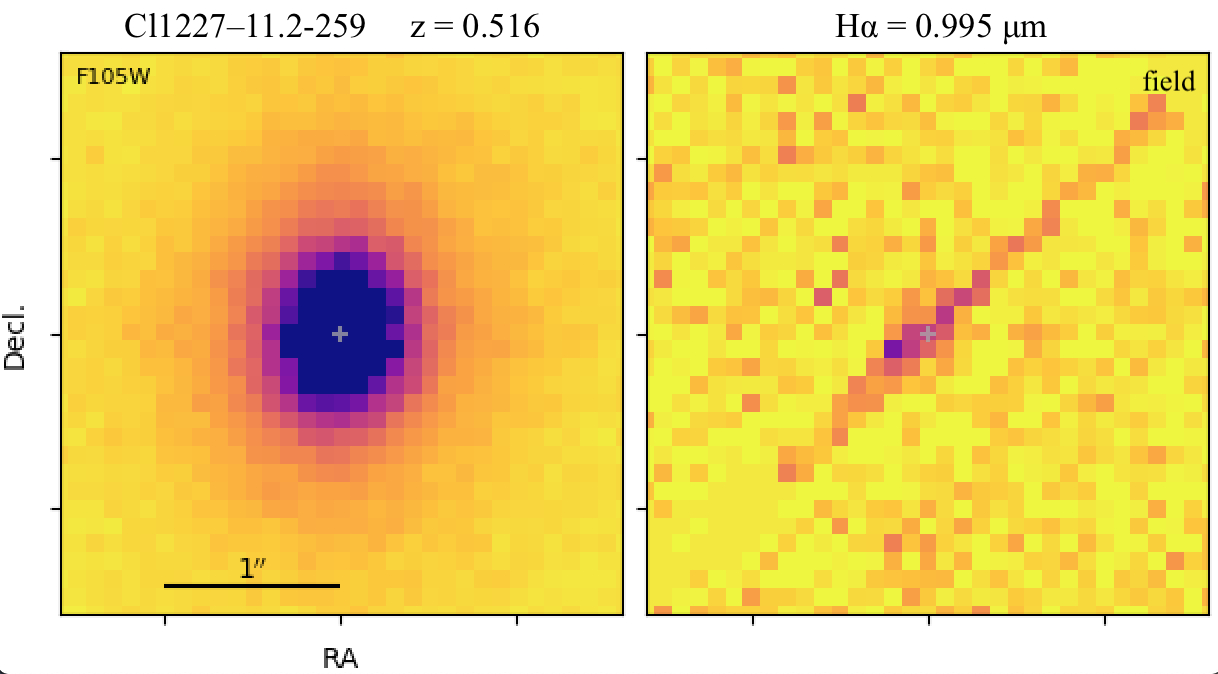}
\includegraphics[scale=0.25]{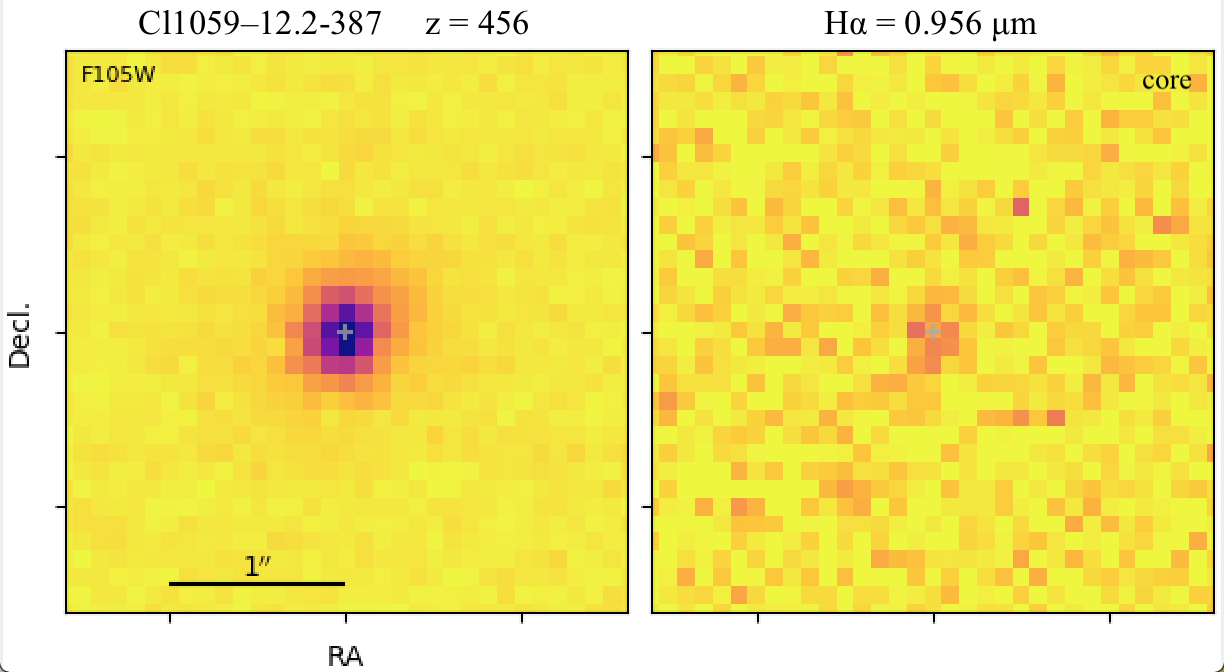}
\includegraphics[scale=0.25]{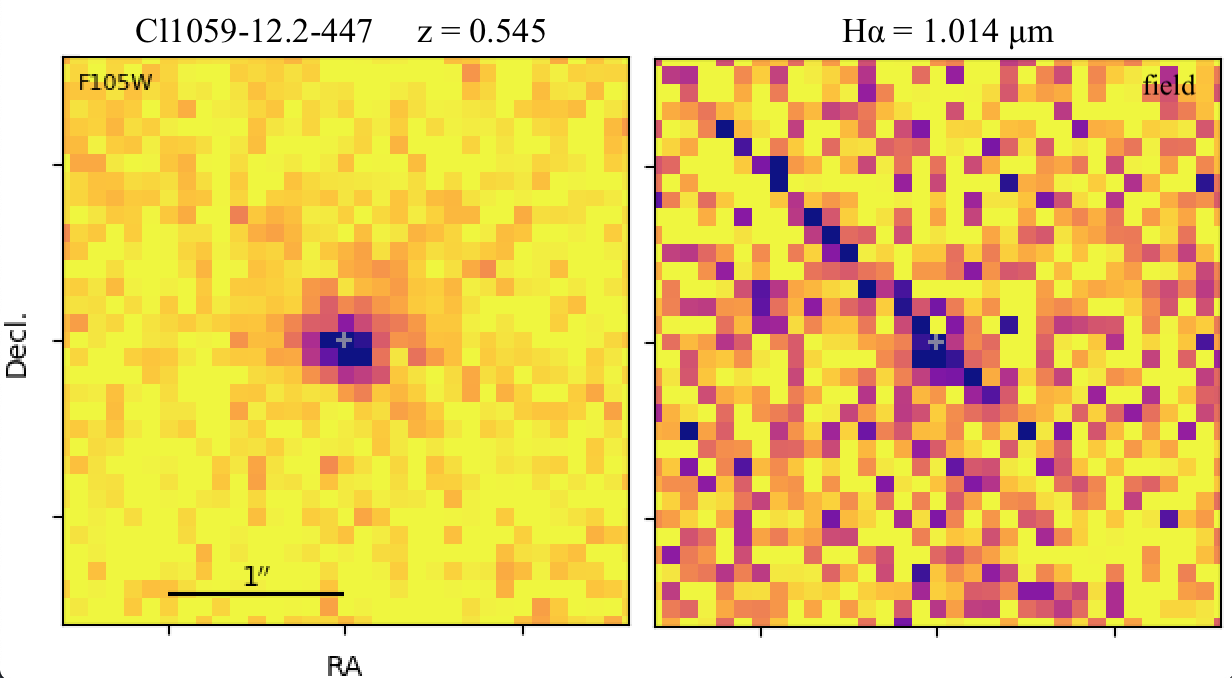}
\includegraphics[scale=0.25]{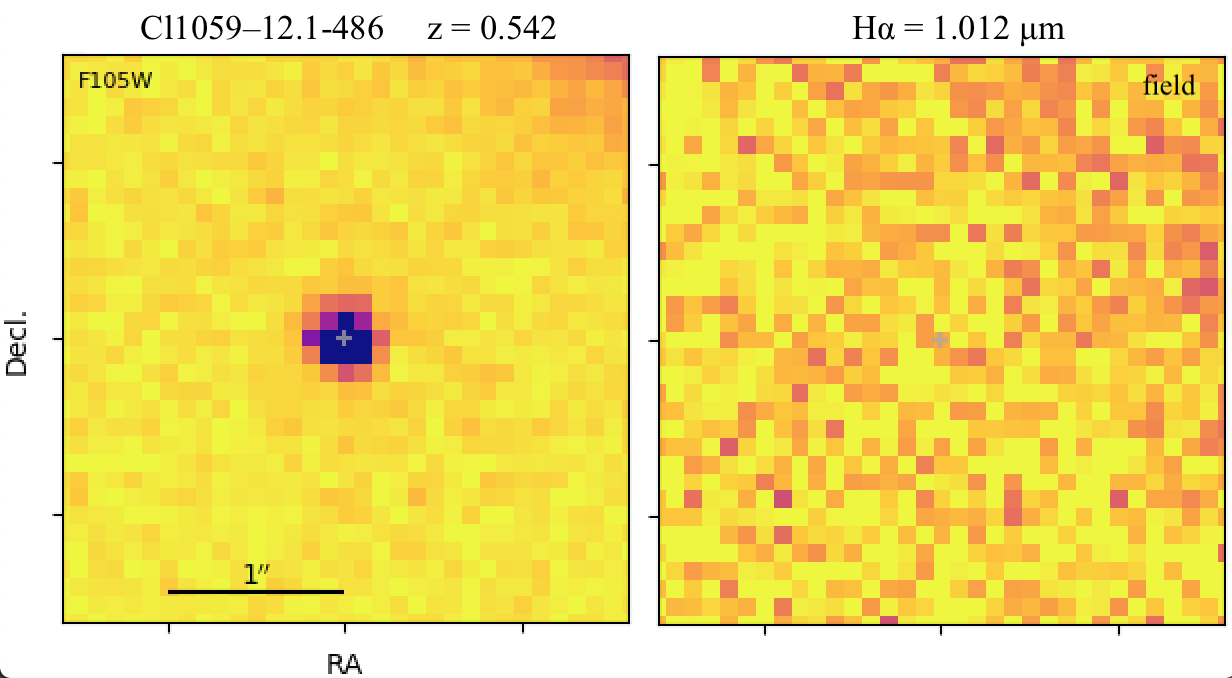}
\includegraphics[scale=0.25]{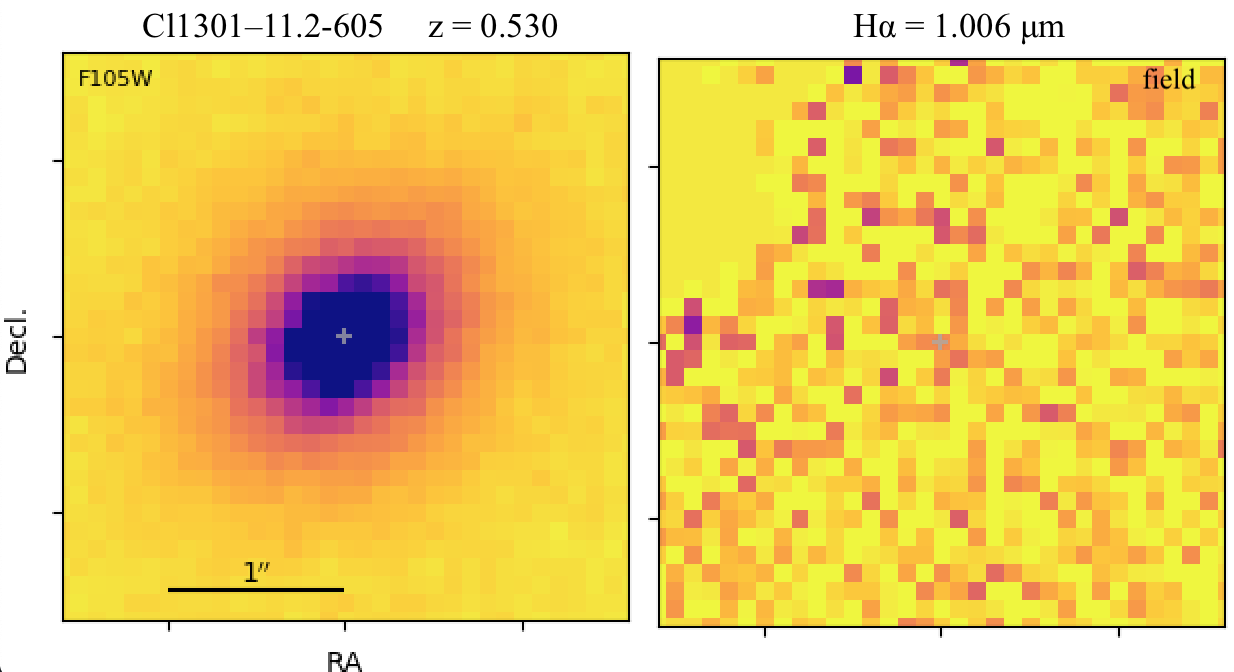}
\includegraphics[scale=0.25]{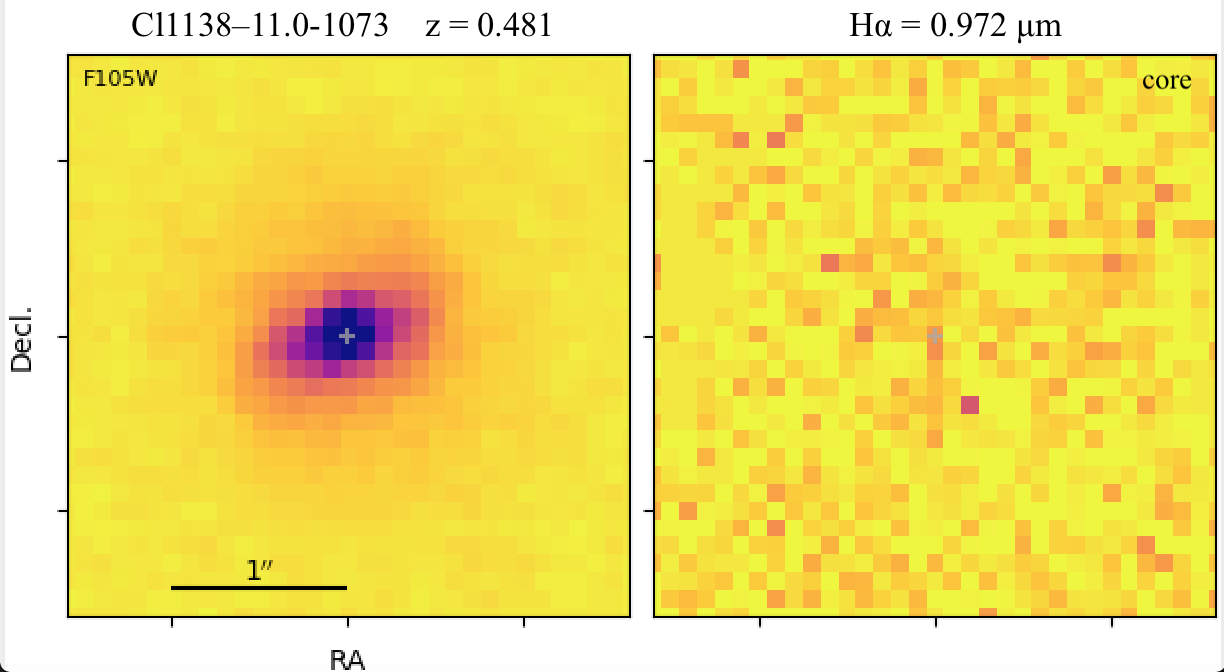}
\includegraphics[scale=0.25]{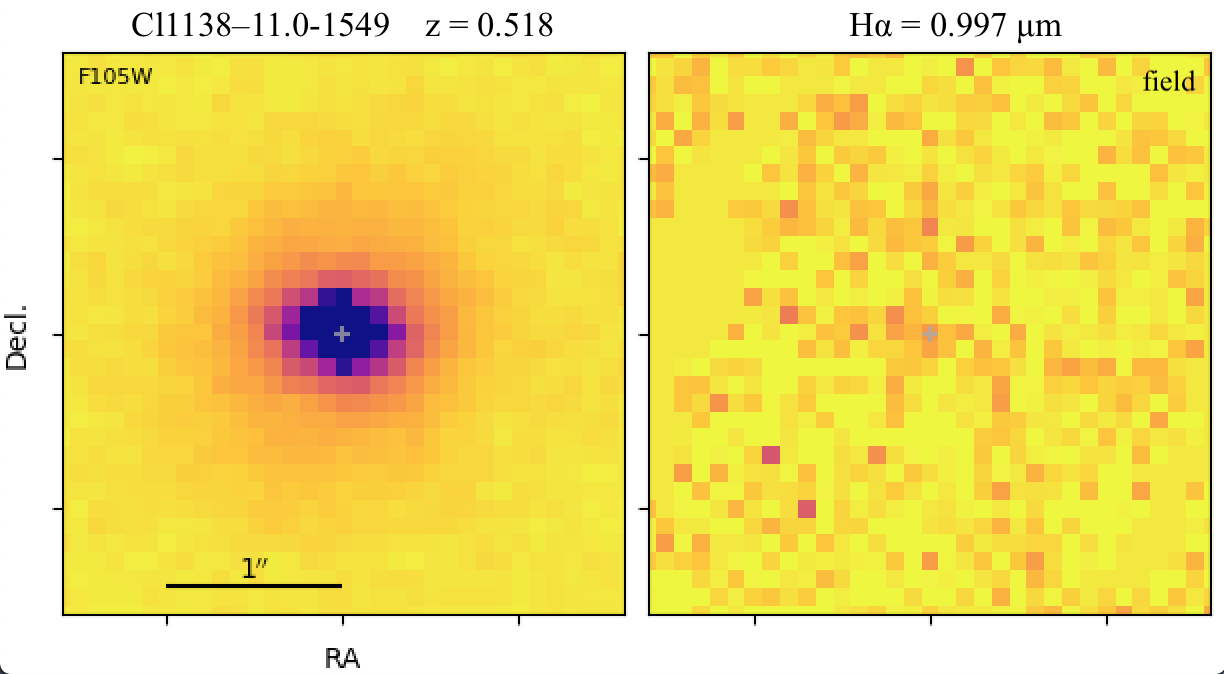}
\includegraphics[scale=0.25]{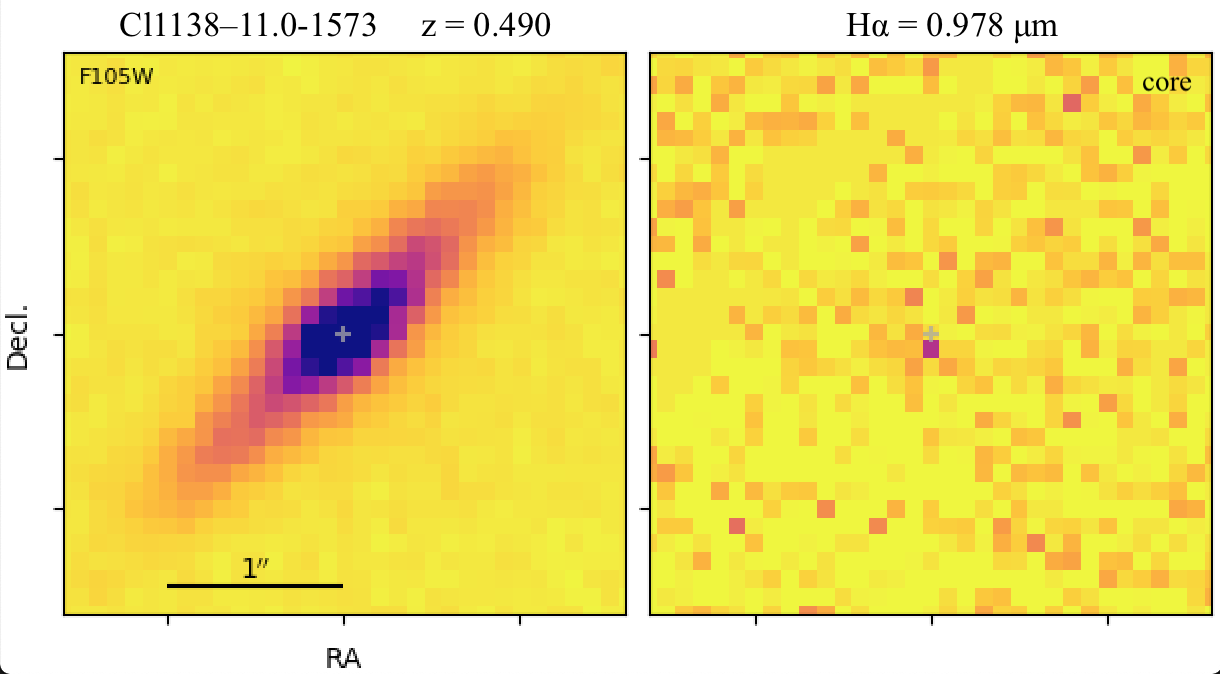}
\caption{The stellar continuum (left) and H$_{\alpha}$ line emission map (right) for each UVJ quiescent galaxy shown as a pair of images. The direct F105W images for each galaxy show signatures of early-type galaxies with a lack of spiral arms or clumpy morphology. The H$\alpha$ emission maps are typically low intensity or diffuse distributions, which makes inferences about the gas morphology difficult. }
\label{fig:app2}
\end{figure*}

In Figure~\ref{fig:UVJ}, 21 galaxies are identified as quiescent based upon the \cite{Just_2019} {\em UVJ} rest-frame colors, with five of them being in the core, two in the infall, and nine in the field. These are identified on the right hand side of Figure~\ref{fig:MS} as salmon triangles. GRIZLI produces a stellar continuum and emission line map for each observed galaxy, which is shown in Figure~\ref{fig:app2}. Many of the galaxies appear to have diffuse H$_{\alpha}$ with little-to-no stellar structure, indicating that these may be early-type galaxies. 

\section{Derivation for Extinction at H$\alpha$}
\label{sec:app4}
\begin{figure}
\centering
\includegraphics[scale=0.45]{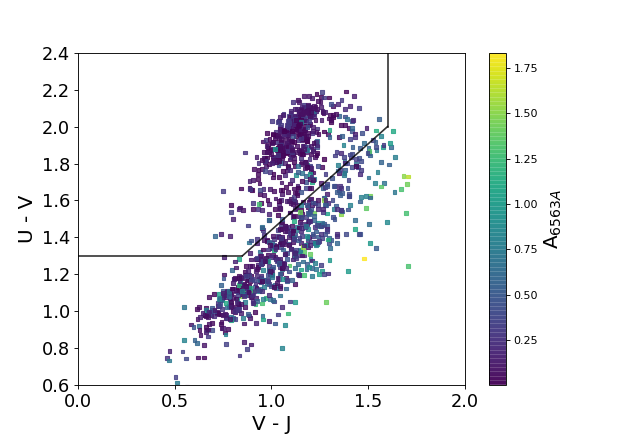}
\includegraphics[scale=0.45]{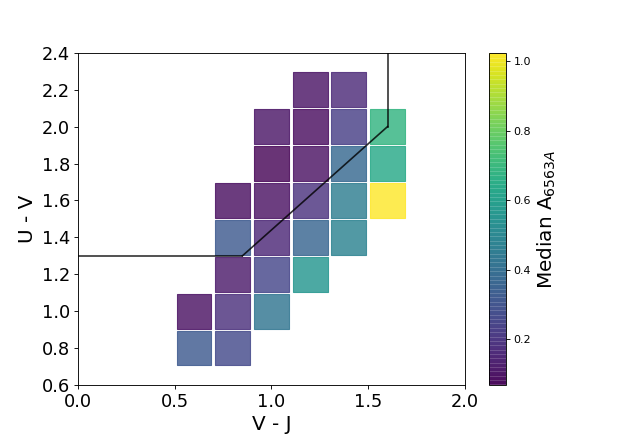}
\includegraphics[scale=0.45]{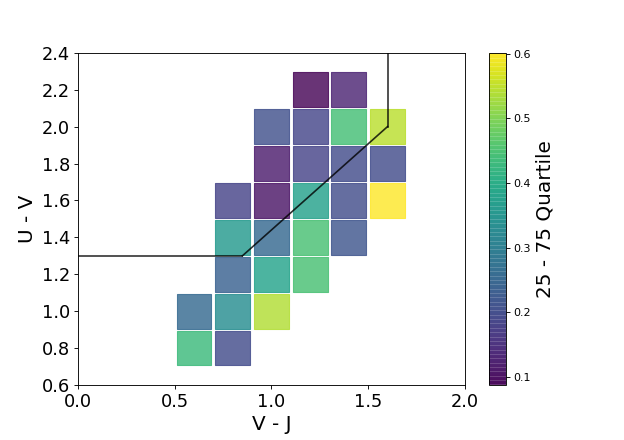}
\caption{(Top) A {\em UVJ} diagram color-coded by A$_{6563\AA}$ is shown for a subset of the ULTRAVISTA catalog that is similar in redshift and mass distribution to the {\em HST} sample in this study. (Middle) The median in 0.2 bins in $U-V$ and $V-J$ from the top plot is shown, which is used to match in color-color space to the {\em HST} sample. If a galaxy falls outside the distribution of ULTRAVISTA, it is matched to the nearest bin. {\em UVJ}-quiescent galaxies have low A$_v$ as expected and redder star-forming galaxies have elevated values. (Bottom) The spread of each bin across the 75 -- 25 percent quartile. The spread is noticeably small in the passive region. Black lines in all plots are the \protect\cite{will09} boundaries.}
\label{fig:uvjtau}
\end{figure}

The narrow spectral range of the {\em HST} G102 grism window does not allow us to determine the attenuation at H$\alpha$ from the grism data alone, and our SED calibration problems prevent us from using direct SED fits to do so. We therefore run MAGPHYS \citep{dacun08} on the ULTRAVISTA catalog \citep{muzzin13} to determine the continuum attenuation at 6563\AA. We run MAGPHYS without the UV and narrow-band filters as the SED fits were poorer using those filters. Our conclusions are unchanged if we include the UV and narrow-band filters, but the scatter of continuum attenuation increases. As described in the text, we use the ratio of the attenuated and unattenuated stellar continuum to derive the optical depth of the continuum at 6563\AA, which we convert to the attenuation in magnitudes at 6563\AA\ using $A_{6563\AA} = 1.086 \times \tau_{6563\AA}$. In Figure~\ref{fig:uvjtau} we show the median and interquartile range of $A_{6563\AA}$ for the ULTRAVISTA sample. We use this distribution of $A_{6563\AA}$ in $UVJ$ space, to infer the $A_{6563\AA}$ of our target galaxies by matching our galaxies to the nearest grid cell in $UVJ$-space. As described in the text, we then use $A_{6563\AA}$ to derive the line attenuation at H$\alpha$.

\section{Main Sequence Comparisons}
\label{sec:app3}
\begin{figure}[h!]
\centering
\includegraphics[scale=0.43]{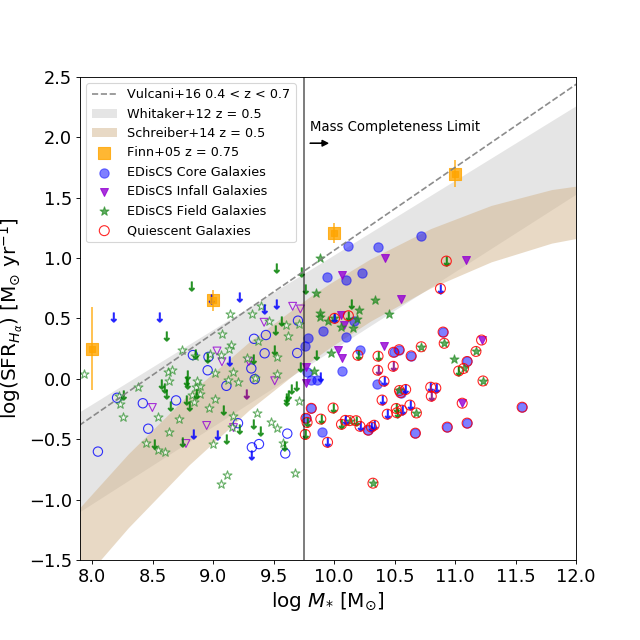}
\caption{The main sequence for the galaxies with S/N $>$ 3 in the core (blue circle), infall (purple triangle) and field (green stars) regions, where {\em UVJ} quiescent galaxies are circled in red and galaxies with less than S/N<3 are shown as the down arrows. Galaxies with S/N $<$ are shown at their 3$\sigma$ value as down arrows with a corresponding environment color. We show different determinations of the star-forming main sequence. As in Figure~\ref{fig:MS}, we show the cluster main sequence relations from \protect\cite{vul16} as a grey dashed line and \protect\cite{finn05} as orange squares. We also show UV$+$IR-based determinations for field galaxies from \protect\cite{whitaker12} in gray and \protect\cite{schr15} in brown, both at $z=0.5$ and with 0.3~dex scatter. The star-forming galaxies in our sample are in good agreement with both \protect\cite{whitaker12} and \protect\cite{schr15}. The $UVJ$-quiescent galaxies are mostly below these relations by $\sim$0.5 dex.}
\label{fig:dist1}
\end{figure}

In Figure~\ref{fig:dist1} we show a version of Figure~\ref{fig:MS}, but now with main sequence determinations from \cite{whitaker12} and \cite{schr15} at the median redshift of 0.487 for our sample. These determinations of the main sequence use SFRs determined from UV$+$IR and are systematically below the H$\alpha$ determinations of \cite{vul10}. They are consistent with the bulk of our $UVJ$ star-forming galaxies, but it is worth noting that our SFRs are systematically below the IR-based SFRs of \cite{finn10}. Taken together, this illustrates the significant systematic offsets between different SFR indicators and the importance of internal comparisons.

\clearpage
\section{{\em HST} catalogs}
This section contains tables with galaxy-specific information separated by each of the three environments.
\begin{table*}

\begin{center}
 \begin{tabular}[t]{lccccccccccc}
  \hline
 Object ID & R.A. & Dec & z  & Distance & Stellar Mass  & flux$_{H\alpha}$ & SFR$_{H\alpha}$  & {\em U -- V} & {\em V -- J} & {\em UVJ}\\
   & Deg. & Deg. & & Mpc & $\log_{10}$(M$_*$/$M\textsubscript{\(\odot\)}$) & 10$^{-16}${\rm erg~s$^{-1}$cm$^{-2}$} &   $M\textsubscript{\(\odot\)}$yr$^{-1}$  & AB & AB & Classification\\
  (1)       &   (2)     & (3)       & (4)       & (5)             & (6)     & (7)  & (8) & (9) & (10) & (11)   \\
  \hline
Cl1059-12.1-382	&	164.7492	&	-12.8713	&	0.4555	&	0.366	&	8.04	&	1.11	$\pm	0.12	$	&	0.70	$\pm	0.08	$	&	0.73	&	0.73	&	sf	\\
Cl1059-12.1-165	&	164.7544	&	-12.8858	&	0.4617	&	0.051	&	10.80	&	43.50	$\pm	0.4	$	&	15.11	$\pm	0.15	$	&	0.99	&	0.92	&	sf	\\
Cl1059-12.1-392	&	164.7546	&	-12.8709	&	0.4602	&	0.368	&	9.83	&	3.33	$\pm	0.16	$	&	1.15	$\pm	0.05	$	&	1.14	&	0.75	&	sf	\\
Cl1059-12.1-087	&	164.7554	&	-12.8909	&	0.4583	&	0.085	&	10.12	&	8.54	$\pm	0.26	$	&	3.00	$\pm	0.09	$	&	1.2	&	0.93	&	sf	\\
Cl1059-12.1-334	&	164.7563	&	-12.8743	&	0.4596	&	0.294	&	10.28	&	5.02	$\pm	0.32	$	&	1.73	$\pm	0.11	$	&	1.25	&	0.97	&	sf	\\
Cl1059-12.1-217	&	164.7583	&	-12.8821	&	0.4565	&	0.124	&	10.46	&	5.48	$\pm	0.34	$	&	1.74	$\pm	0.11	$	&	1.63	&	1.2	&	q	\\
Cl1059-12.1-431	&	164.7654	&	-12.8686	&	0.4601	&	0.407	&	9.65	&	1.02	$\pm	0.14	$	&	0.35	$\pm	0.05	$	&	1.22	&	0.74	&	sf	\\
Cl1059-12.1-461	&	164.7692	&	-12.8664	&	0.4595	&	0.453	&	10.91	&	1.13	$\pm	0.33	$	&	0.43	$\pm	0.13	$	&	1.95	&	1.23	&	q	\\
Cl1059-12.2-435*	&	164.7708	&	-12.8683	&	0.451	&	0.287	&	9.31	&	0.75	&	0.23	&	-0.63	&	1.08	&	q	\\						
Cl1059-12.2-320	&	164.7787	&	-12.8982	&	0.4615	&	0.227	&	10.71	&	1.03	$\pm	0.227	$	&	0.40	$\pm	0.10	$	&	1.88	&	1.09	&	q	\\
Cl1059-12.2-244	&	164.7819	&	-12.9035	&	0.4538	&	0.342	&	11.58	&	1.73	$\pm	0.49	$	&	0.58	$\pm	0.17	$	&	2.18	&	1.41	&	q	\\
Cl1059-12.2-231	&	164.7829	&	-12.905	&	0.4554	&	0.374	&	10.96	&	3.72	$\pm	0.24	$	&	1.39	$\pm	0.09	$	&	1.97	&	1.38	&	q	\\
Cl1059-12.2-272*	&	164.7835	&	-12.9017	&	0.4573	&	0.338	&	10.43	&	1.54	&	0.51	&	2.07	&	1.27	&	q	\\						
Cl1059-12.2-261	&	164.7852	&	-12.9025	&	0.4733	&	0.32	&	9.63	&	0.54	$\pm	0.17	$	&	0.29	$\pm	0.09	$	&	1.06	&	0.69	&	sf	\\
Cl1059-12.2-192	&	164.7861	&	-12.907	&	0.4609	&	0.416	&	10.60	&	4.79	$\pm	0.226	$	&	1.66	$\pm	0.09	$	&	1.43	&	1.07	&	sf	\\
Cl1059-12.2-364*	&	164.7864	&	-12.8947	&	0.4544	&	0.331	&	10.55	&	1.69	& 0.55	&	1.97	&	1.15	&	q	\\						
Cl1059-12.2-325	&	164.7884	&	-12.8979	&	0.4528	&	0.224	&	9.85	&	2.93	$\pm	0.36	$	&	0.97	$\pm	0.12	$	&	1.09	&	0.76	&	sf	\\
Cl1059-12.2-434	&	164.7897	&	-12.8874	&	0.4598	&	0.005	&	10.78	&	5.67	$\pm	0.39	$	&	2.44	$\pm	0.17	$	&	1.81	&	1.3	&	q	\\
Cl1059-12.2-311	&	164.7908	&	-12.8992	&	0.4587	&	0.253	&	9.80	&	0.95	$\pm	0.16	$	&	0.47	$\pm	0.08	$	&	1.42	&	1.09	&	sf	\\
Cl1059-12.2-452	&	164.7911	&	-12.8849	&	0.4577	&	0.059	&	5.77	&	0.99	$\pm	0.14	$	&	1.46	$\pm	0.21	$	&	0.86	&	0.51	&	sf	\\
Cl1059-12.2-350	&	164.7912	&	-12.8954	&	0.4573	&	0.172	&	8.88	&	0.79	$\pm	0.17	$	&	0.39	$\pm	0.08	$	&	0.8	&	-0.46	&	sf	\\
Cl1059-12.2-441	&	164.794	&	-12.8867	&	0.4505	&	0.021	&	10.44	&	2.35	$\pm	0.33	$	&	0.77	$\pm	0.11	$	&	1.62	&	1.19	&	sf	\\
Cl1059-12.2-460*	&	164.7944	&	-12.884	&	0.4571	&	0.371	&	10.32	&	1.25	&	0.41	&	2.0	&	1.12	&	q	\\						
Cl1059-12.2-466	&	164.7959	&	-12.8802	&	0.4613	&	0.161	&	9.31	&	0.78	$\pm	0.23	$	&	0.27	$\pm	0.08	$	&	1.0	&	0.80	&	sf	\\
Cl1059-12.2-449	&	164.7962	&	-12.8853	&	0.4524	&	0.053	&	8.62	&	1.31	$\pm	0.19	$	&	0.63	$\pm	0.09	$	&	0.58	&	0.19	&	sf	\\
Cl1059-12.2-463	&	164.7962	&	-12.8828	&	0.4564	&	0.106	&	9.39	&	3.32	$\pm	0.16	$	&	1.63	$\pm	0.08	$	&	0.85	&	0.68	&	sf	\\
Cl1059-12.2-387	&	164.8008	&	-12.8917	&	0.4561	&	0.099	&	9.81	&	1.17	$\pm	0.16	$	&	0.57	$\pm	0.08	$	&	1.37	&	0.95	&	sf	\\
Cl1059-12.2-395	&	164.8048	&	-12.8905	&	0.455	&	0.079	&	9.35	&	2.96	$\pm	0.13	$	&	0.99	$\pm	0.04	$	&	0.94	&	0.83	&	sf	\\
Cl1059-12.2-157*	&	164.8058	&	-12.909	&	0.4569	&	2.486	&	10.83	&	2.54	&	0.84	&	2.07	&	1.25	&	q	\\						
Cl1059-12.2-172	&	164.8059	&	-12.9082	&	0.4613	&	0.455	&	9.85	&	2.84	$\pm	0.18	$	&	0.98	$\pm	0.06	$	&	1.10	&	0.96	&	sf	\\
Cl1059-12.2-109*	&	164.8092	&	-12.9133	&	0.4473	&	0.398	&	9.94	&	0.00	&	0.30	&	1.88	&	1.01	&	q	\\						
Cl1138-11.0-1623*	&	174.5321	&	-11.5609	&	0.4778	&	0.392	&	10.79	&	2.32	&	0.85 &	2.05	&	1.17	&	q	\\						
Cl1138-11.0-1772	&	174.5345	&	-11.5584	&	0.4829	&	0.035	&	9.32	&	2.17	$\pm	0.29	$	&	1.22	$\pm	0.17	$	&	0.87	&	0.36	&	sf	\\
Cl1138-11.0-1460	&	174.5358	&	-11.5635	&	0.4781	&	0.047	&	8.83	&	3.16	$\pm	0.15	$	&	1.57	$\pm	0.07	$	&	0.69	&	0.85	&	sf	\\
Cl1138-11.0-199	&	174.5399	&	-11.5897	&	0.4799	&	0.452	&	9.70	&	3.72	$\pm	0.24	$	&	1.63	$\pm	0.10	$	&	1.08	&	0.72	&	sf	\\
Cl1138-11.0-1203	&	174.5431	&	-11.5686	&	0.4785	&	0.124	&	9.11	&	1.60	$\pm	0.29	$	&	0.87	$\pm	0.16	$	&	0.76	&	0.34	&	sf	\\
Cl1138-11.0-1754*	&	174.5435	&	-11.559	&	0.4994	&	2.79	&	10.40	&	2.33	&	0.96	&	1.99	&	1.25	&	sf	\\						
Cl1138-11.0-1545*	&	174.5447	&	-11.5621	&	0.4771	&	0.617	&	10.62	&	1.45	&	0.61	&	2.17	&	1.37	&	q	\\						
Cl1138-11.0-1073	&	174.545	&	-11.5705	&	0.481	&	0.155	&	10.28	&	1.04	$\pm	0.3	$	&	0.38	$\pm	0.11	$	&	1.82	&	0.92	&	q	\\
Cl1138-11.0-1230	&	174.5452	&	-11.5681	&	0.478	&	0.117	&	8.96	&	2.78	$\pm	0.46	$	&	1.52	$\pm	0.25	$	&	0.74	&	0.01	&	sf	\\
Cl1138-11.0-1573	&	174.5459	&	-11.5616	&	0.4902	&	0.016	&	10.67	&	0.88	$\pm	0.26	$	&	0.35	$\pm	0.10	$	&	2.11	&	1.22	&	q	\\
Cl1138-11.0-375	&	174.5464	&	-11.5843	&	0.4787	&	0.368	&	9.76	&	4.83	$\pm	0.63	$	&	1.88	$\pm	0.24	$	&	1.17	&	0.77	&	sf	\\
Cl1138-11.0-298	&	174.5464	&	-11.5867	&	0.4852	&	0.406	&	9.60	&	0.66	$\pm	0.16	$	&	0.24	$\pm	0.06	$	&	1.85	&	0.98	&	q	\\
Cl1138-11.0-1516	&	174.5496	&	-11.5626	&	0.4637	&	0.032	&	8.05	&	0.49	$\pm	0.1	$	&	0.25	$\pm	0.05	$	&	0.22	&	-0.12	&	sf	\\
Cl1138-11.0-955*	&	174.5528	&	-11.5727	&	0.4895	&	0.404	&	10.30	&	1.02	&	0.40	&	2.10	&	1.12	&	sf	\\						
Cl1138-11.0-1409	&	174.5531	&	-11.5647	&	0.4875	&	0.066	&	9.70	&	6.64	$\pm	0.4	$	&	3.03	$\pm	0.18	$	&	0.95	&	0.79	&	sf	\\
Cl1138-11.0-990*	&	174.5531	&	-11.5719	&	0.4796	&	0.936	&	8.82	&	0.69	&	0.35	&	0.68	&	0.95	&	sf	\\						
Cl1138-11.0-1897	&	174.5574	&	-11.5566	&	0.4822	&	0.064	&	9.34	&	5.56	$\pm	0.49	$	&	2.14	$\pm	0.19	$	&	1.02	&	0.48	&	sf	\\
Cl1138-11.0-668	&	174.5624	&	-11.5779	&	0.4988	&	0.275	&	8.69	&	1.09	$\pm	0.19	$	&	0.66	$\pm	0.11	$	&	0.29	&	-0.38	&	sf	\\
Cl1138-11.0-979	&	174.5641	&	-11.5727	&	0.4851	&	0.195	&	10.12	&	25.10	$\pm	0.78	$	&	12.65	$\pm	0.39	$	&	1.17	&	0.99	&	sf	\\
Cl1138-11.0-1265	&	174.5651	&	-11.5673	&	0.4697	&	0.113	&	9.90	&	0.39	$\pm	0.093	$	&	0.36	$\pm	0.09	$	&	1.93	&	1.62	&	sf	\\
Cl1138-11.0-348*	&	174.5681	&	-11.5848	&	0.4944	&	0.834	&	9.02	&	0.84	&	0.35	&	1.16	&	-0.34	&	sf	\\						
Cl1227-11.0-126	&	186.972	&	-11.6148	&	0.6375	&	0.709	&	10.10	&	4.65	$\pm	0.25	$	&	6.61	$\pm	0.35	$	&	1.37	&	1.41	&	sf	\\
Cl1227-11.0-244*	&	186.9794	&	-11.6056	&	0.6375	&	2.328	&	9.88	&	0.73	&	1.03	&	1.42	&	1.53	&	sf	\\						
Cl1227-11.0-332	&	186.9794	&	-11.5997	&	0.626	&	0.324	&	10.07	&	0.76	$\pm	0.23	$	&	1.17	$\pm	0.36	$	&	1.28	&	1.22	&	sf	\\
Cl1227-11.0-217	&	186.9838	&	-11.6078	&	0.6321	&	0.524	&	10.64	&	1.44	$\pm	0.23	$	&	1.55	$\pm	0.24	$	&	1.44	&	0.57	&	sf	\\
Cl1227-11.0-482*	&	186.9877	&	-11.5889	&	0.6429	&	0.05	&	10.39	&	0.86	&	0.67	&	2.15	&	1.29	&	q	\\							

\hline
 \end{tabular} \par
 \bigskip
\caption{Information for core galaxies. 1. Pointing ID - GRIZLI Object ID. IDs ending with * are galaxies with S/N$_{H\alpha}$ $<$ 3. 2. Right Ascension 3. Declination 4. Redshift 5. Cluster-centric distance in Mpc. 6. Stellar Mass 7. Uncorrected H$\alpha$ flux in cgs units. S/N $<3$ detections are listed at the 3$\sigma$ upper limit. 8. Star-formation rate. S/N $<3$ detections are listed at the 3$\sigma$ upper limit. 9. {\em U -- V} rest-frame color 10. {\em V -- J} rest-frame color 11. {\em UVJ} classification based on \protect\cite{will09} where sf and q represent star-forming and quiescent, respectively.}
 \label{tab:continued}
\end{center}
\end{table*}

\begin{table*}
\begin{center}
 \contcaption{Core Galaxies}
 \label{tab:continued}
 \begin{tabular}[t]{lcccccccccc}
  \hline
 Object ID & R.A. & Dec & z  & Distance & Stellar Mass  & flux$_{H\alpha}$ & SFR$_{H\alpha}$  & {\em U -- V} & {\em V -- J} & {\em UVJ}\\
   & Deg. & Deg. & & Mpc & $\log_{10}$(M$_*$/$M\textsubscript{\(\odot\)}$) & 10$^{-16}${\rm erg~s$^{-1}$cm$^{-2}$} &   $M\textsubscript{\(\odot\)}$yr$^{-1}$  & AB & AB & Classification\\
  (1)       &   (2)     & (3)       & (4)       & (5)             & (6)     & (7)  & (8) & (9) & (10) & (11)   \\
  \hline
Cl1227-11.0-348	& 	186.988	& 	11.5988	& 	0.6335	&	0.298	&	10.36	&	9.63	$\pm	0.27	$	&	1.09	$\pm	0.54	$	&	0.78	&	1.08	&	sf	\\
Cl1227-11.0-368*	&	186.9907	&	-11.5979	&	0.6277	&	0.741	&	10.20	&	0.56	&	0.40	&	1.61	&	0.37	&	q	\\						
Cl1227-11.0-527*	&	186.9947	&	-11.586	&	0.6429	&	0.688	&	10.58	&	0.84	& 0.82	&	1.91	&	1.32	&	q	\\						
Cl1301-11.0-134	&	195.3694	&	-11.6311	&	0.5003	&	0.458	&	9.44	&	3.78	$\pm	0.54	$	&	2.30	$\pm	0.33	$	&	0.87	&	0.48	&	sf	\\
Cl1301-11.0-304	&	195.3789	&	-11.6149	&	0.4681	&	0.714	&	9.21	&	11.60	$\pm	0.52	$	&	0.43	$\pm	0.08	$	&	0.49	&	0.29	&	sf	\\
Cl1301-11.0-159*	&	195.3722	&	-11.6286	&	0.4911	&	0.094	&	9.21	&	11.80	&	4.74	&	0.93	&	0.37	&	sf	\\						
Cl1301-11.0-193*	&	195.3731	&	-11.6251	&	0.465	&	0.065	&	8.55	&	6.36	&	3.25	&	0.57	&	0.18	&	sf	\\						
Cl1301-11.0-384*	&	195.3745	&	-11.6091	&	0.4655	&	0.062	&	9.12	&	3.86	&	61.41	&	1.21	&	0.82	&	sf	\\						
Cl1301-11.0-092*	&	195.3748	&	-11.6342	&	0.493	&	0.024	&	10.08	&	1.10	&	0.45	&	2.23	&	0.93	&	q	\\						
Cl1301-11.0-023*	&	195.3766	&	-11.6408	&	0.4896	&	0.006	&	10.87	&	13.80	&	5.58	&	2.00	&	1.04	&	q	\\						
Cl1301-11.0-059*	&	195.3806	&	-11.6374	&	0.483	&	0.033	&	9.99	&	8.08	&	3.17	&	1.92	&	0.92	&	q	\\						
Cl1301-11.0-408*	&	195.383	&	-11.6074	&	0.4703	&	0.412	&	9.51	&	10.50	&	4.19	&	1.46	&	0.94	&	q	\\						
Cl1301-11.0-298*	&	195.3837	&	-11.6154	&	0.488	&	0.304	&	8.97	&	8.02	&	4.60	&	0.6	&	0.44	&	sf	\\						
Cl1301-11.0-143*	&	195.3846	&	-11.63	&	0.4631	&	0.155	&	8.16	&	6.36	&	3.22	&	0.21	&	0.64	&	sf	\\						
Cl1301-11.0-157	&	195.386	&	-11.6292	&	0.4871	&	0.468	&	9.90	&	4.86	$\pm	0.48	$	&	2.48	$\pm	0.24	$	&	1.14	&	0.79	&	sf	\\
Cl1301-11.0-187	&	195.3861	&	-11.6258	&	0.4823	&	0.524	&	9.94	&	13.90	$\pm	0.44	$	&	6.94	$\pm	0.22	$	&	1.16	&	0.86	&	sf	\\
Cl1301-11.0-060*	&	195.3888	&	-11.6373	&	0.5015	&	0.568	&	9.41	&	4.18	&	3.78	&	1.14	&	1.19	&	sf	\\						
Cl1301-11.1-502	&	195.3762	&	-11.6977	&	0.482	&	0.715	&	8.96	&	2.11	$\pm	0.2	$	&	1.18	$\pm	0.11	$	&	1.21	&	0.96	&	sf	\\
Cl1301-11.1-491	&	195.3847	&	-11.6989	&	0.481	&	0.725	&	10.23	&	19.10	$\pm	0.52	$	&	7.54	$\pm	0.21	$	&	1.16	&	1.01	&	sf	\\
Cl1301-11.1-525	&	195.3862	&	-11.6954	&	0.4855	&	0.665	&	10.1	&	3.30	$\pm	0.3	$	&	2.22	$\pm	0.20	$	&	1.08	&	0.94	&	sf	\\
Cl1301-11.1-560	&	195.3924	&	-11.6929	&	0.4832	&	0.617	&	9.78	&	3.27	$\pm	0.23	$	&	2.18	$\pm	0.15	$	&	0.93	&	0.92	&	sf	\\
Cl1301-11.1-324	&	195.3935	&	-11.7106	&	0.4862	&	0.911	&	10.35	&	1.78	$\pm	0.25	$	&	0.90	$\pm	0.13	$	&	1.28	&	1.00	&	sf	\\

  \hline

 \end{tabular} \par
 \bigskip
\contcaption{Information for core galaxies. 1. Pointing ID - GRIZLI Object ID. IDs ending with * are galaxies with S/N$_{H\alpha}$ $<$ 3. 2. Right Ascension 3. Declination 4. Redshift 5. Cluster-centric distance in Mpc. 6. Stellar Mass 7. Uncorrected H$\alpha$ flux in cgs units. S/N $<3$ detections are listed at the 3$\sigma$ upper limit. 8. Star-formation rate. S/N $<3$ detections are listed at the 3$\sigma$ upper limit. 9. {\em U -- V} rest-frame color 10. {\em V -- J} rest-frame color 11. {\em UVJ} classification based on \protect\cite{will09} where sf and q represent star-forming and quiescent, respectively. }
 \label{tab:continued}
\end{center}
\end{table*}

\begin{table*}
\begin{center}
 \begin{tabular}[t]{lccccccccccc}
  \hline
 Object ID & R.A. & Dec & z  & Distance & Stellar Mass  & flux$_{H\alpha}$ & SFR$_{H\alpha}$ & {\em U -- V} & {\em V =- J} & {\em UVJ}\\
   & Deg. & Deg. & & Mpc & $\log_{10}$(M$_*$/$M\textsubscript{\(\odot\)}$) & 10$^{-16}${\rm erg~s$^{-1}$cm$^{-2}$} &   $M\textsubscript{\(\odot\)}$yr$^{-1}$  & AB & AB & Classification\\
  (1)       &   (2)     & (3)       & (4)       & (5)             & (6)     & (7)  & (8) & (9) & (10) & (11) \\
  \hline
Cl1138-11.2-166	&	174.3821	&	-11.4293	&	0.4812	&	2.44	&	10.03	&	1.04	$\pm	0.166	$	&	1.73	$\pm	0.27	$	&	1.4	&	1.94	&	sf	\\
Cl1138-11.2-222	&	174.3956	&	-11.4266	&	0.4818	&	2.41	&	10.08	&	3.90	$\pm	0.219	$	&	2.81	$\pm	0.16	$	&	1.17	&	1.82	&	sf	\\
Cl1138-11.2-213	&	174.3994	&	-11.4272	&	0.4654	&	2.39	&	10.42	&	3.33	$\pm	0.314	$	&	1.86	$\pm	0.18	$	&	1.7	&	1.45	&	sf	\\
Cl1138-11.2-490	&	174.401	&	-11.4085	&	0.4735	&	2.67	&	9.77	&	3.37	$\pm	0.486	$	&	1.24	$\pm	0.18	$	&	1.08	&	0.67	&	sf	\\
Cl1138-11.2-434*	&	174.4048	&	-11.4127	&	0.4878	&	0.387	&	10.48	&	3.41	&	1.27	&	1.85	&	1.04	&	q	\\						
Cl1138-11.1-077	&	174.6872	&	-11.5849	&	0.4869	&	0.7	&	9.77	&	2.26	$\pm	0.624	$	&	0.92	$\pm	0.25	$	&	1.16	&	0.69	&	sf	\\
Cl1138-11.1-137	&	174.7016	&	-11.5811	&	0.4894	&	0.71	&	9.07	&	2.39	$\pm	0.589	$	&	1.38	$\pm	0.34	$	&	0.66	&	0.27	&	sf	\\
Cl1138-11.1-388	&	174.7105	&	-11.5616	&	0.4945	&	0.45	&	9.72	&	6.39	$\pm	0.571	$	&	3.79	$\pm	0.34	$	&	0.81	&	0.5	&	sf	\\
Cl1138-11.1-263	&	174.7122	&	-11.5714	&	0.4903	&	0.61	&	11.09	&	5.51	$\pm	0.973	$	&	9.56	$\pm	1.69	$	&	1.67	&	2.55	&	sf	\\
Cl1227-11.1-088	&	186.9564	&	-11.6694	&	0.6166	&	2.1	&	9.30	&	1.37	$\pm	0.172	$	&	1.39	$\pm	0.17	$	&	0.76	&	0.32	&	sf	\\
Cl1227-11.1-226	&	186.9569	&	-11.6607	&	0.6383	&	1.88	&	10.07	&	5.49	$\pm	0.254	$	&	7.17	$\pm	0.33	$	&	0.8	&	1.06	&	sf	\\
Cl1227-11.1-309	&	186.9807	&	-11.6551	&	0.6358	&	1.71	&	10.06	&	3.87	$\pm	0.253	$	&	3.36	$\pm	0.22	$	&	1	&	0.87	&	sf	\\
Cl1227-11.2-225	&	187.0602	&	-11.5263	&	0.6246	&	1.62	&	11.22	&	2.32	$\pm	0.385	$	&	2.09	$\pm	0.35	$	&	1.98	&	1.37	&	q	\\
Cl1227-11.2-163	&	187.0621	&	-11.5297	&	0.6428	&	1.54	&	9.66	&	3.55	$\pm	0.162	$	&	3.98	$\pm	0.18	$	&	0.79	&	0.61	&	sf	\\
Cl1227-11.2-347	&	187.0682	&	-11.5193	&	0.6312	&	1.82	&	9.03	&	1.58	$\pm	0.159	$	&	1.70	$\pm	0.17	$	&	0.5	&	0.37	&	sf	\\
Cl1301-11.3-220	&	195.2301	&	-11.5203	&	0.4683	&	2.84	&	8.50	&	1.11	$\pm	0.247	$	&	0.58	$\pm	0.13	$	&	0.13	&	0.61	&	sf	\\
Cl1301-11.3-337	&	195.2334	&	-11.513	&	0.4941	&	2.95	&	10.07	&	3.64	$\pm	0.244	$	&	1.49	$\pm	0.10	$	&	1.01	&	0.57	&	sf	\\
Cl1301-11.3-268*	&	195.2339	&	-11.5175	&	0.4892	&	0.223	&	9.27	&	1.87	&	0.75	&	1.02	&	0.09	&	sf	\\						
Cl1301-11.3-144	&	195.2358	&	-11.5266	&	0.4926	&	2.71	&	10.55	&	4.38	$\pm	0.403	$	&	4.58	$\pm	0.42	$	&	1.91	&	1.78	&	sf	\\
Cl1301-11.3-223	&	195.2426	&	-11.5201	&	0.4947	&	2.77	&	8.78	&	0.49	$\pm	0.151	$	&	0.29	$\pm	0.09	$	&	0.17	&	-0.11	&	sf	\\
Cl1301-11.3-226*	&	195.2557	&	-11.5202	&	0.4652	&	0.19	&	10.79	&	2.00	&	0.72	&	2.18	&	1.01	&	q	\\						
Cl1301-11.3-106*	&	195.2643	&	-11.5294	&	0.4957	&	0.175	&	10.35	&	2.29	&	1.18	&	1.97	&	1.42	&	q	\\						
Cl1301-11.1-300	&	195.3445	&	-11.5418	&	0.4974	&	1.99	&	9.51	&	1.35	$\pm	0.214	$	&	0.96	$\pm	0.15	$	&	0.88	&	1.25	&	sf	\\
Cl1301-11.1-421	&	195.3487	&	-11.5294	&	0.4976	&	2.19	&	9.17	&	0.95	$\pm	0.249	$	&	0.39	$\pm	0.10	$	&	0.94	&	0.13	&	sf	\\
Cl1301-11.1-061	&	195.3502	&	-11.5586	&	0.4672	&	1.7	&	9.42	&	5.68	$\pm	0.35	$	&	2.94	$\pm	0.18	$	&	0.73	&	0.53	&	sf	\\
Cl1301-11.1-354	&	195.3503	&	-11.5372	&	0.4944	&	2.06	&	10.42	&	18.80	$\pm	0.828	$	&	9.96	$\pm	0.44	$	&	1.13	&	0.91	&	sf	\\
Cl1301-11.1-361	&	195.3542	&	-11.5363	&	0.4933	&	2.06	&	8.95	&	0.77	$\pm	0.22	$	&	0.41	$\pm	0.12	$	&	0.42	&	1.35	&	sf	\\
Cl1301-11.1-208	&	195.3585	&	-11.5476	&	0.501	&	1.87	&	11.06	&	1.48	$\pm	0.229	$	&	0.63	$\pm	0.10	$	&	2.11	&	1.28	&	q	\\
  \hline

 \end{tabular} \par
 \bigskip
\caption{Information for infall galaxies. 1. Pointing ID - GRIZLI Object ID. IDs ending with * are galaxies with S/N$_{H\alpha}$ $<$ 3. 2. Right Ascension 3. Declination 4. Redshift 5. Cluster-centric distance in Mpc. 6. Stellar Mass 7. Uncorrected H$\alpha$ flux in cgs units. S/N $<3$ detections are listed at the 3$\sigma$ upper limit. 8. Star-formation rate. S/N $<3$ detections are listed at the 3$\sigma$ upper limit. 9. {\em U -- V} rest-frame color 10. {\em V -- J} rest-frame color 11. {\em UVJ} classification based on \protect\cite{will09} where sf and q represent star-forming and quiescent, respectively.}
 \label{tab:infall}
\end{center}
\end{table*}

\begin{table*}

\begin{center}
 \begin{tabular}[t]{lccccccccc}
  \hline
 Object ID & R.A. & Dec & z   & Stellar Mass  & flux$_{H\alpha}$ & SFR$_{H\alpha}$ & {\em U-V} & {\em V-J} & {\em UVJ} \\
  & Deg. & Deg. &  & $\log_{10}$(M$_*$/$M\textsubscript{\(\odot\)}$) & 10$^{-16}${\rm erg~s$^{-1}$cm$^{-2}$}  &   $M\textsubscript{\(\odot\)}$yr$^{-1}$ & AB & AB & Classification \\
  (1)       &   (2)     & (3)       & (4)       & (5)             & (6) &(7) & (8) & (9) & (10)  \\
  \hline
Cl1059-12.1-326	&	164.7441	&	-12.8751	&	0.6477	&	10.19	&	1.32	$\pm	0.21	$	&	4.50	$\pm	0.73	$	&	1.04	&	1.99	&	sf	\\
Cl1059-12.1-413	&	164.7442	&	-12.8697	&	0.666	&	10.46	&	1.39	$\pm	0.3	$	&	5.07	$\pm	1.10	$	&	1.57	&	2.09	&	sf	\\
Cl1059-12.1-505	&	164.7475	&	-12.8625	&	0.6323	&	9.15	&	0.85	$\pm	0.14	$	&	2.74	$\pm	0.45	$	&	1.13	&	1.97	&	sf	\\
Cl1059-12.1-408	&	164.7506	&	-12.8697	&	0.6582	&	8.54	&	0.65	$\pm	0.12	$	&	0.77	$\pm	0.14	$	&	0.41	&	0.19	&	sf	\\
Cl1059-12.1-192	&	164.7514	&	-12.8837	&	0.6705	&	9.12	&	0.85	$\pm	0.1	$	&	1.55	$\pm	0.18	$	&	0.88	&	1.70	&	sf	\\
Cl1059-12.1-162	&	164.7514	&	-12.8859	&	0.6716	&	8.54	&	0.37	$\pm	0.12	$	&	0.42	$\pm	0.13	$	&	0.55	&	0.86	&	sf	\\
Cl1059-12.1-517	&	164.752	&	-12.8617	&	0.4064	&	10.05	&	4.08	$\pm	0.32	$	&	2.23	$\pm	0.17	$	&	1.19	&	1.35	&	sf	\\
Cl1059-12.1-496	&	164.7523	&	-12.8632	&	0.6979	&	10.99	&	0.9	$\pm	0.24	$	&	2.19	$\pm	0.57	$	&	1.98	&	2.24	&	sf	\\
Cl1059-12.1-486	&	164.7528	&	-12.8638	&	0.6544	&	10.32	&	0.22	$\pm	0.06	$	&	0.20	$\pm	0.06	$	&	2.22	&	1.58	&	sf	\\
Cl1059-12.1-086	&	164.7536	&	-12.8908	&	0.6076	&	8.45	&	0.47	$\pm	0.11	$	&	0.41	$\pm	0.10	$	&	0.23	&	0.74	&	sf	\\
Cl1059-12.1-183	&	164.758	&	-12.8842	&	0.6428	&	9.16	&	0.41	$\pm	0.11	$	&	0.59	$\pm	0.16	$	&	1.19	&	1.64	&	sf	\\
Cl1059-12.1-101	&	164.7626	&	-12.8893	&	0.6107	&	8.22	&	1.14	$\pm	0.35	$	&	1.03	$\pm	0.31	$	&	0.23	&	1.01	&	sf	\\
Cl1059-12.1-333	&	164.7639	&	-12.874	&	0.5182	&	8.6	&	0.4	$\pm	0.09	$	&	0.27	$\pm	0.06	$	&	0.50	&	0.46	&	sf	\\
Cl1059-12.1-241	&	164.7699	&	-12.8805	&	0.4865	&	9.85	&	3.02	$\pm	0.17	$	&	5.14	$\pm	0.29	$	&	1.37	&	2.06	&	sf	\\
Cl1059-12.2-285	&	164.773	&	-12.8781	&	0.5736	&	8.63	&	1.86	$\pm	0.09	$	&	1.44	$\pm	0.07	$	&	0.44	&	0.70	&	sf	\\
Cl1059-12.2-220	&	164.778	&	-12.9053	&	0.4291	&	9.71	&	0.45	$\pm	0.14	$	&	0.57	$\pm	0.17	$	&	1.43	&	1.75	&	q	\\
Cl1059-12.2-405	&	164.7856	&	-12.8899	&	0.5032	&	8.72	&	0.87	$\pm	0.13	$	&	0.49	$\pm	0.07	$	&	0.62	&	0.90	&	sf	\\
Cl1059-12.2-289	&	164.7867	&	-12.8999	&	0.4245	&	9.53	&	0.66	$\pm	0.19	$	&	0.34	$\pm	0.10	$	&	1.58	&	1.45	&	sf	\\
Cl1059-12.2-310	&	164.7926	&	-12.8992	&	0.4127	&	9.83	&	0.84	$\pm	0.25	$	&	0.96	$\pm	0.29	$	&	1.56	&	1.78	&	sf	\\
Cl1059-12.2-447	&	164.7978	&	-12.8854	&	0.545	&	10.13	&	0.99	$\pm	0.12	$	&	0.51	$\pm	0.06	$	&	2.37	&	1.15	&	sf	\\
Cl1059-12.2-396	&	164.7999	&	-12.8906	&	0.5936	&	9.46	&	3.29	$\pm	0.17	$	&	3.92	$\pm	0.21	$	&	0.88	&	1.92	&	sf	\\
Cl1059-12.2-399	&	164.8028	&	-12.8902	&	0.4168	&	8.64	&	0.64	$\pm	0.13	$	&	0.30	$\pm	0.06	$	&	1.00	&	0.92	&	sf	\\
Cl1059-12.2-390	&	164.805	&	-12.8911	&	0.4245	&	9.53	&	0.76	$\pm	0.22	$	&	0.94	$\pm	0.27	$	&	1.33	&	1.80	&	sf	\\
Cl1138-11.2-158	&	174.38	&	-11.4304	&	0.4139	&	9.21	&	1.36	$\pm	0.42	$	&	0.78	$\pm	0.24	$	&	1.16	&	1.44	&	sf	\\
Cl1138-11.2-090	&	174.3832	&	-11.4337	&	0.4552	&	9.32	&	4.38	$\pm	0.45	$	&	3.14	$\pm	0.32	$	&	0.52	&	1.78	&	sf	\\
Cl1138-11.2-060	&	174.3834	&	-11.4351	&	0.6938	&	9.89	&	2.82	$\pm	0.22	$	&	5.62	$\pm	0.44	$	&	1.22	&	1.19	&	sf	\\
Cl1138-11.2-504	&	174.3887	&	-11.4074	&	0.6761	&	9.98	&	1.85	$\pm	0.22	$	&	2.51	$\pm	0.30	$	&	1.32	&	1.20	&	sf	\\
Cl1138-11.2-376	&	174.3972	&	-11.4166	&	0.6875	&	9.28	&	1.26	$\pm	0.19	$	&	2.46	$\pm	0.36	$	&	0.72	&	1.50	&	q	\\
Cl1138-11.2-246	&	174.3987	&	-11.4245	&	0.5759	&	9.02	&	1.1	$\pm	0.24	$	&	0.86	$\pm	0.19	$	&	0.80	&	0.76	&	sf	\\
Cl1138-11.2-079	&	174.4013	&	-11.4342	&	0.5195	&	10.34	&	2.4	$\pm	0.46	$	&	4.78	$\pm	0.92	$	&	1.31	&	2.03	&	sf	\\
Cl1138-11.0-1481	&	174.5333	&	-11.5634	&	0.641	&	8.82	&	0.99	$\pm	0.19	$	&	1.10	$\pm	0.21	$	&	0.63	&	-0.27	&	sf	\\
Cl1138-11.0-1299*	&	174.5355	&	-11.5666	&	0.6558	&	8.60	&	0.64	&	0.75	&	0.26	&	-0.36	&	sf	\\						
Cl1138-11.0-1136*	&	174.5373	&	-11.5692	&	0.5856	&	9.33	&	0.68	&	0.42	&	1.90	&	0.85	&	q	\\						
Cl1138-11.0-789	&	174.5375	&	-11.5756	&	0.5063	&	9.58	&	0.73	$\pm	0.11	$	&	0.31	$\pm	0.04	$	&	1.51	&	0.59	&	sf	\\
Cl1138-11.0-1650*	&	174.538	&	-11.5602	&	0.5333	&	10.05	&	0.84	&	0.42	&	2.00	&	0.89	&	q	\\						
Cl1138-11.0-1779*	&	174.5386	&	-11.5584	&	0.6437	&	9.71	&	1.03	&	1.24	&	1.43	&	1.19	&	sf	\\						
Cl1138-11.0-1231*	&	174.539	&	-11.5681	&	0.5607	&	9.61	&	1.31	&	0.73	&	1.32	&	0.03	&	q	\\						
Cl1138-11.0-558*	&	174.5391	&	-11.5797	&	0.6363	&	8.56	&	0.83	&	0.9	&	0.75	&	-0.05	&	sf	\\						
Cl1138-11.0-1404*	&	174.54	&	-11.5647	&	0.562	&	10.17	&	0.8	&	0.45	&	2.20	&	0.90	&	q	\\						
Cl1138-11.0-1549	&	174.5401	&	-11.562	&	0.5182	&	10.68	&	1.2	$\pm	0.24	$	&	0.56	$\pm	0.11	$	&	1.98	&	1.00	&	sf	\\
Cl1138-11.0-1000*	&	174.5411	&	-11.5718	&	0.5795	&	8.85	&	0.86	&	0.75	&	0.71	&	-0.45	&	sf	\\						
Cl1138-11.0-1645*	&	174.5423	&	-11.5607	&	0.6136	&	10.92	&	11.80	&	9.43	&	1.60	&	1.14    &	q	\\						
Cl1138-11.0-1403	&	174.5429	&	-11.5647	&	0.5186	&	8.23	&	1.01	$\pm	0.1	$	&	0.67	$\pm	0.07	$	&	0.68	&	0.26	&	sf	\\
Cl1138-11.0-1138*	&	174.5438	&	-11.5694	&	0.6076	&	9.50	&	2.85	&	2.53	&	0.67	&	1.00 &	sf	\\						
Cl1138-11.0-1420*	&	174.5453	&	-11.5646	&	0.4503	&	9.60	&	1.93	&	0.69	&	1.39	&	1.07	&	sf	\\						
Cl1138-11.0-180	&	174.547	&	-11.59	&	0.6205	&	8.86	&	1.7	$\pm	0.11	$	&	1.21	$\pm	0.08	$	&	0.91	&	0.30	&	q	\\
Cl1138-11.0-1308*	&	174.5483	&	-11.5664	&	0.6046	&	8.74	&	0.45	&	0.3	&	0.91	&	0.24	&	sf	\\						
Cl1138-11.0-590*	&	174.5484	&	-11.5793	&	0.4568	&	10.51	&	1.53	&    0.52	&	1.90	& 1.06 & q	\\							
Cl1138-11.0-297*	&	174.5487	&	-11.5866	&	0.593	&	9.21	&	0.61	&	0.39	&	1.51	&	-0.20	&	q	\\						
Cl1138-11.0-1973	&	174.549	&	-11.5557	&	0.6994	&	8.38	&	1.12	$\pm	0.23	$	&	1.54	$\pm	0.32	$	&	0.21	&	0.59	&	sf	\\
Cl1138-11.0-1291*	&	174.5505	&	-11.5668	&	0.5875	&	9.75	&	0.57	&	0.35	&	1.79	&	0.38	&	q	\\						
Cl1138-11.0-824*	&	174.5524	&	-11.5749	&	0.5214	&	8.24	&	0.93	&	0.74	&	0.65	&	1.34    &	sf	\\						
Cl1138-11.0-264*	&	174.553	&	-11.588	&	0.5208	&	9.98	&	1.24	&	0.57	&	1.71	&	1.16	&	q	\\						
Cl1138-11.0-1791*	&	174.5531	&	-11.5581	&	0.4146	&	7.66	&	1.48	&	0.58	&	0.20	&	0.01	&	sf	\\						
Cl1138-11.0-429*	&	174.5533	&	-11.582	&	0.6441	&	8.78	&	0.96	&	1.08	&	0.51	&	0.05	&	sf	\\						
Cl1138-11.0-2141	&	174.5566	&	-11.5758	&	0.4094	&	8.26	&	0.78	$\pm	0.23	$	&	0.38	$\pm	0.11	$	&	1.02	&	1.98	&	sf	\\
Cl1138-11.0-778	&	174.5613	&	-11.5758	&	0.6987	&	8.66	&	1.51	$\pm	0.44	$	&	2.08	$\pm	0.60	$	&	0.48	&	0.24	&	sf	\\
Cl1138-11.0-2084*	&	174.5631	&	-11.5543	&	0.564	&	8.64	&	0.81	&	0.6	&	0.69	&	0.73  &	sf	\\						
Cl1138-11.0-411*	&	174.5635	&	-11.5825	&	0.5627	&	9.10	&	0.56	&	0.32	&	1.16	&	0.23   &	sf	\\																					
															
\hline

 \end{tabular} \par
 \bigskip
 \caption{Information for field galaxies. 1. Pointing ID - GRIZLI Object ID. IDs ending with * are galaxies with S/N$_{H\alpha}$ $<$ 3. 2. Right Ascension 3. Declination 4. Redshift 5. Stellar Mass 6. Uncorrected H$\alpha$ flux in cgs units. S/N $<3$ detections are listed at the 3$\sigma$ upper limit. 7. Star-formation rate. S/N $<3$  detections are listed at the 3$\sigma$ upper limit. 8. {\em U -- V} rest-frame color 9. {\em V -- J} rest-frame color 10. {\em UVJ} classification based on \protect\cite{will09} where sf and q represent star-forming and quiescent, respectively.}
 \label{tab:field}
\end{center}
\end{table*}

\begin{table*}
 \contcaption{Field Galaxies}
 \label{tab:continued}
 \begin{tabular}{lccccccccc}
   \hline

 Object ID & R.A. & Dec & z   & Stellar Mass  & flux$_{H\alpha}$ & SFR$_{H\alpha}$ & {\em U -- V} & {\em V -- J} & {\em UVJ} \\
  & Deg. & Deg. &  & $\log_{10}$(M$_*$/$M\textsubscript{\(\odot\)}$) & 10$^{-16}${\rm erg~s$^{-1}$cm$^{-2}$}  &   $M\textsubscript{\(\odot\)}$yr$^{-1}$ & AB & AB & Classification \\
  (1)       &   (2)     & (3)       & (4)       & (5)             & (6) &(7) & (8) & (9) & (10)  \\
  \hline
Cl1138-11.0-851*	&	174.5652	&	-11.5746	&	0.4397	&			9.88	&	1.6	&	0.46	&	1.74	&	0.97	&	q	\\				
Cl1138-11.0-430*	&	174.5659	&	-11.5819	&	0.6222	&			8.59	&	0.9	&	0.84	&	0.20	&	0.93	&	sf	\\				
Cl1138-11.0-673	&	174.5676	&	-11.5779	&	0.6286	&	9.4	&	3.6	$\pm	0.19	$	&	5.63	$\pm	0.29	$	&	1.17	&	1.20	&	q	\\
Cl1138-11.1-409*	&	174.6811	&	-11.5603	&	0.502	&			9.32	&	1.74 &	1.45 &	1.28   & 1.05	&	sf	\\					
Cl1138-11.1-120*	&	174.6827	&	-11.5823	&	0.6227	&			9.49	&	2.44	&	1.74	&	1.31	&	0.03	&	q	\\				
Cl1138-11.1-047	&	174.6884	&	-11.5869	&	0.5494	&	9.89	&	5.12	$\pm	0.29	$	&	11.68	$\pm	0.66	$	&	1.32	&	2.73	&	q	\\
Cl1138-11.1-157*	&	174.6907	&	-11.5797	&	0.5831	&			8.6	&	2.56	&	2.26	&	0.21	&	0.60	&	sf	\\				
Cl1138-11.1-364	&	174.6985	&	-11.5631	&	0.4051	&	8.75	&	2.88	$\pm	0.71	$	&	1.37	$\pm	0.34	$	&	0.64	&	2.30	&	sf	\\
Cl1138-11.1-411	&	174.7007	&	-11.56	&	0.4531	&	9	&	0.6	$\pm	0.17	$	&	0.26	$\pm	0.07	$	&	0.80	&	0.98	&	sf	\\
Cl1138-11.1-193	&	174.703	&	-11.577	&	0.6282	&	10	&	2.8	$\pm	0.54	$	&	4.38	$\pm	0.84	$	&	1.23	&	1.13	&	sf	\\
Cl1138-11.1-273	&	174.704	&	-11.5697	&	0.6198	&	9.89	&	1.45	$\pm	0.24	$	&	4.44	$\pm	0.73	$	&	1.42	&	1.91	&	sf	\\
Cl1138-11.1-162*	&	174.7157	&	-11.5794	&	0.5771	&			9.72	&	8.97	&	7.73	&	0.84	&	0.68	&	sf	\\				
Cl1227-11.1-330*	&	186.9512	&	-11.6522	&	0.5491	&			8.78	&	1.19	&	0.91	&	0.38	&	-0.69	&	sf	\\				
Cl1227-11.0-382	&	186.959	&	-11.6475	&	0.5286	&	9.14	&	2.8	$\pm	0.26	$	&	1.95	$\pm	0.18	$	&	0.72	&	-0.22	&	q	\\
Cl1227-11.0-402	&	186.9637	&	-11.6452	&	0.5272	&	8.97	&	2.59	$\pm	0.13	$	&	1.79	$\pm	0.09	$	&	0.69	&	0.56	&	sf	\\
Cl1227-11.1-419*	&	186.9669	&	-11.6437	&	0.443	&			8.5	&	0.64	&	0.29	&	0.21	&	0.43	&	sf	\\				
Cl1227-11.0-519*	&	186.9669	&	-11.5863	&	0.5679	&			8.77	&	1.16	&	0.96	&	0.59	&	0.04	&	sf	\\				
Cl1227-11.0-413	&	186.9706	&	-11.6444	&	0.4342	&	9.48	&	0.68	$\pm	0.21	$	&	0.38	$\pm	0.12	$	&	1.36	&	1.38	&	q	\\
Cl1227-11.0-196	&	186.971	&	-11.6092	&	0.5473	&	9.17	&	1.76	$\pm	0.24	$	&	1.33	$\pm	0.18	$	&	0.82	&	0.24	&	q	\\
Cl1227-11.0-166	&	186.9742	&	-11.6632	&	0.5755	&	10.2	&	1.76	$\pm	0.18	$	&	4.50	$\pm	0.46	$	&	1.51	&	1.62	&	sf	\\
Cl1227-11.0-305*	&	186.9755	&	-11.6016	&	0.5086	&			9.77	&	1	&	0.43	&	1.81	&	1.22	&	q	\\				
Cl1227-11.1-396*	&	186.9784	&	-11.6464	&	0.5519	&			11.02	&	2.23	&	1.16	&	2.03	&	1.22	&	q	\\				
Cl1227-11.1-138	&	186.9822	&	-11.6656	&	0.4939	&	10.54	&	2	$\pm	0.45	$	&	0.81	$\pm	0.18	$	&	1.71	&	1.18	&	sf	\\
Cl1227-11.1-162	&	186.9869	&	-11.6119	&	0.5512	&	10.91	&	2.67	$\pm	0.32	$	&	2.24	$\pm	0.27	$	&	1.79	&	1.32	&	sf	\\
Cl1227-11.1-322	&	186.989	&	-11.6003	&	0.4108	&	8.85	&	1.51	$\pm	0.37	$	&	0.52	$\pm	0.13	$	&	0.73	&	0.84	&	sf	\\
Cl1227-11.1-113	&	186.9907	&	-11.6159	&	0.5745	&	11.07	&	1.97	$\pm	0.59	$	&	1.45	$\pm	0.43	$	&	2.04	&	1.46	&	q	\\
Cl1227-11.2-177*	&	187.052	&	-11.5288	&	0.5643	&			8.94	&	1.81	&	1.47	&	0.54	&	-0.34	&	sf	\\				
Cl1227-11.2-136	&	187.0564	&	-11.5315	&	0.5183	&	9.37	&	1.24	$\pm	0.25	$	&	0.57	$\pm	0.11	$	&	1.13	&	0.57	&	sf	\\
Cl1227-11.2-362	&	187.0578	&	-11.5183	&	0.4995	&	9.12	&	0.27	$\pm	0.09	$	&	0.16	$\pm	0.05	$	&	0.89	&	-0.19	&	sf	\\
Cl1227-11.2-279*	&	187.0604	&	-11.5238	&	0.5857	&			10.13	&	1.56	&	4.15	&	1.63	&	1.50	&	sf	\\				
Cl1227-11.2-411*	&	187.0614	&	-11.5141	&	0.6019	&			8.79	&	0.42	& 0.60	& 0.21 & 0.15 &	sf	\\					
Cl1227-11.2-035	&	187.0636	&	-11.5409	&	0.5963	&	9.62	&	3.47	$\pm	0.49	$	&	2.95	$\pm	0.42	$	&	0.89	&	0.81	&	sf	\\
Cl1227-11.2-247	&	187.0677	&	-11.5249	&	0.5311	&	9.02	&	1.77	$\pm	0.19	$	&	1.25	$\pm	0.13	$	&	0.69	&	0.29	&	sf	\\
Cl1227-11.2-152	&	187.0722	&	-11.5304	&	0.4723	&	8.63	&	1.73	$\pm	0.28	$	&	0.92	$\pm	0.15	$	&	0.62	&	0.32	&	sf	\\
Cl1227-11.2-208	&	187.0804	&	-11.5276	&	0.5349	&	9.37	&	3.25	$\pm	0.52	$	&	2.33	$\pm	0.37	$	&	0.84	&	0.41	&	sf	\\
Cl1227-11.2-259	&	187.0843	&	-11.5248	&	0.5124	&	11.17	&	4.07	$\pm	0.57	$	&	1.77	$\pm	0.25	$	&	2.05	&	1.23	&	sf	\\
Cl1227-11.2-146	&	187.0855	&	-11.5309	&	0.4908	&	9.34	&	2.56	$\pm	0.54	$	&	1.03	$\pm	0.22	$	&	0.96	&	0.53	&	sf	\\
Cl1227-11.2-204	&	187.0903	&	-11.5276	&	0.5165	&	10.72	&	2.73	$\pm	0.3	$	&	1.96	$\pm	0.22	$	&	1.81	&	1.31	&	sf	\\
Cl1301-11.3-418*	&	195.2246	&	-11.507	&	0.63	&			10.19	&	1.47	&	1.56	&	1.48	&	0.66	&	q	\\				
Cl1301-11.3-278*	&	195.2269	&	-11.5167	&	0.4351	&			9.63	&	1.80	& 0.82	& 1.02  & 0.60		&	sf	\\					
Cl1301-11.3-197*	&	195.2335	&	-11.5223	&	0.4574	&			9.59	&	1.93	&	0.65	&	1.57	&	1.08	&	q	\\				
Cl1301-11.3-410	&	195.2339	&	-11.5075	&	0.5476	&	9.16	&	0.76	$\pm	0.16	$	&	0.58	$\pm	0.12	$	&	0.77	&	0.45	&	q	\\
Cl1301-11.3-311	&	195.2408	&	-11.5145	&	0.6833	&	9.07	&	2.95	$\pm	0.23	$	&	3.84	$\pm	0.30	$	&	0.25	&	0.70	&	sf	\\
Cl1301-11.3-176	&	195.2446	&	-11.5234	&	0.5353	&	8.85	&	2.46	$\pm	0.2	$	&	1.77	$\pm	0.14	$	&	0.97	&	-0.31	&	sf	\\
Cl1301-11.3-095	&	195.2452	&	-11.5302	&	0.4281	&	8.89	&	1.78	$\pm	0.39	$	&	0.75	$\pm	0.16	$	&	0.83	&	0.40	&	sf	\\
Cl1301-11.3-091	&	195.2487	&	-11.5304	&	0.5213	&	8.89	&	1.33	$\pm	0.15	$	&	0.82	$\pm	0.09	$	&	0.53	&	0.80	&	sf	\\
Cl1301-11.3-123	&	195.2489	&	-11.5279	&	0.5358	&	9.53	&	2.37	$\pm	0.42	$	&	1.70	$\pm	0.30	$	&	0.75	&	0.66	&	sf	\\
Cl1301-11.3-470	&	195.2491	&	-11.5031	&	0.5178	&	10.15	&	4.3	$\pm	0.54	$	&	2.85	$\pm	0.36	$	&	0.89	&	0.41	&	sf	\\
Cl1301-11.3-446*	&	195.2521	&	-11.5054	&	0.5228	&			9.57	&	0.59	&	0.28	&	1.15	&	0.33	&	sf	\\				
Cl1301-11.3-344	&	195.2549	&	-11.5127	&	0.5986	&	9.71	&	4.32	$\pm	0.35	$	&	3.70	$\pm	0.30	$	&	0.73	&	0.84	&	sf	\\
Cl1301-11.3-085*	&	195.2631	&	-11.5306	&	0.5219	&			8.77	&	1.09	&	0.67	&	0.53	&	0.83	&	sf	\\				
Cl1301-11.1-176	&	195.3304	&	-11.549	&	0.4339	&	9.06	&	0.39	$\pm	0.1	$	&	0.12	$\pm	0.03	$	&	0.29	&	-0.15	&	sf	\\
Cl1301-11.1-242	&	195.3365	&	-11.5452	&	0.5238	&	9.07	&	2.54	$\pm	0.18	$	&	1.73	$\pm	0.12	$	&	0.50	&	0.63	&	sf	\\
Cl1301-11.1-441	&	195.3429	&	-11.5251	&	0.5795	&	9.6	&	4.43	$\pm	0.13	$	&	3.51	$\pm	0.10	$	&	0.88	&	0.74	&	sf	\\
Cl1301-11.3-446*	&	195.345	&	-11.5216	&	0.4046	&			9.07	&	1.5	&	0.55	&	0.78	&	0.09	&	sf	\\				
Cl1301-11.0-044	&	195.351	&	-11.5609	&	0.4224	&	8.87	&	1.98	$\pm	0.52	$	&	0.81	$\pm	0.21	$	&	0.69	&	0.28	&	sf	\\
Cl1301-11.1-262	&	195.3524	&	-11.5436	&	0.5149	&	8.97	&	0.95	$\pm	0.09	$	&	0.62	$\pm	0.06	$	&	0.80	&	0.50	&	sf	\\
 \hline

 \end{tabular} \par
 \bigskip
\contcaption{Information for field galaxies. 1. Pointing ID - GRIZLI Object ID. IDs ending with * are galaxies with S/N$_{H\alpha}$ $<$ 3. 2. Right Ascension 3. Declination 4. Redshift 5. Stellar Mass 6. Uncorrected H$\alpha$ flux in cgs units. S/N $<3$ detections are listed at the 3$\sigma$ upper limit. 7. Star-formation rate. S/N $<3$  detections are listed at the 3$\sigma$ upper limit. 8. {\em U -- V} rest-frame color 9. {\em V -- J} rest-frame color 10. {\em UVJ} classification based on \protect\cite{will09} where sf and q represent star-forming and quiescent, respectively.}
 \label{tab:continued}
\end{table*}

\begin{table*}
 \contcaption{Field Galaxies}
 \label{tab:continued}
 \begin{tabular}{lccccccccc}
   \hline

 Object ID & R.A. & Dec & z   & Stellar Mass  & flux$_{H\alpha}$ & SFR$_{H\alpha}$ & {\em U -- V} & {\em V -- J} & {\em UVJ} \\
  & Deg. & Deg. &  & $\log_{10}$(M$_*$/$M\textsubscript{\(\odot\)}$) & 10$^{-16}${\rm erg~s$^{-1}$cm$^{-2}$}  &   $M\textsubscript{\(\odot\)}$yr$^{-1}$ & AB & AB & Classification \\
  (1)       &   (2)     & (3)       & (4)       & (5)             & (6) &(7) & (8) & (9) & (10)  \\
  \hline
Cl1301-11.1-356	&	195.3558	&	-11.5368	&	0.451	&	7.94	&	2.26	$\pm	0.30	$	&	0.98	$\pm	0.13	$	&	-0.11	&	0.91	&	sf	\\
Cl1301-11.2-587*	&	195.3655	&	-11.691	&	0.6237	&			9.84	&	1.90	&	1.57	&	1.10	&	0.78	&	sf	\\				
Cl1301-11.0-067*	&	195.3666	&	-11.6366	&	0.5524	&			8.84	&	2.07	& 1.60	&	0.58	&	0.47	&	sf	\\					
Cl1301-11.2-605	&	195.3697	&	-11.6895	&	0.5304	&	11.23	&	2.14	$\pm	0.61	$	&	1.04	$\pm	0.30	$	&	2.13	&	1.23	&	q	\\
Cl1301-11.0-151*	&	195.37	&	-11.6294	&	0.5449	&			8.81	&	8.67	&	5.91	&	0.55	&	1.23	&	sf	\\				
Cl1301-11.2-486*	&	195.3727	&	-11.699	&	0.4046	&			9.68	&	3.52	&	0.87	&	2.01	&	1.29	&	q	\\				
Cl1301-11.0-265*	&	195.3755	&	-11.6184	&	0.6343	&			9.51	&	7.66	&	8.31	&	0.61	&	0.53	&	sf	\\				
Cl1301-11.2-319*	&	195.3757	&	-11.711	&	0.5816	&			10.35	&	2.60	&	1.54	&	1.98	&	1.20	&	q	\\				
Cl1301-11.2-641	&	195.3771	&	-11.6856	&	0.5177	&	9.95	&	3.77	$\pm	0.33	$	&	3.23	$\pm	0.28	$	&	1.48	&	1.35	&	sf	\\
Cl1301-11.0-231*	&	195.3782	&	-11.6208	&	0.5656	&			10.11	&	5.89	&	3.33	&	1.64	&	0.94	&	q	\\				
Cl1301-11.2-603*	&	195.3819	&	-11.6895	&	0.6242	&			10.5	&	0.78	&   0.57	&	1.95	&	1.02	&	q	\\					
Cl1301-11.2-514*	&	195.383	&	-11.6966	&	0.6524	&			9.38	&	0.32	&	0.37	&	0.59	&	-0.59	&	sf	\\				
Cl1301-11.0-096*	&	195.3865	&	-11.634	&	0.4045	&			9.75	&	15.00	&	5.49	&	1.45	&	0.72	&	q	\\				
Cl1301-11.2-685	&	195.3878	&	-11.6797	&	0.423	&	9.68	&	0.50	$\pm	0.16	$	&	0.14	$\pm	0.04	$	&	1.57	&	-0.39	&	sf	\\
Cl1301-11.0-099*	&	195.3879	&	-11.6337	&	0.5451	&			9.55	&	5.82	&	3.02	&	1.08	&	0.48	&	sf	\\				
Cl1301-11.2-262	&	195.388	&	-11.6186	&	0.5329	&	9.14	&	5.42	$\pm	0.22	$	&	3.85	$\pm	0.16	$	&	0.54	&	0.02	&	sf	\\
																									
  \hline

 \end{tabular} \par
 \bigskip
\contcaption{Information for field galaxies. 1. Pointing ID - GRIZLI Object ID. IDs ending with * are galaxies with S/N$_{H\alpha}$ $<$ 3. 2. Right Ascension 3. Declination 4. Redshift 5. Stellar Mass 6. Uncorrected H$\alpha$ flux in cgs units. S/N $<3$ detections are listed at the 3$\sigma$ upper limit. 7. Star-formation rate. S/N $<3$  detections are listed at the 3$\sigma$ upper limit. 8. {\em U -- V} rest-frame color 9. {\em V -- J} rest-frame color 10. {\em UVJ} classification based on \protect\cite{will09} where sf and q represent star-forming and quiescent, respectively.}
 \label{tab:continued}
\end{table*}


\bsp	
\label{lastpage}
\end{document}